\documentclass[annual]{acmsiggraph}

\usepackage{mathptmx}
\usepackage{amsmath}
\usepackage{courier}
\usepackage{graphicx}	
\usepackage{subfigure} 
\usepackage{color} 		
\usepackage{listings} 
\usepackage{textcomp} 
\usepackage{times}
\usepackage[T1]{fontenc}
\usepackage{ae,aecompl}
\usepackage{textcomp}
\usepackage{aeguill}

\TOGonlineid{45678}
\TOGvolume{0}
\TOGnumber{0}
\TOGarticleDOI{1111111.2222222}
\TOGprojectURL{}
\TOGvideoURL{}
\TOGdataURL{}
\TOGcodeURL{}

\title{Sparse Norm Filtering}

\author{Chengxi Ye$^{\dag}$, Dacheng Tao$^{\ddag}$, Mingli Song$^{\S}$, David W. Jacobs$^{\dag}$, Min Wu$^{\dag}$\\$^{\dag}$Department of Computer Science, University of Maryland\\ $^{\ddag}$Centre for Quantum Computation and Intelligent Systems, University of Technology Sydney\\ $^{\S}$College of Computer Science, Zhejiang University}
\pdfauthor{Anonymous}

\begin{document}

\teaser{
\centering
\subfigure[]
{\includegraphics[height=1.5in]{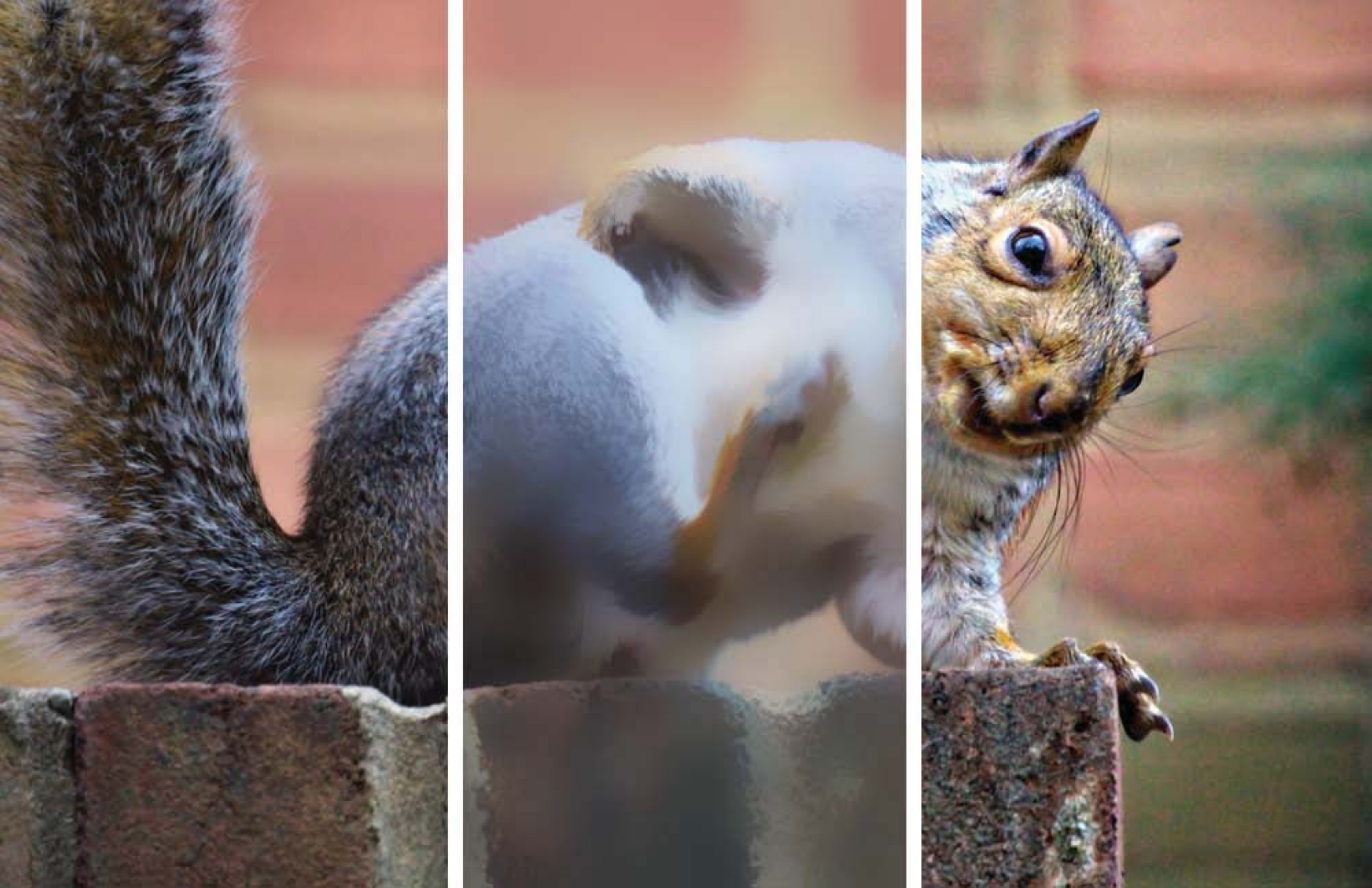}}
\subfigure[]{\includegraphics[height=1.5in]{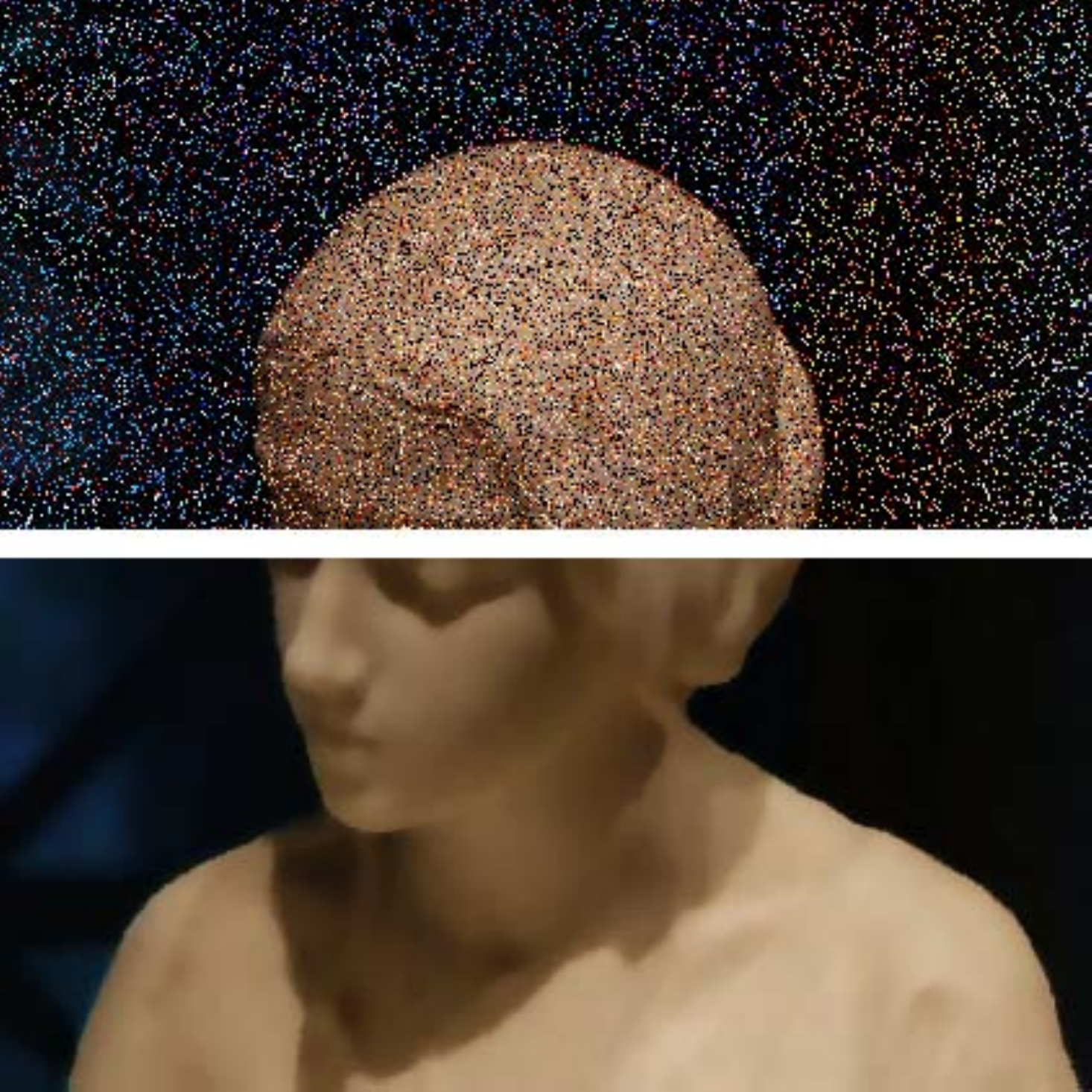}}
\subfigure[]{\includegraphics[height=1.5in]{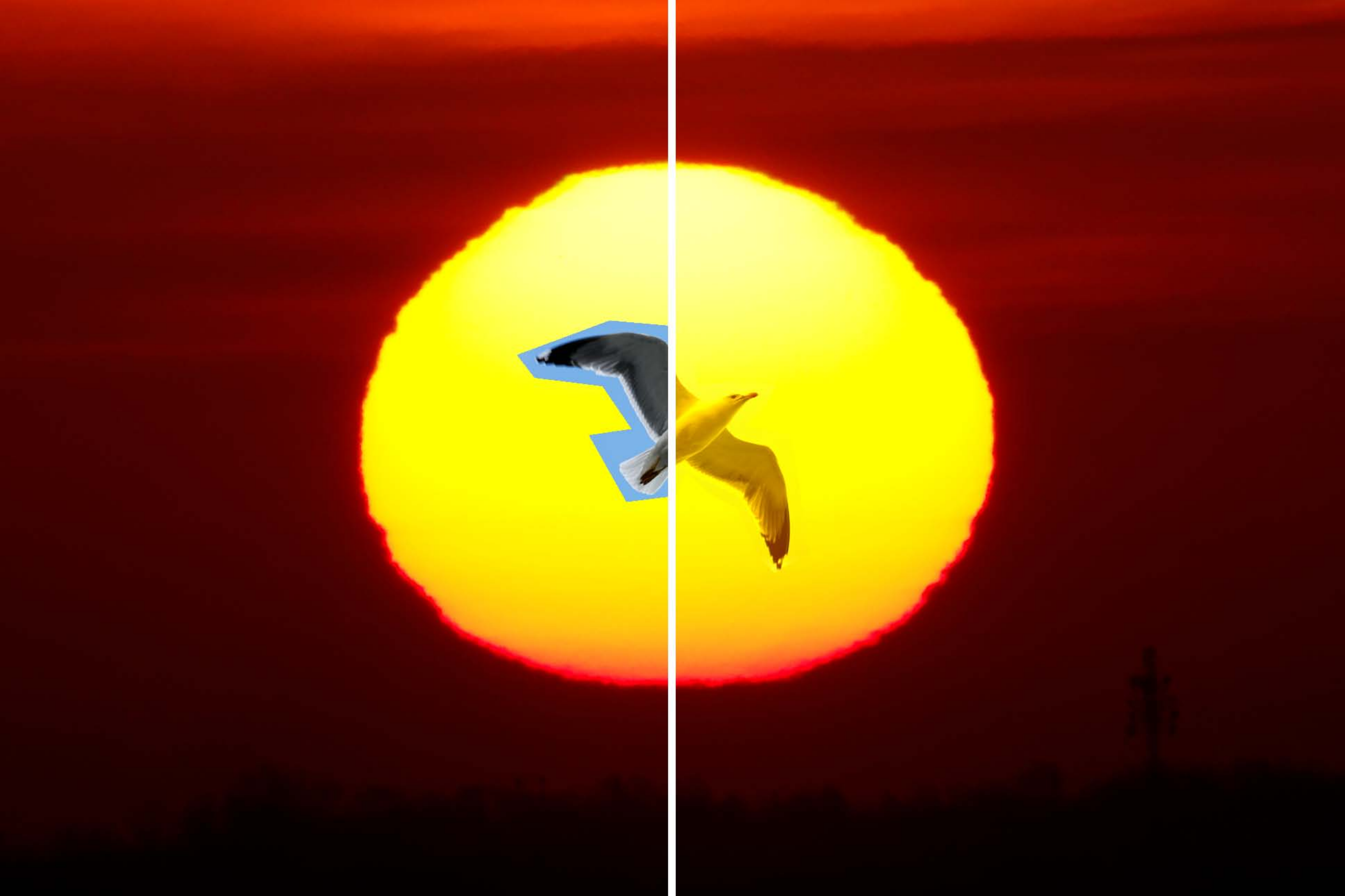}}
\caption{Examples of filtering results using different norms. (a) Left: original image Middle: smoothed image via minimizing $l^{0}$ energy. Right: sharpened image. (b) Up: image with pepper and salt noise. Down: smoothed result by minimizing the $l^{1}$  norm. (c) Left: drag-and-drop editing. Right: seamlessly editing using  $l^{2}$ norm filtering.}\label{fig1}
}
\maketitle

\begin{abstract}

Optimization-based filtering smoothes an image by minimizing a fidelity function and simultaneously preserves edges by exploiting a sparse norm penalty over gradients. It has obtained promising performance in practical problems, such as detail manipulation, HDR compression and deblurring, and thus has received increasing attentions in fields of graphics, computer vision and image processing. This paper derives a new type of image filter called sparse norm filter (SNF) from optimization-based filtering. SNF has a very simple form, introduces a general class of filtering techniques, and explains several classic filters as special implementations of SNF, e.g. the averaging filter and the median filter. It has advantages of being halo free, easy to implement, and low time and memory costs (comparable to those of the bilateral filter). Thus, it is more generic than a smoothing operator and can better adapt to different tasks. We validate the proposed SNF by a wide variety of applications including edge-preserving smoothing, outlier tolerant filtering, detail manipulation, HDR compression, non-blind deconvolution, image segmentation, and colorization.

\end{abstract}

\begin{CRcatlist}
  \CRcat{I.3.3}{Computer Graphics}{Pciture/Image Generation}{Display Algorithms}
  \CRcat{I.4.3}{Image Processing and Computer Vision}{Enhancement}{Grayscale Manipulation}
  \CRcat{I.4.9}{Image Processing and Computer Vision}{Application}{};
\end{CRcatlist}

\keywordlist


\section{Introduction}
\label{sec1}

Image filtering plays a fundamental role in image processing, computer graphics and computer vision, and has been widely used to reduce noise and extract useful image structures. In particular, edge-preserving smoothing operations have been studied for decades and have been proven to be critical for a wide variety of applications including blurring, sharpening, stylization and edge detection.

In general, existing edge-preserving filtering techniques can be classified into the following two groups: weighted average filtering and optimization-based filtering.

Well-known techniques of weighted average filtering includes anisotropic diffusion~\cite{Malik:1990:TPAMI,Sapiro:1998:TIP} and bilateral filtering~\cite{Manduchi:1998:ICCV}. Anisotropic diffusion uses the gradients of each pixel to guide a diffusion process and avoids blurring across edges. The bilateral filter can be regarded as a non-local diffusion process that uses pixel intensities within a neighborhood to guide the diffusion. Both approaches can be implemented using explicit weighted averaging. Acceleration of weighted filtering has been a research hotspot in recent years~\cite{Durand:2007:IJCV,Porikli:2008:CVPR,Ahuja:2009:CVPR,Tang:2010:ECCV,Oliveira:2012:TOG}.

Optimization-based filtering formulates edge preserving filtering as an optimization problem that consists of a fidelity term and a penalty term~\cite{Fatemi:1992:PD,Szeliski:2008:TOG,Lu:2011:IRN}. Edge preserving is enforced by introducing a sparse norm penalty on the gradients, thus the cost function is usually non-quadratic, and solving the system is more time consuming~\cite{Zhang:2008:SIAM} compared with weighted average filtering. Nevertheless, this framework often produces high quality results.

In this paper, we present a novel type of edge-preserving filter, called sparse norm filter~(SNF), derived from a sparse optimization problem. For each pixel, the filtering output minimizes its difference with its neighboring pixels; the penalty is defined by a sparse norm. Although SNF is closely related to and produces results as excellent as optimization-based filters, it is conceptually and computationally simpler than optimization-based filters.

SNF naturally preserves edges through the use of the sparse norm, and is capable of producing halo-free filtering effects, which is a desirable but lacking property of current weighted average filtering techniques. We demonstrate many of the other favorable properties of this simple and versatile approach to filtering via a wide variety of applications. Fig.~\ref{fig1} shows some applications of our filtering technique. Fig.~\ref{fig1}(a) demonstrates our smoothing and sharpening results that approximate the $l^{0}$  energy. Note that the filtering result preserves edges and does not introduce halos. Fig.~\ref{fig1}(b) shows our $l^{1}$  norm filtering effect to remove the pepper and salt noise. Fig.~\ref{fig1}(c) shows a new way of seamless editing enabled by the $l^{2}$ norm filtering. More detailed discussions on applications will be presented in the Applications section.

\section{Background}
\label{sec2}

One simple and classic way to smooth an image is to minimize the difference of each pixel with nearby ones, which can be formulated as

\begin{equation}
\min_{I_{i}^{new}}\sum_{j\in N_{i}}\left(I_{i}^{new}-I_{j}\right)^2.
\label{eq1}
\end{equation}

The solution of this optimization can be found by averaging the nearby pixels and is known as the box filter when we consider a square neighborhood

\begin{equation}
I_{i}^{new}=\frac{1}{|N_{i}|}\sum_{j\in N_{i}}{I_j}.
\label{eq2}
\end{equation}

The box filter can be calculated in linear time with the integral image technique~\cite{Porikli:2008:CVPR}. However, it does not preserve the salient structures or edges in an image.

Modern filtering techniques solve this problem by taking the weighted average of nearby pixels~\cite{Malik:1990:TPAMI,Sapiro:1998:TIP,Manduchi:1998:ICCV}. In the anisotropic diffusion framework, the neighborhood consists of the adjacent pixels, and the system has to be iterated tens of times to produce good smoothing result. Most recent filtering techniques consider a larger neighborhood consisting of tens or hundreds of neighboring pixels and the filtering is solved in one or a few rounds. Edges are preserved by constructing the weight matrix with the criterion that similar nearby pixels shall be given higher weights. As an example, the bilateral filter~\cite{Manduchi:1998:ICCV} uses the intensity to measure similarity and assigns weights by

\begin{equation}
I_{i}^{new}=\frac{\sum_{j\in N_i}w_{ij}I_{j}}{\sum_{j\in N_{N_{i}}}w_{ij}}, w_{ij}=\exp\left(-\frac{\left(j-i\right)^{2}}{2\sigma_{s}^{2}}\right)\exp\left(-\frac{\left(I_{j}-I_{i}\right)^{2}}{2\sigma_{r}^{2}}\right).
\label{eq3}
\end{equation}

Edge-preserving image smoothing can also be achieved by solving the following optimization problem

\begin{equation}
\min_{B}\|B-I\|^{q}+\lambda\|\nabla B\|^{p}.
\label{eq4}
\end{equation}

The penalty term $\lambda\|\nabla B\|^{p}$  controls the amount of smoothness of the output $B$ and the fidelity term $\|B-I\|^{q}$  controls the similarity with the input $I$. When $p=q=2$, the optimization problem is the well-known Tikhonov regularization~\cite{Tikhonov:1995:NMIP}. The explicit solution can be found by

\begin{equation}
B=\left(Id+\lambda\nabla ^{T}\nabla\right)^{-1}I.
\label{eq5}
\end{equation}

Since sparse norms have better tolerance for outliers than the $l^{2}$ norm, the optimization was later extended to total variation regularization with $p=1$~\cite{Fatemi:1992:PD} and even sparser versions~\cite{Szeliski:2008:TOG}\cite{Lu:2011:IRN} for edge-preserving purposes. Solving these non-quadratic optimizations is more time-consuming. Thus, variable splitting~\cite{Zhang:2008:SIAM} is usually exploited  to cast the original large optimization problem into several small sub-problems and alternatively minimize each of these sub-problems.

\section{The Sparse Norm Filter}
\label{sec3}

\subsection{Definition}
\label{sec3-1}

We propose SNF by generalizing~(\ref{eq1}) through allowing the original $l^{2}$ norm to be a fractional-norm. To preserve strong edges, we need to smooth the image while to tolerate the outlier pixels by assigning lower weights to them. This type of adaptive weighting ideas have been well explored in robust statistics~\cite{Sapiro:1998:TIP} and we achieve this by exploiting sparse norms. Then, SNF is defined by

\begin{equation}
\min_{I_{i}^{new}}\sum_{j\in N_{i}}|I_{i}^{new}-I_{j}|^{p}, 0<p<=2.
\label{eq6}
\end{equation}

Minimizing this non-quadratic cost function when $p<2$ is difficult. Especially when $p<1$, the cost function is non-convex and conventional gradient descent-based algorithms are easily trapped into local minima.

In this paper, we consider two approximation strategies. The first strategy iteratively exploits the weighted least square technique

\begin{eqnarray}
\sum_{j\in N_{i}}|I_{i}^{new}-I_{j}|^{p}&&\approx \sum_{j\in N_{i}}|I_{i}-I_{j}|^{p-2}\left(I_{i}^{new}-I_{j}\right)^{2}\nonumber\\
&&=\sum_{j\in N_{i}}w_{ij}\left(I_{i}^{new}-I_{j}\right)^{2}.
\label{eq7}
\end{eqnarray}

By taking the derivative, we find that the solution can be approximated by using weighted average filtering $I_{i}^{new}=\frac{\sum_{j\in N_i}w_{ij}I_{j}}{\sum_{j\in N_{i}}w_{ij}}$, with $w_{ij}=|I_{i}-I_{j}|^{p-2}$. This solution can be understood as one iteration of the anisotropic diffusion process, with the diffusity $w_{ij}$ calculated based at the current pixel intensity. This way of weighting naturally enforces fidelity with the input image. Similar to the anisotropic diffusion, we can iteratively update the diffusity once we update the image with this weighted average filtering result. In practice, like the bilateral filter, one iteration is usually sufficient because the diffusion is non-local. It is noteworthy that when  $I_{i}=I_{j}$, the weight goes to infinity. In practice, we can avoid this by setting a threshold and raising the pixel differences by/to the threshold. We can also modify the optimization by weighting pixels according to distance using a Gaussian-like weight. However, we observe that treating all neighboring pixels equally is good enough in practice.

Another strategy quantizes the solution into a set of discrete values. For each of these discrete values $Q_b$, we calculate $\sum_{j\in N_i}|Q_{b}-I_{j}|^{p}$ for each pixel $i$, which can be done efficiently by using the box filter. We compare the energy at each of these discrete values and select the minimum. Similar technique is used to approximate the median filter~\cite{Ahuja:2009:CVPR}. In this strategy, only discrete solutions at certain quantization levels are allowed because this approximation is based on brute force searching in the solution space. In practice, this strategy is more preferable when images are contaminated by outliers, e.g., the salt and pepper noise, we need a large number of iterations of the first strategy (if possible) to reach a suitable solution. For example, if the center pixel is noised and we conduct one iteration of filtering, we will assign high weights to similar pixels that are potentially noised. Thus, the obtained solution can be far away from suitable.

Both strategies are valuable. The first strategy makes the results look natural to the eye and its effect is similar to the bilateral filter, while the second strategy can filter out outliers and its effect is similar to the median filter. In all experiments except the outlier-tolerant-filtering, we choose the first strategy.

\subsection{Complexity}
\label{sec3-2}

The sparse norm filter benefits from off-the-shelf acceleration methods~\cite{Ahuja:2009:CVPR,Oliveira:2012:TOG}, and can be calculated in linear time $O(BN)$, where $B$ is the number of bins for quantization and the pixel number $N$. For a grayscale image, the brute force solution can be calculated with $B$ box filters if we quantize the intensities into $B$ bins. For the weighted average solution, we can similarly quantize the center pixel intensities (in the weight term) into $B$ bins~\cite{Ahuja:2009:CVPR}.  The weighted sum (numerator), and the sum of weights (denominator) can also be calculated using $B$ box filters, respectively. In comparison, an excellent state-of-art filtering technique~\cite{Tang:2010:ECCV} uses 7 box filters. In our Matlab implementation, the box filter takes 0.04 seconds per mega-pixel. The weighted average implementation of the sparse norm filter takes 0.5 seconds per mega-pixel when $B=4$, and 1 second with $B=8$. Experiments are carried out on an Intel i7 3610QM CPU with 8G memory. The pixel level operations will experience significant speedups in C++ implementations. For example, our direct single thread implementation of the box filter in C++ took 0.01 seconds per mega-pixel. Filtering based methods are faster than optimization-based methods~\cite{Szeliski:2008:TOG,Lu:2011:IRN}, as the latter take 2-4 seconds per mega-pixel in the same environment.

\subsection{Connections to Related Work}
\label{sec3-3}

Optimization-based filters~\cite{Fatemi:1992:PD,Szeliski:2008:TOG,Lu:2011:IRN,Elad:2002:TIP} have been widely used image enhancement tasks, e.g. denoising, edge preserving smoothing and deconvolution~\cite{Freeman:2007:TOG,Fergus:2009:NIPS}, and share the form of (\ref{eq4}), or in the pixel level notation

\begin{equation}
\min_{I_{i}^{new}}\|I_{i}^{new}-I_{i}\|^{q}+\lambda\|\nabla I_{i}^{new}\|^{p},0\leq p,q\leq2.
\label{eq8}
\end{equation}

The norm in the fidelity term is usually an $l^{2}$  norm in existing works. SNF simplifies~(\ref{eq8}) by integrating the fidelity term and the sparse norm penalty. By setting $\lambda=1$, changing the $l^{q}$ norm in the first term to the $l^{p}$ norm, and defining the neighborhood to contain the current pixel, we can write the first term in~(\ref{eq8}) into the second term and reduce (\ref{eq8}) to the form of (\ref{eq6}). Thus we establish the connection with optimization-based filtering.

It is noteworthy that in most optimization-based filters, the neighborhood is of small size and only contains the adjacent pixels. By contrast, SNF extends the concept of neighborhood in a non-local way to potentially include more pixels. We consider the difference of a pixel with all the pixels, not only those that are horizontal and vertical. SNF has advantages over optimization-based filter: a one pass approximation exists and is less likely to be trapped in poor local minima, thanks to the non-local diffusion.

In addition, SNF has a close relationship with several well-known filters. By setting $p=2$, SNF reduces to the averaging filter or box filter if we consider square neighborhoods. By setting $p=1$, SNF is equivalent to the median filter. This can be proved by taking the derivative on the original cost function. By setting $p=0$, the sparse norm filter is the dominant mode filter~\cite{Solomon:2010:TOG}.

\section{Application}
\label{sec4}

\subsection{Halo Free Edge Preserving Filtering and Detail Manipulation}
\label{sec4-1}

\begin{figure*}
\centering
\subfigure[]
{\includegraphics[height=1.5in]{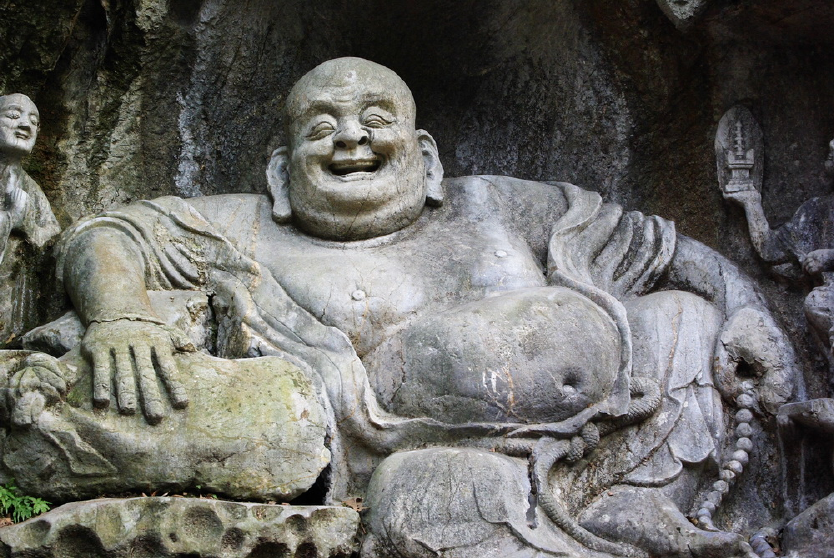}}
\subfigure[]{\includegraphics[height=1.5in]{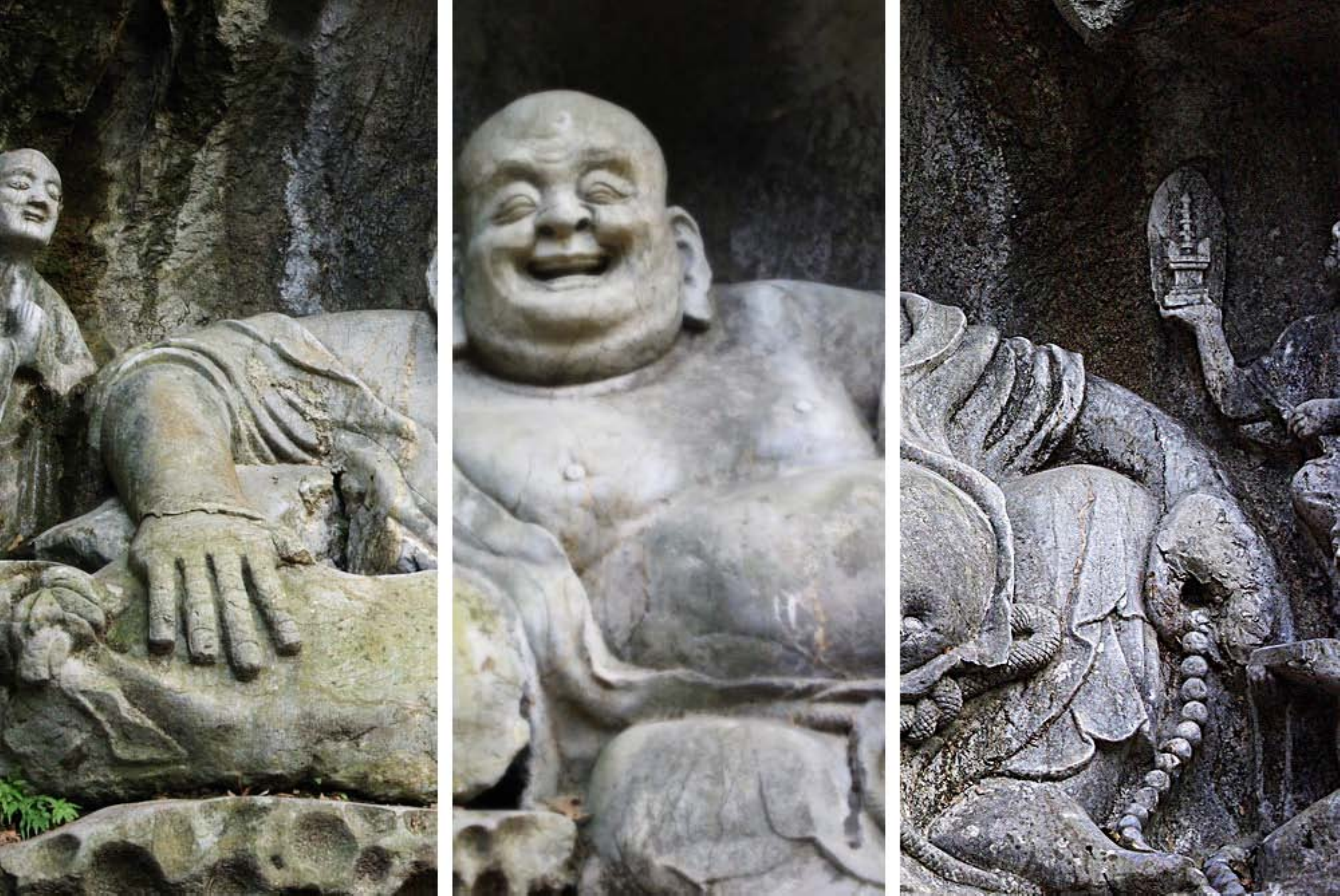}}
\subfigure[]{\includegraphics[height=1.5in]{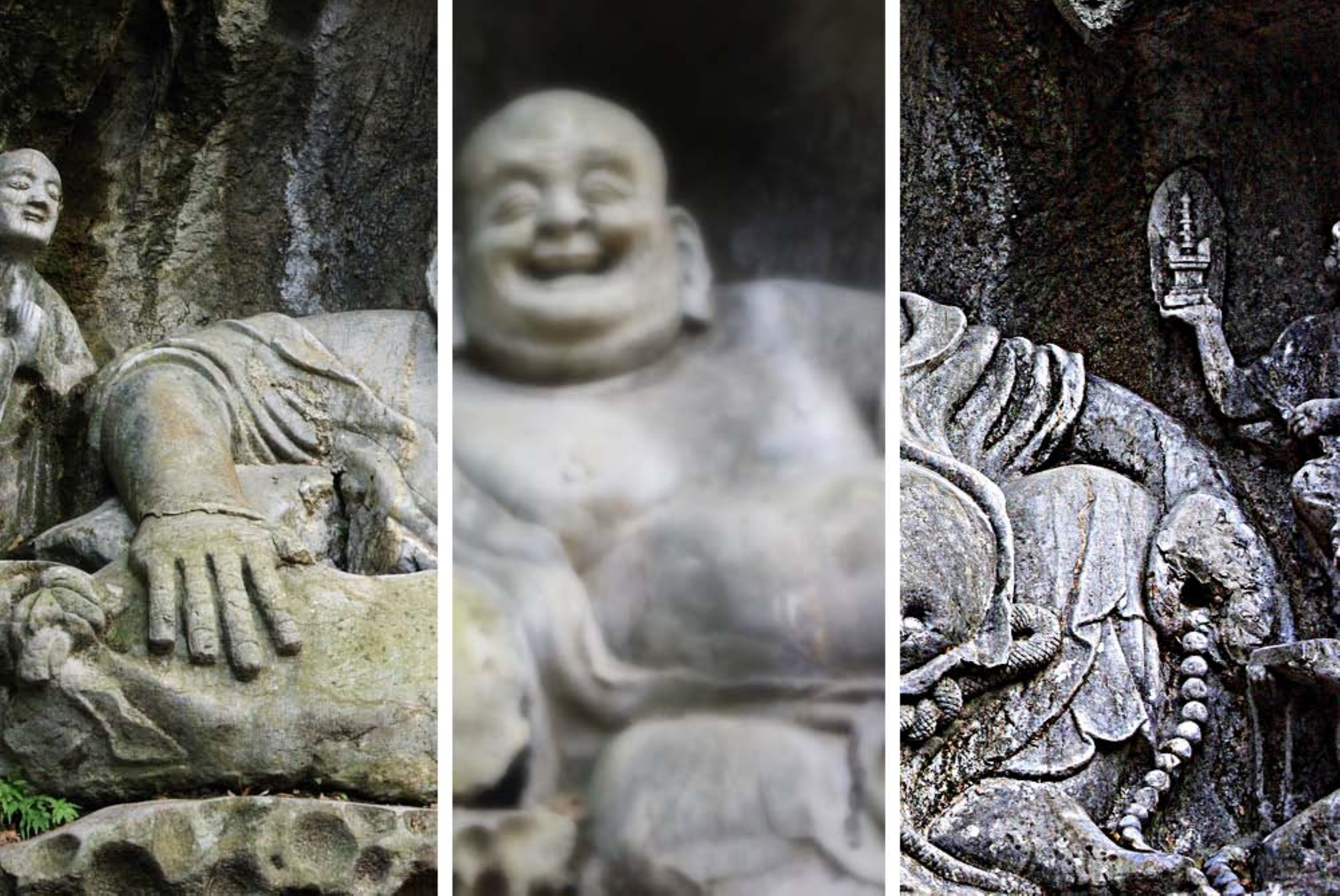}}
\subfigure[]{\includegraphics[height=1.5in]{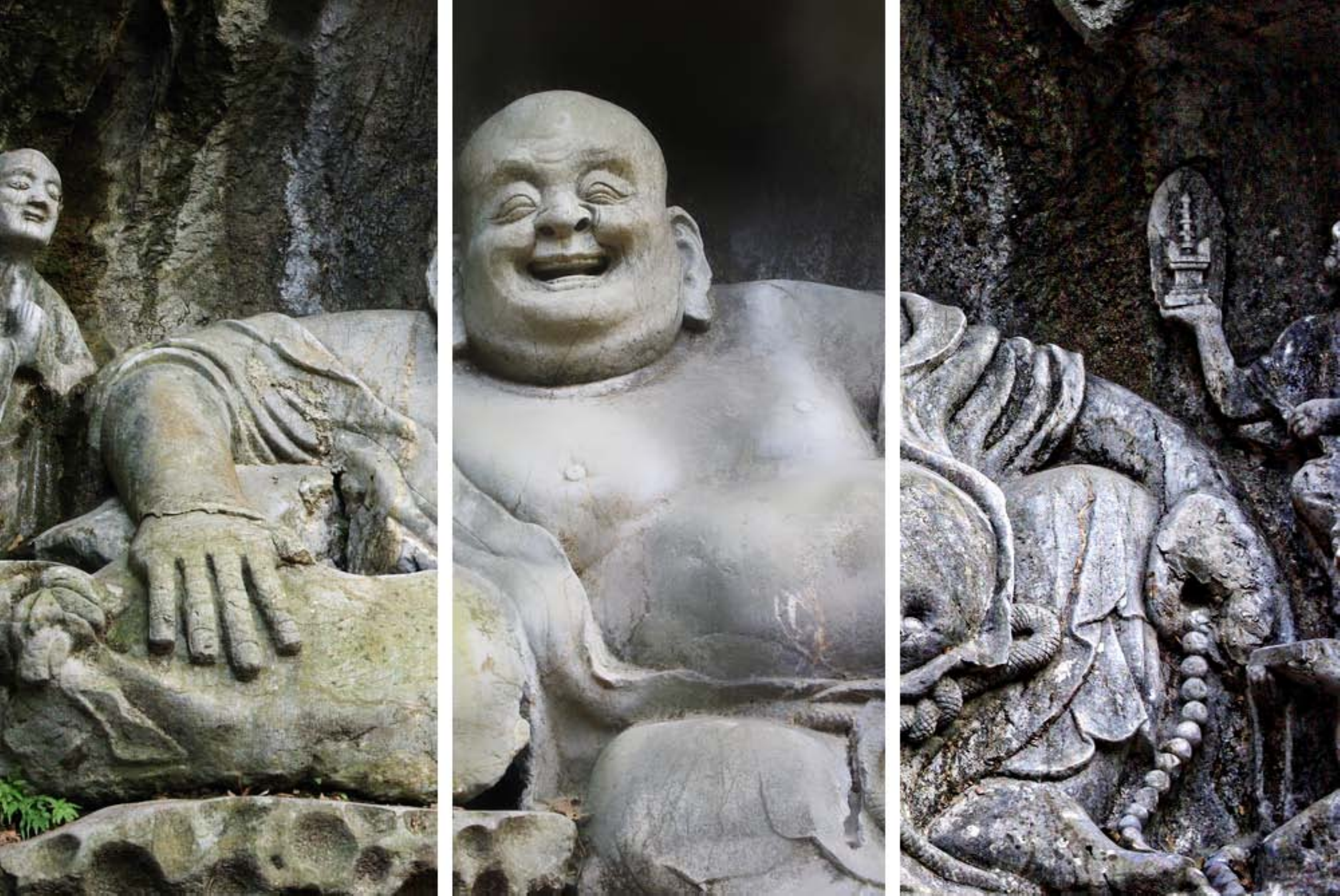}}
\subfigure[]{\includegraphics[height=1.5in]{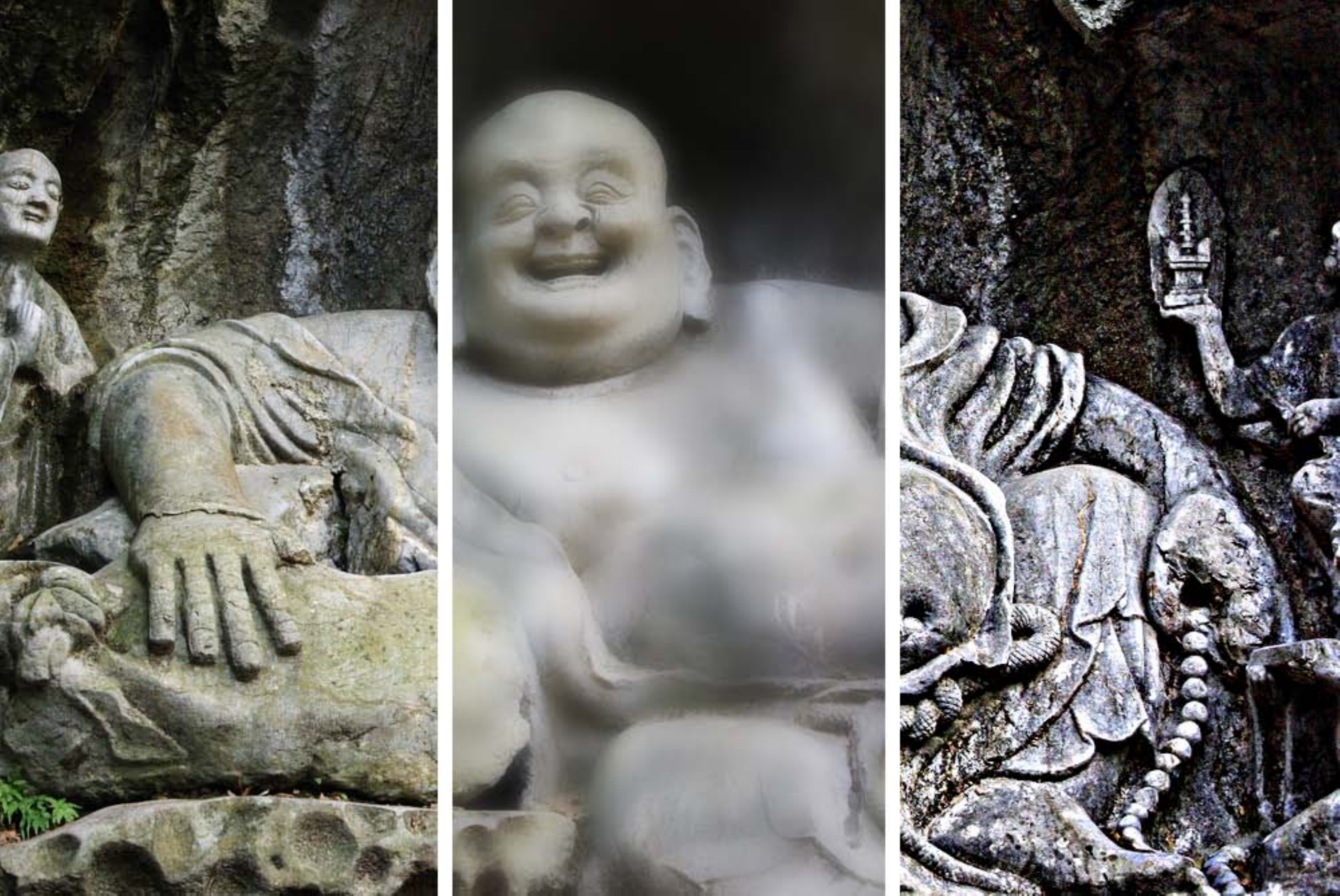}}
\subfigure[]{\includegraphics[height=1.5in]{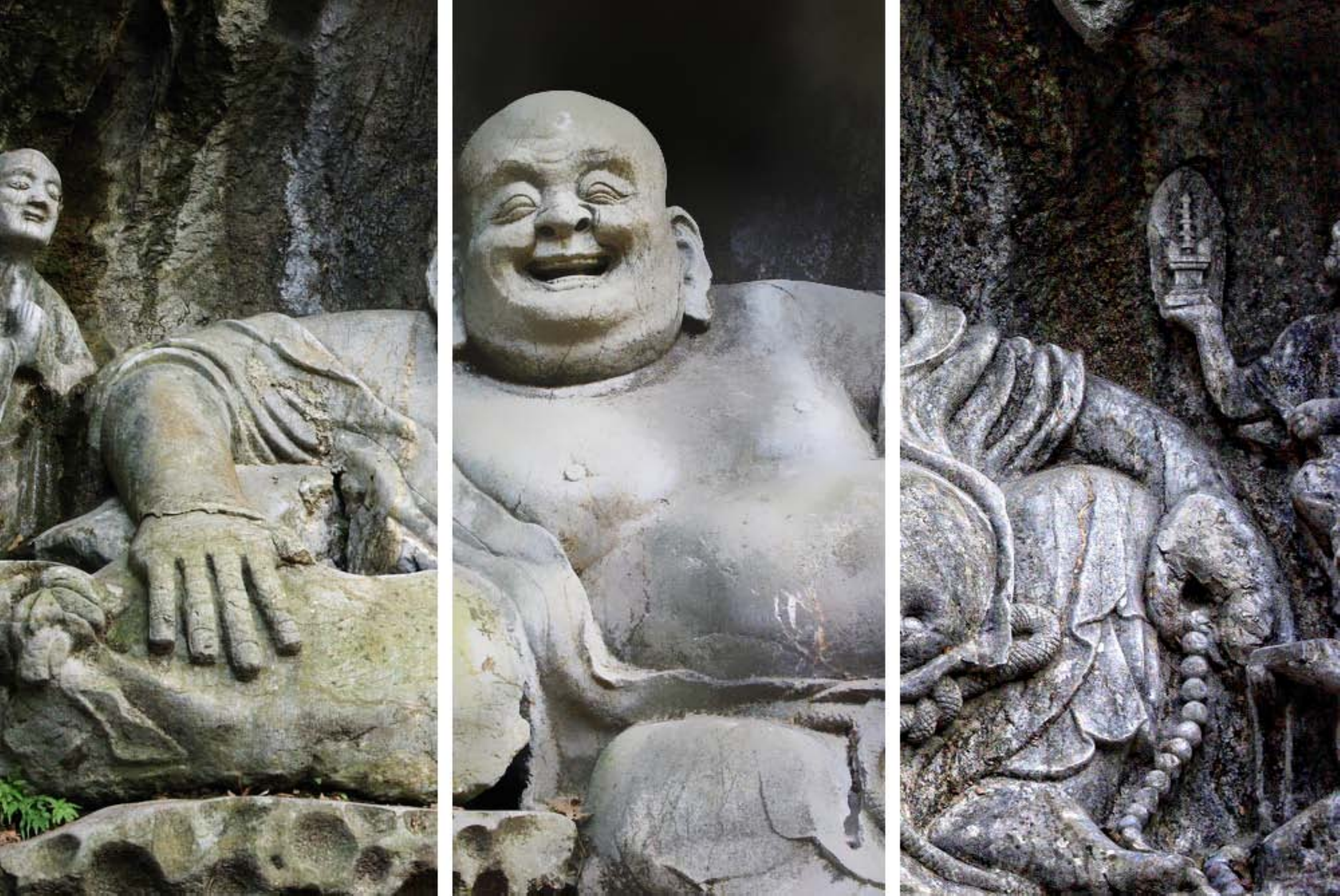}}
\subfigure[]{\includegraphics[height=1.5in]{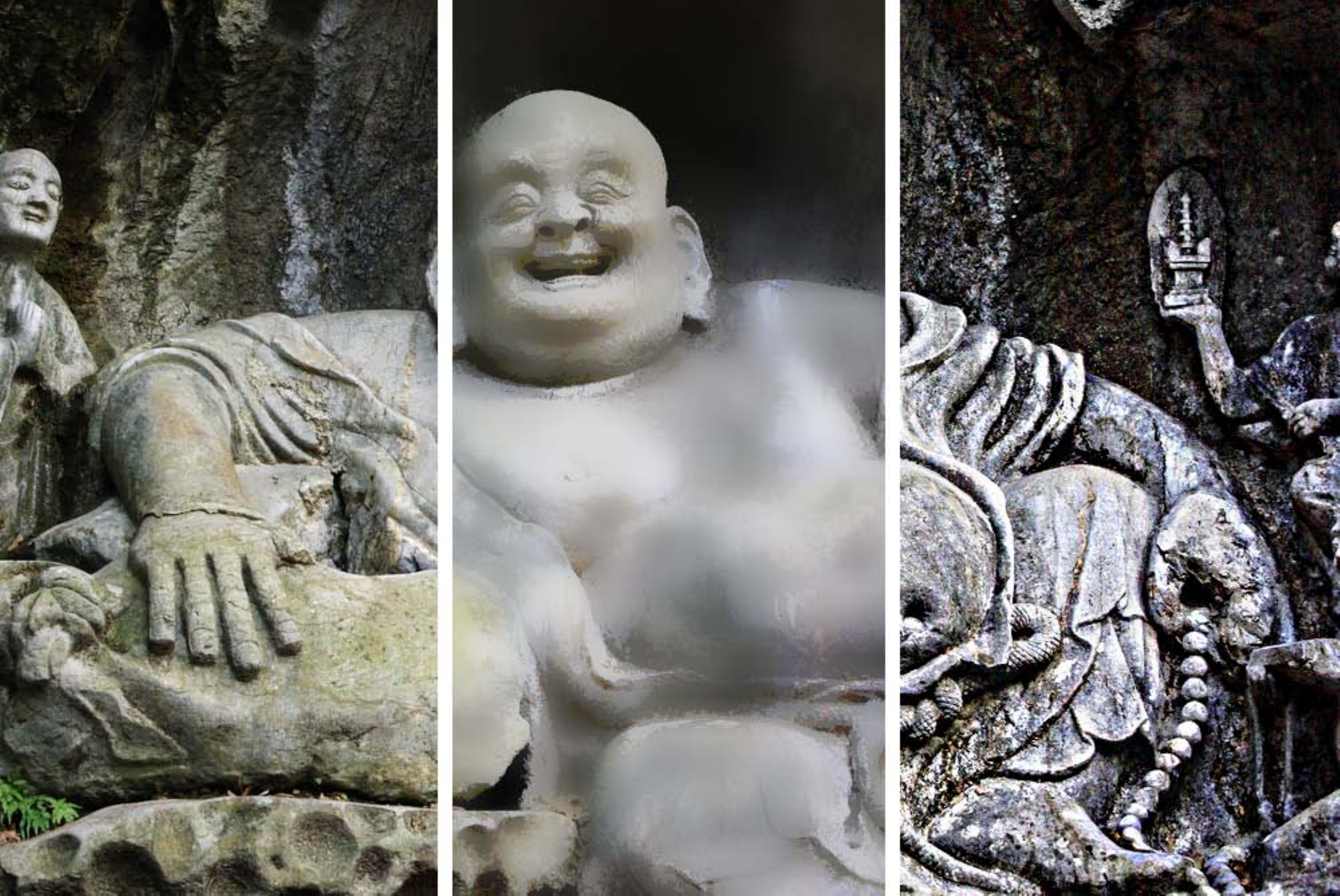}}
\subfigure[]{\includegraphics[height=1.5in]{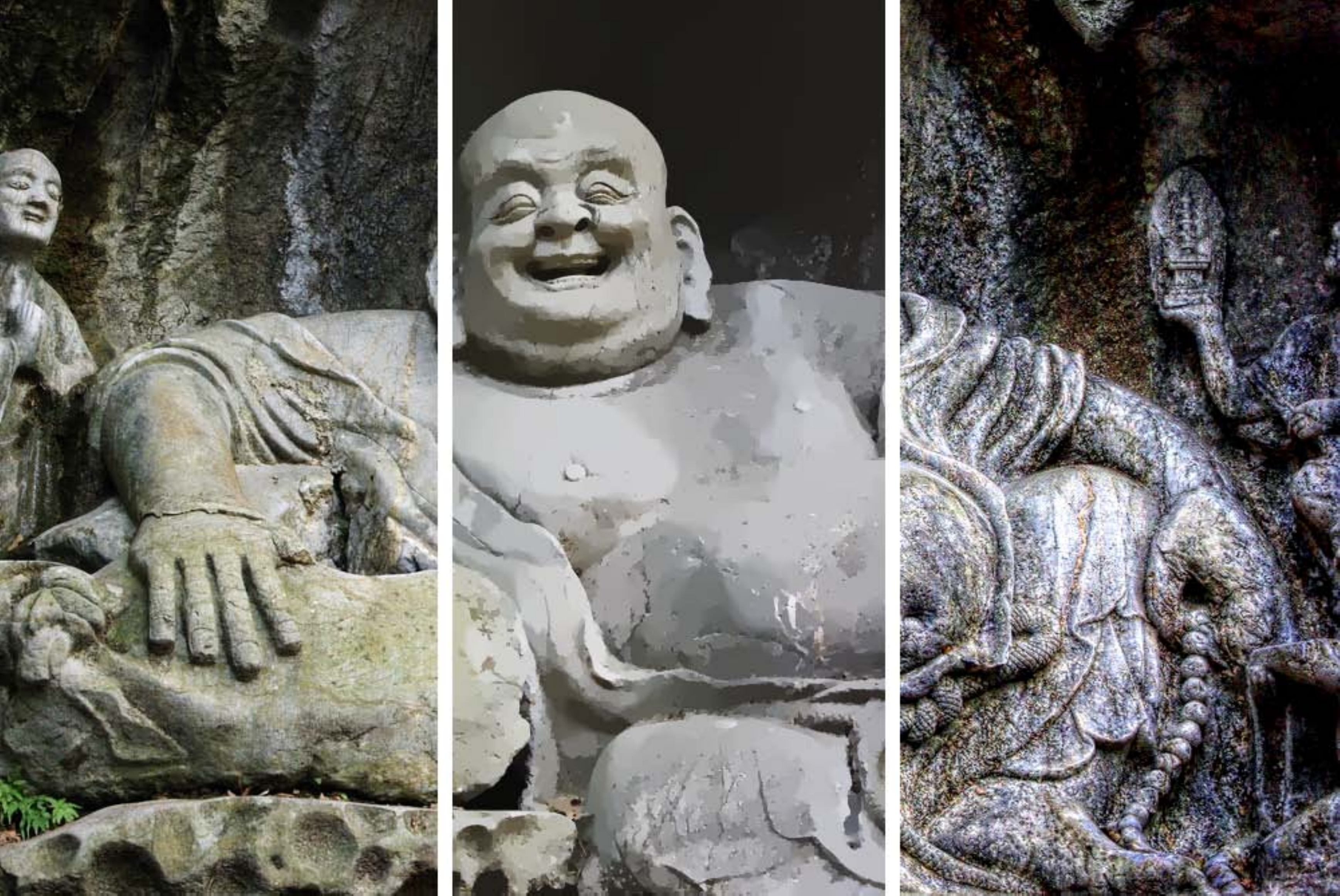}}
\subfigure[]{\includegraphics[height=1.5in]{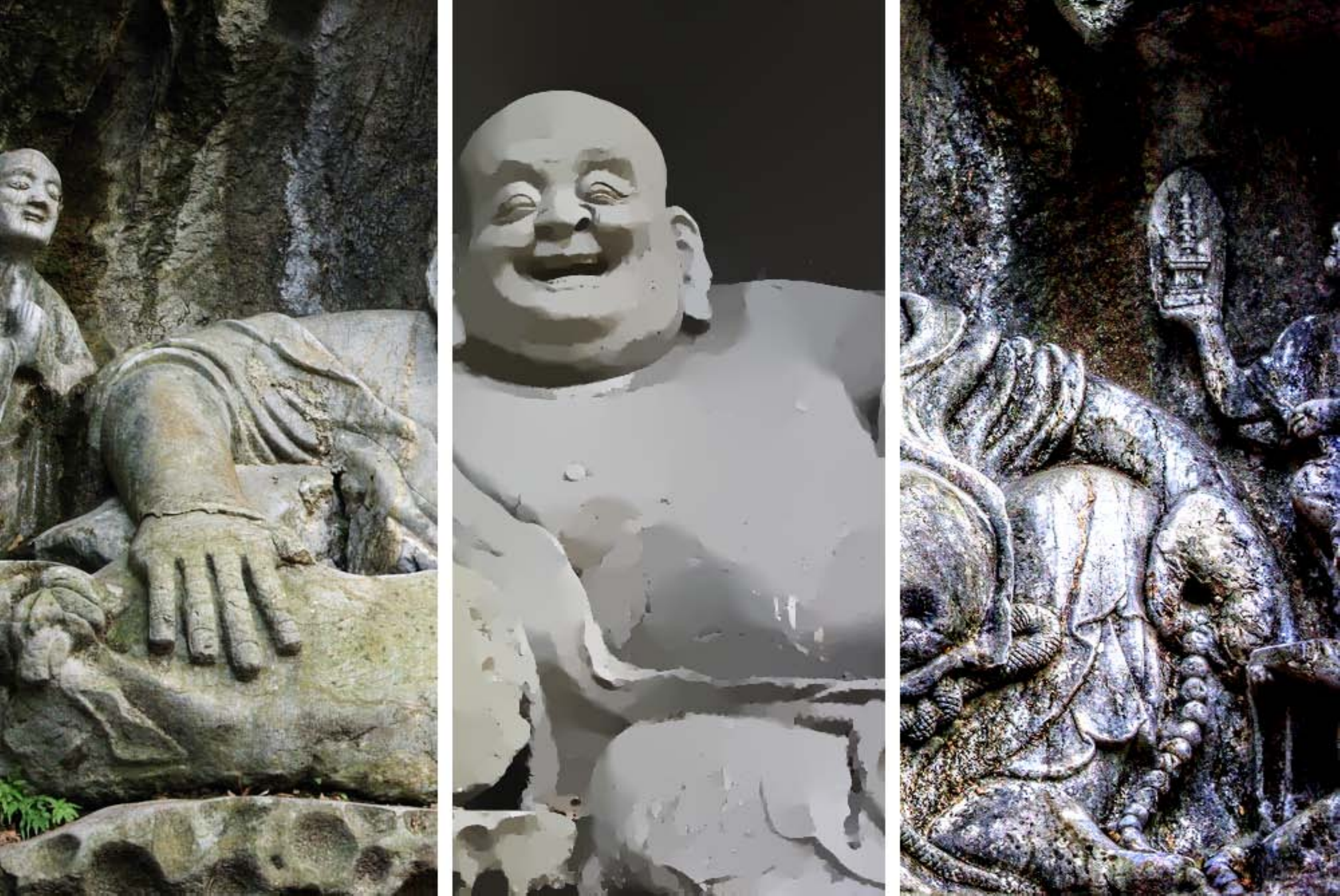}}
\subfigure[]{\includegraphics[height=1.5in]{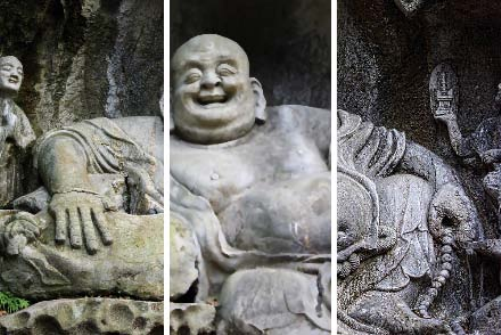}}
\subfigure[]{\includegraphics[height=1.5in]{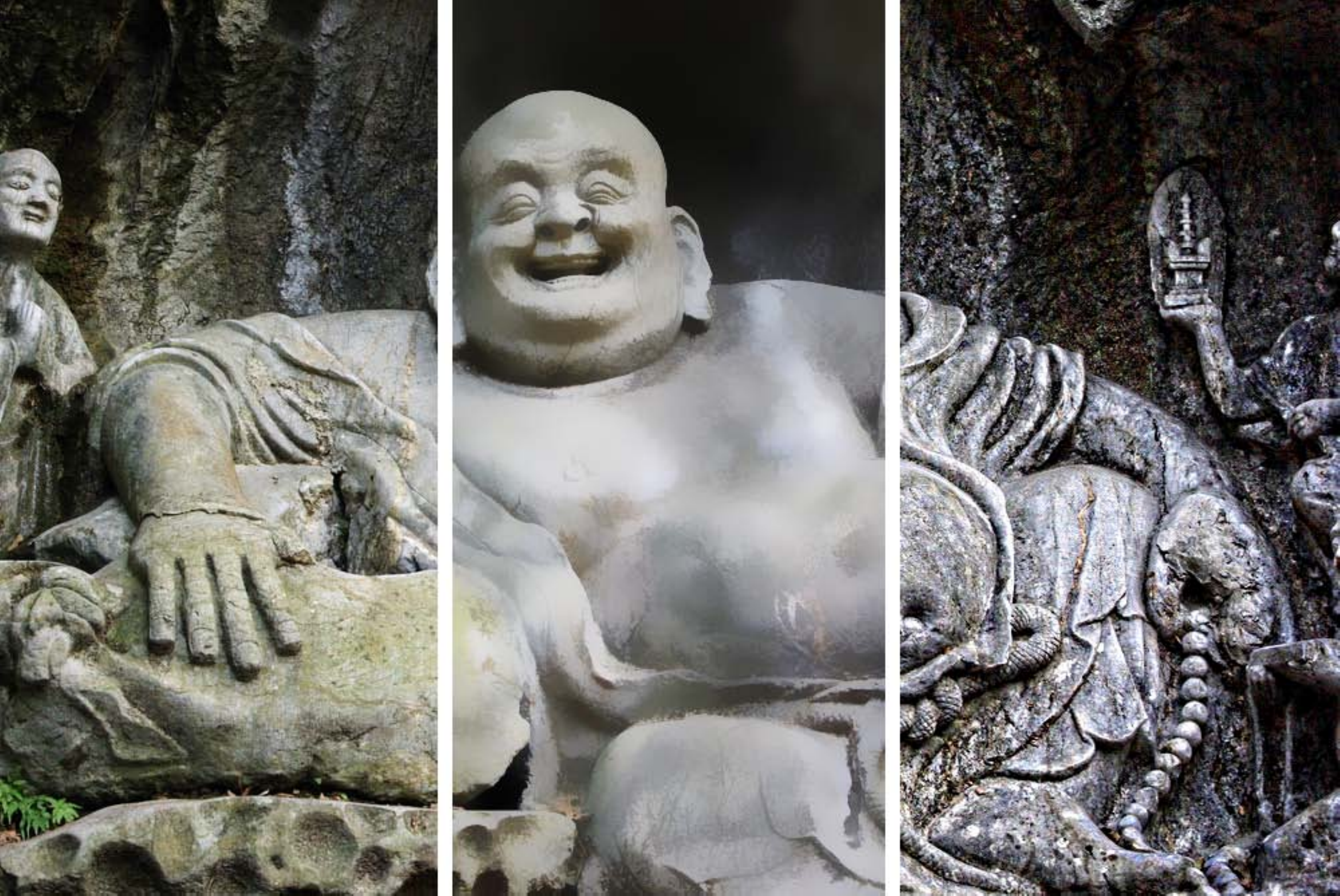}}
\subfigure[]{\includegraphics[height=1.5in]{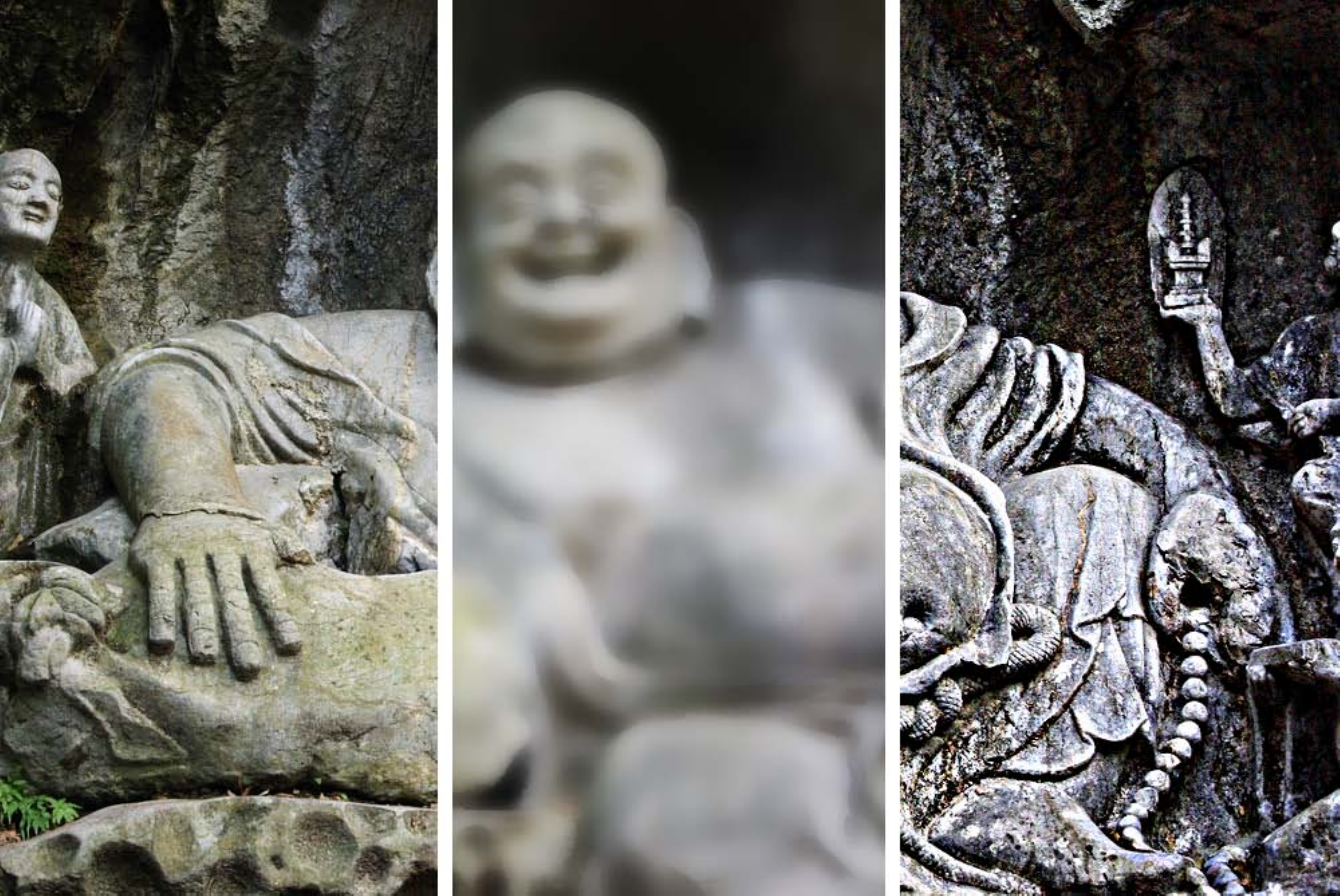}}
\caption{Filtering results. For each result, left: original image. Middle: smoothing result. Right: sharpening result. (a) Original image. (b) WLS [10] result $\lambda=1$, $p=0.1$. (c) WLS result, $\lambda=10$, $p=0.1$. (d) Guided image filter [7], $r=10$, $\epsilon=0.1^{2}$. (e) Guided image filter, $r=10$, $\sigma_{r}=0.1$. (f) Bilateral filter [3], $r=10$, $\sigma_{r}=0.1$. (g) Bilateral filter, $r=10$, $\sigma_{r}=0.3$. (h) $l^{0}$-smoothing [11], $\lambda=0.02$. (i) $l^{0}$-smoothing, $\lambda=0.05$. (j) SNF, $r=2$, $p=0.05$. (k) SNF, $r=10$, $p=0.2$. (l) SNF, $r=10$, $p=1.2$.}\label{fig2}
\end{figure*}

Explicit filtering techniques are known to create faint light rims along strong edges known as halo artifacts. This unrealistic effect has been widely discussed in~\cite{Tang:2010:ECCV,Szeliski:2008:TOG,Lu:2011:IRN}. This section shows that SNF can produce halo-free results. Fig~\ref{fig2} compares representative edge-preserving smoothing techniques. Although all the methods can produce high quality results, we find some tiny differences. Optimization-based smoothing algorithms~\cite{Szeliski:2008:TOG,Lu:2011:IRN} are more capable of producing halo-free looks, but the obtained results can occasionally be unexpected if the optimization is non-convex. In Fig.~\ref{fig2}(c) the edges look overly smoothed; the $l^{0}$-smoothing preserves edges perfectly but it also retains speckles. Traditional weighted average filtering techniques produce smoother looks, but tend to produce halos near strong edges. These halos also lead to unnatural transitions in sharpening.

\begin{figure*}
\centering
\subfigure[]
{\includegraphics[height=1.5in]{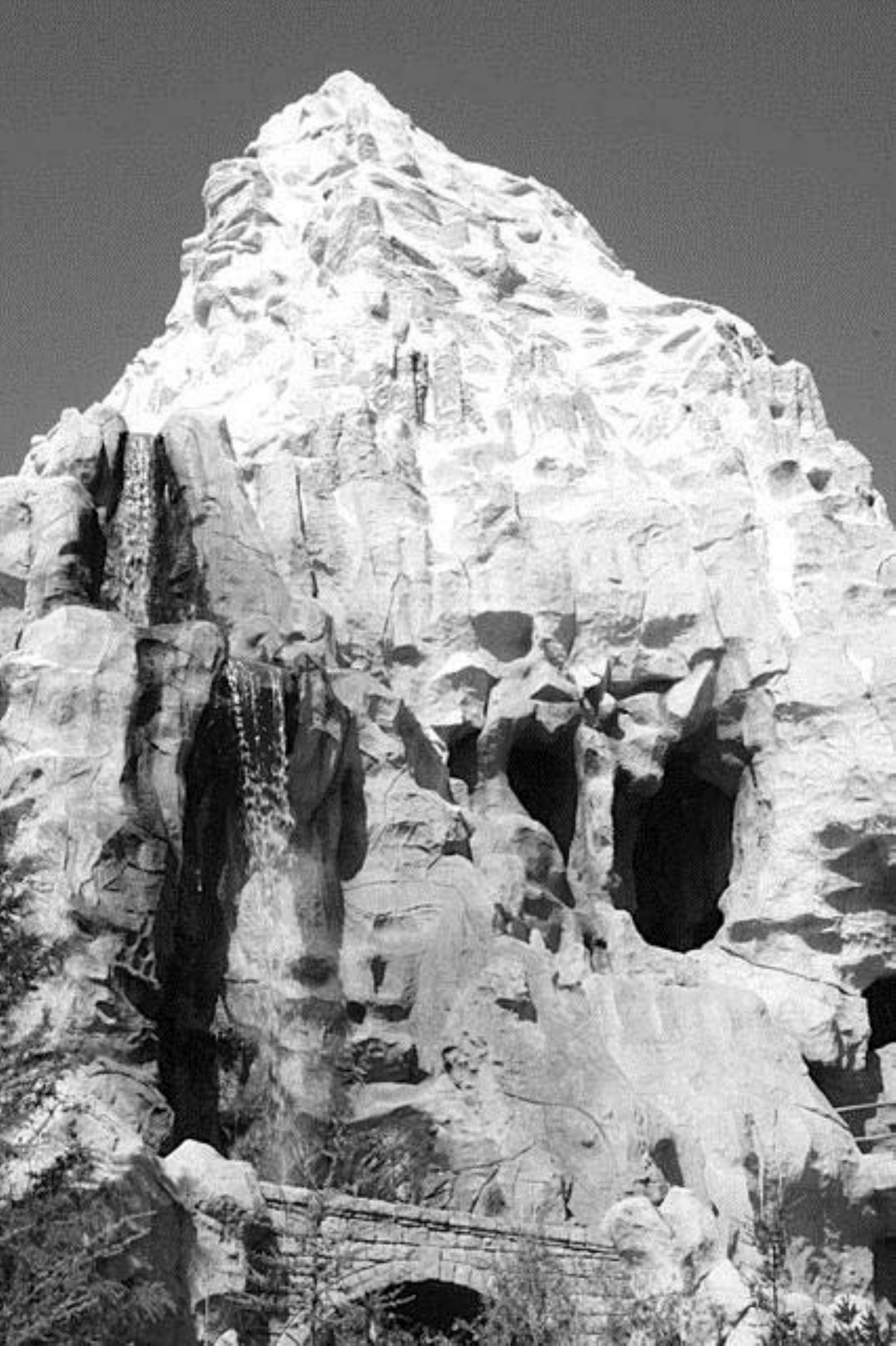}}
\subfigure[]{\includegraphics[height=1.5in]{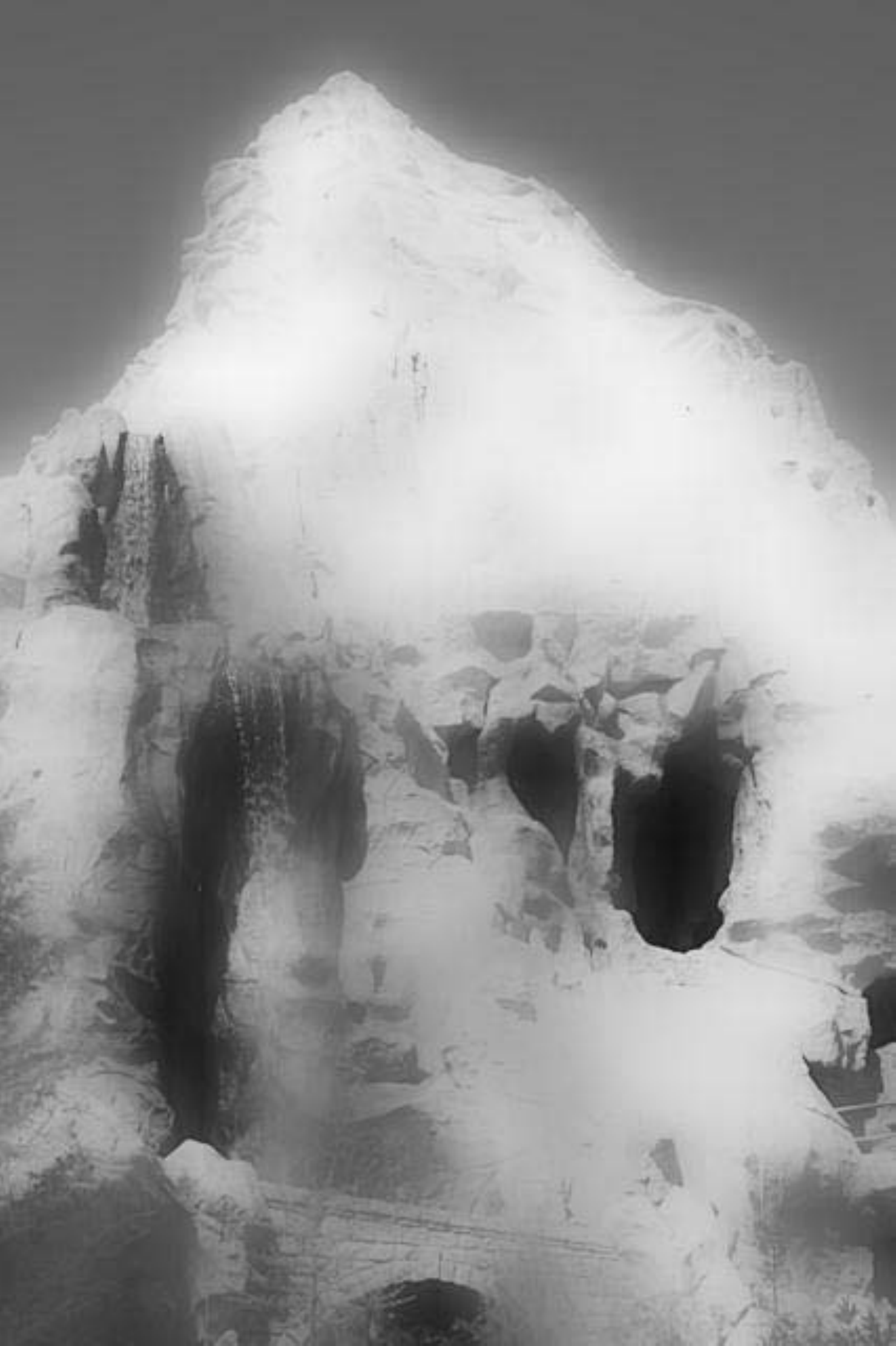}}
\subfigure[]{\includegraphics[height=1.5in]{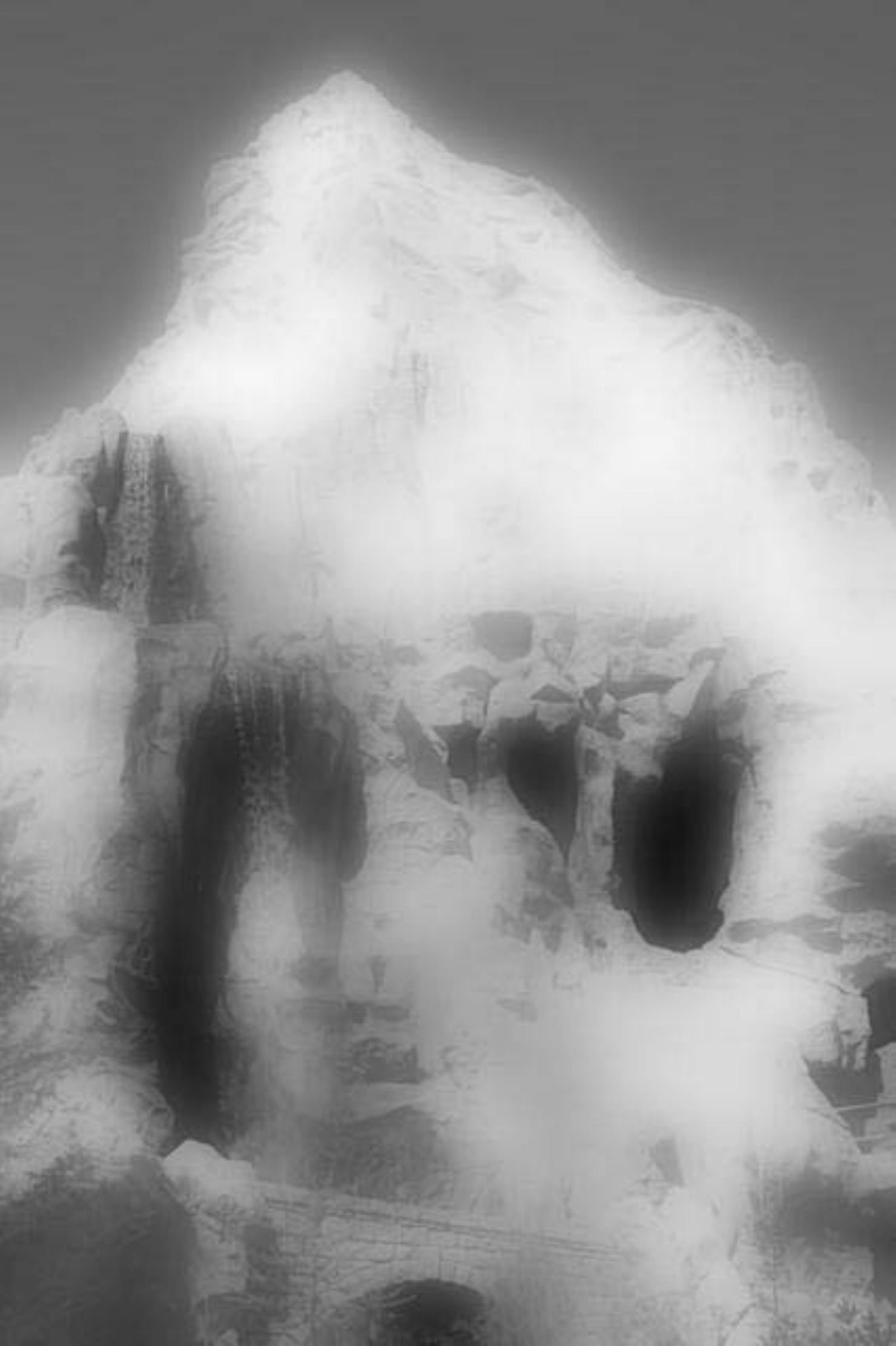}}
\subfigure[]{\includegraphics[height=1.5in]{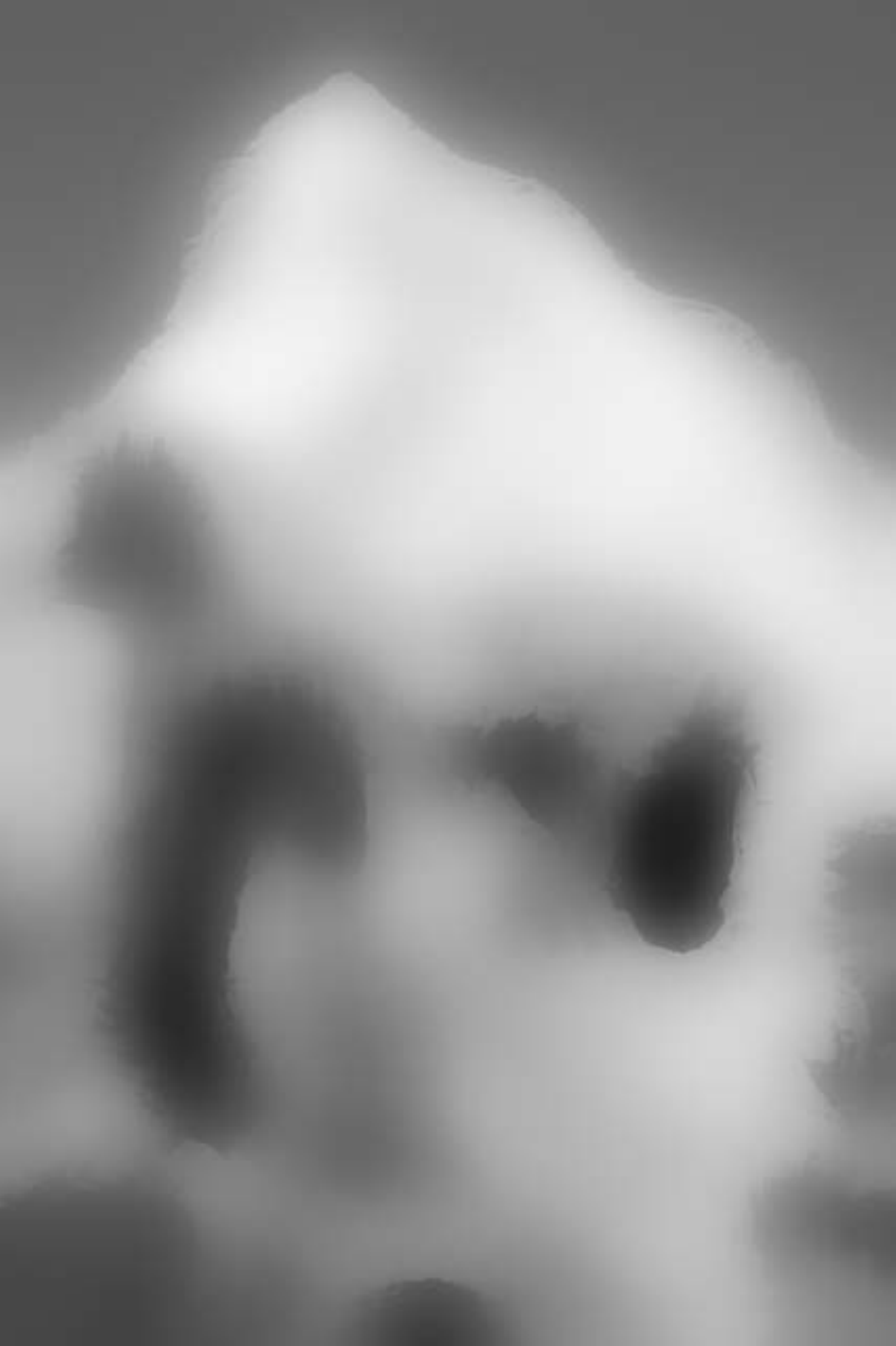}}
\subfigure[]{\includegraphics[height=1.5in]{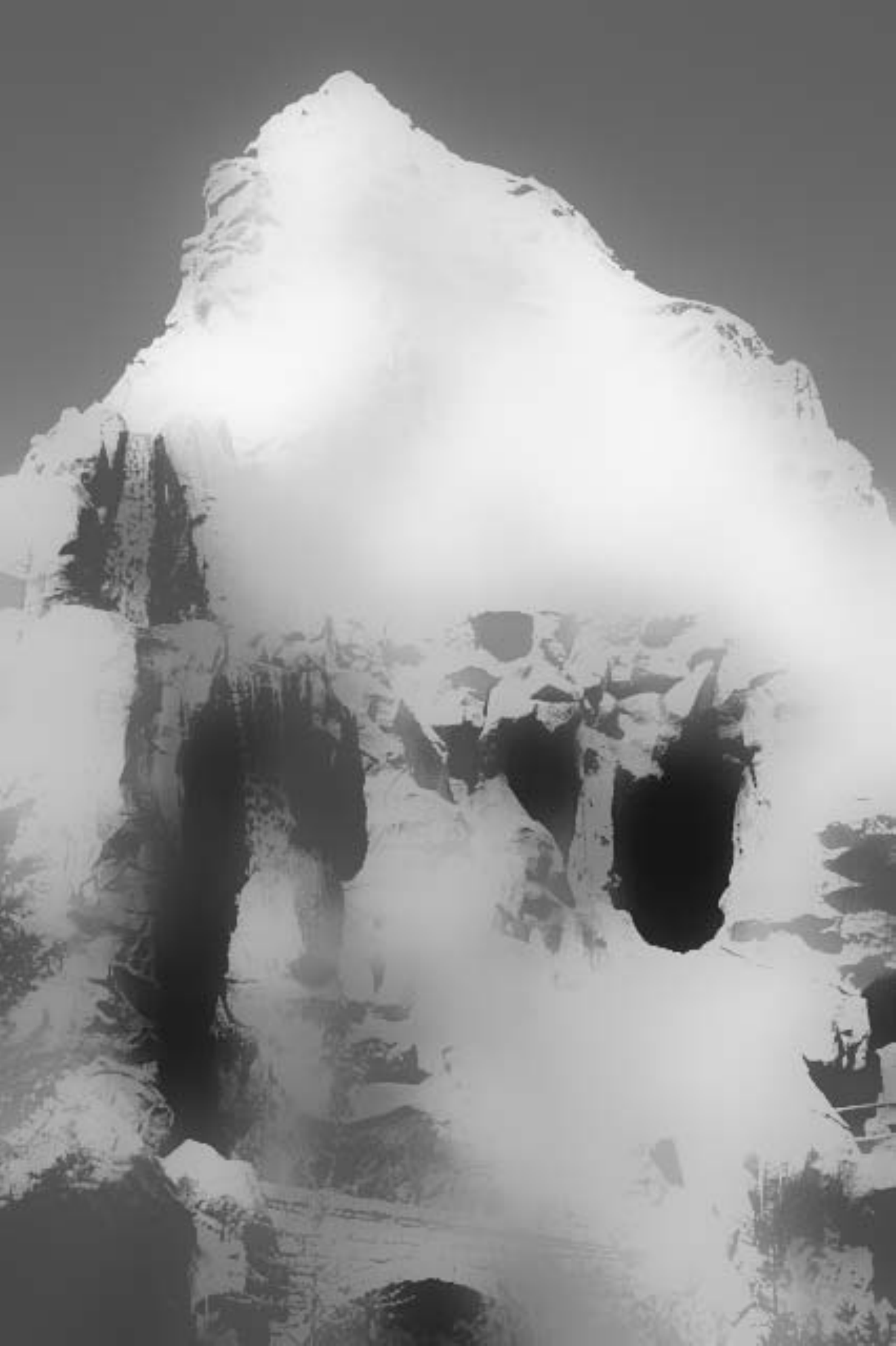}}
\caption{ Halo effects. (a) Original image. (b) Bilateral filter result using   $\sigma_{s} = 16$, $ \sigma_{r}= 0.4$. (c) Guided image filter result using $r = 16$, $\epsilon = 0.4^{2}$. (d) Our result using $p=1.2$, $r=16$. (e) Our result using $p=0.05$, $r=16$.}\label{fig3}
\end{figure*}

\begin{figure*}
\centering
\subfigure[]
{\includegraphics[height=1.8in]{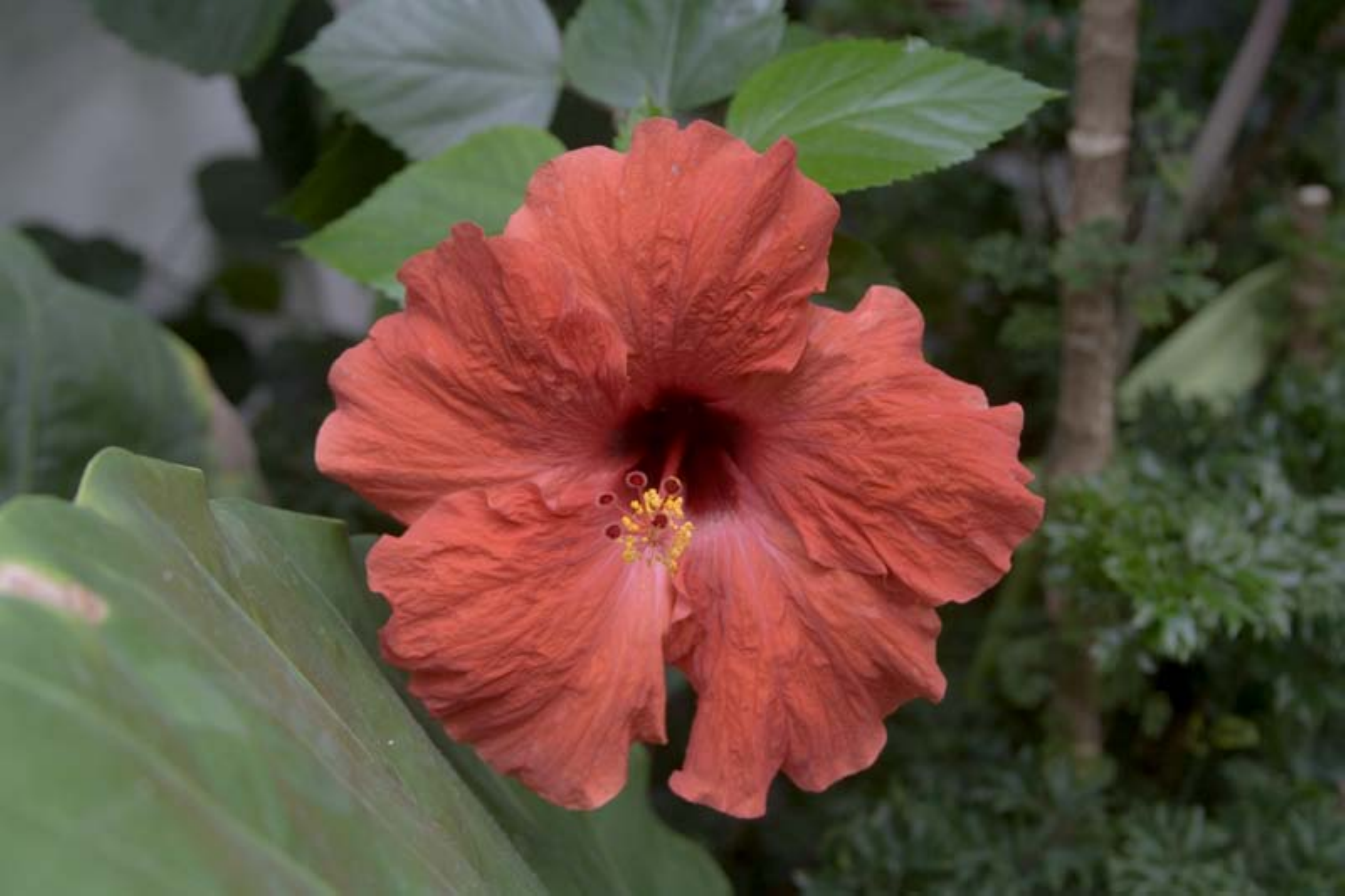}}
\subfigure[]{\includegraphics[height=2in]{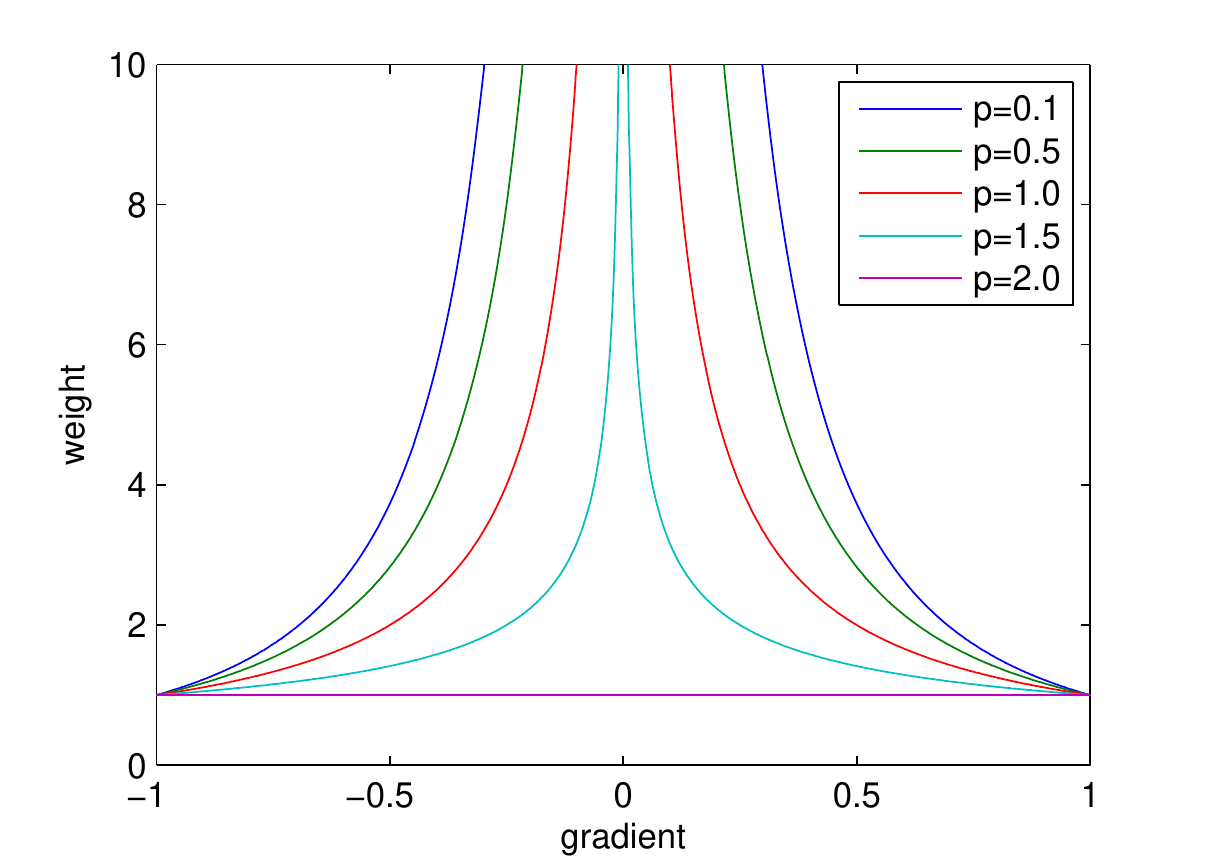}}
\caption{ (a) The original flower image. (b) Weights assigned to different gradients under different norms.}\label{fig4}
\end{figure*}

By using SNF with $p<2$, similar pixels will be assigned larger weights than dissimilar pixels, thus the filter is edge preserving. When is approaching to zero, the sparse norm approximates the $l^{0}$  energy, and the filter result exhibits no visible halo effects (Fig. \ref{fig3}), since pixels with different intensities are assigned much lower weights than the pixels with similar weights (Fig. \ref{fig4}(b)). The idea is also similar to the edge-stopping diffusion in the anisotropic diffusion framework~\cite{Sapiro:1998:TIP}.

\begin{figure*}
\centering
\includegraphics[height=3.7in]{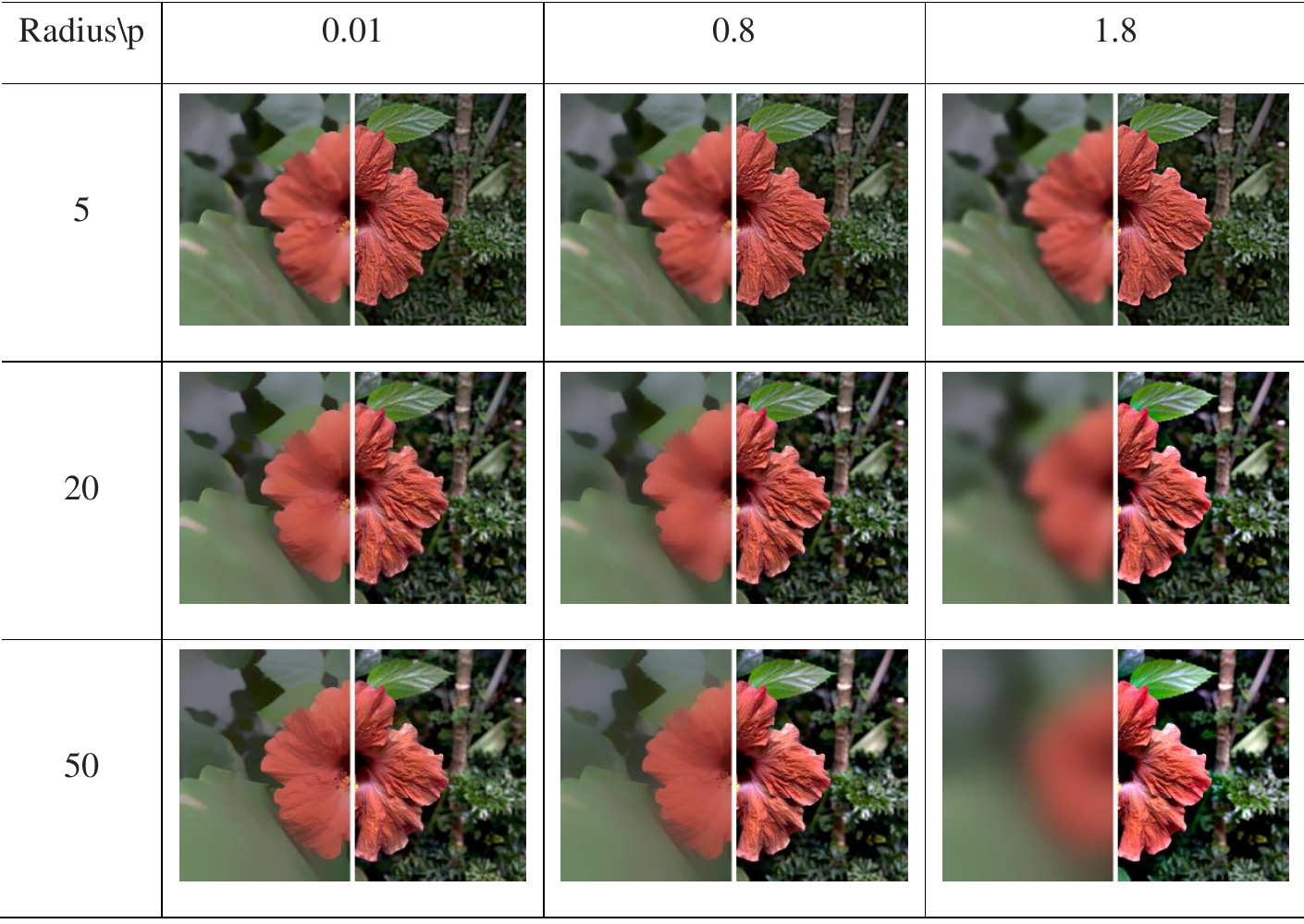}
\caption{Smoothing/sharpening using various radius/norm settings. Left half of each image: smoothing result. Right half: sharpening result by adding the detail layer.}\label{fig5}
\end{figure*}

We use SNF to decompose the image into a base layer and detail layer $I=B+D$. Here the base layer $B$ is the cartoon-like filtering result using the sparse norm filter. Detail enhancement can be achieved by boosting the detail layer $I_{boosted}=B+2\times D$. We demonstrate the results on a flower photo (Fig. \ref{fig4}(a)) by trying different combinations of the filtering radius $r$ and the norm $p$ (Fig. \ref{fig5}).

\subsection{Outlier Tolerant Filtering}
\label{sec4-2}

Standard edge preserving filters~\cite{Malik:1990:TPAMI,Manduchi:1998:ICCV} are very effective for Gaussian-like noise reduction. In the presence of extreme noise, none of them are as robust as the classic median filter. The culprit is the weighting can be misled by noise. In comparison, the sparse norm filter is a whole class of filter that can perform similarly with the median filter.

\begin{figure*}
\centering
\begin{tabular}{c@{\hspace{2mm}}c@{\hspace{2mm}}c}Input image & $p=1$ & $p=0.1$\\
\includegraphics[height=1.2in]{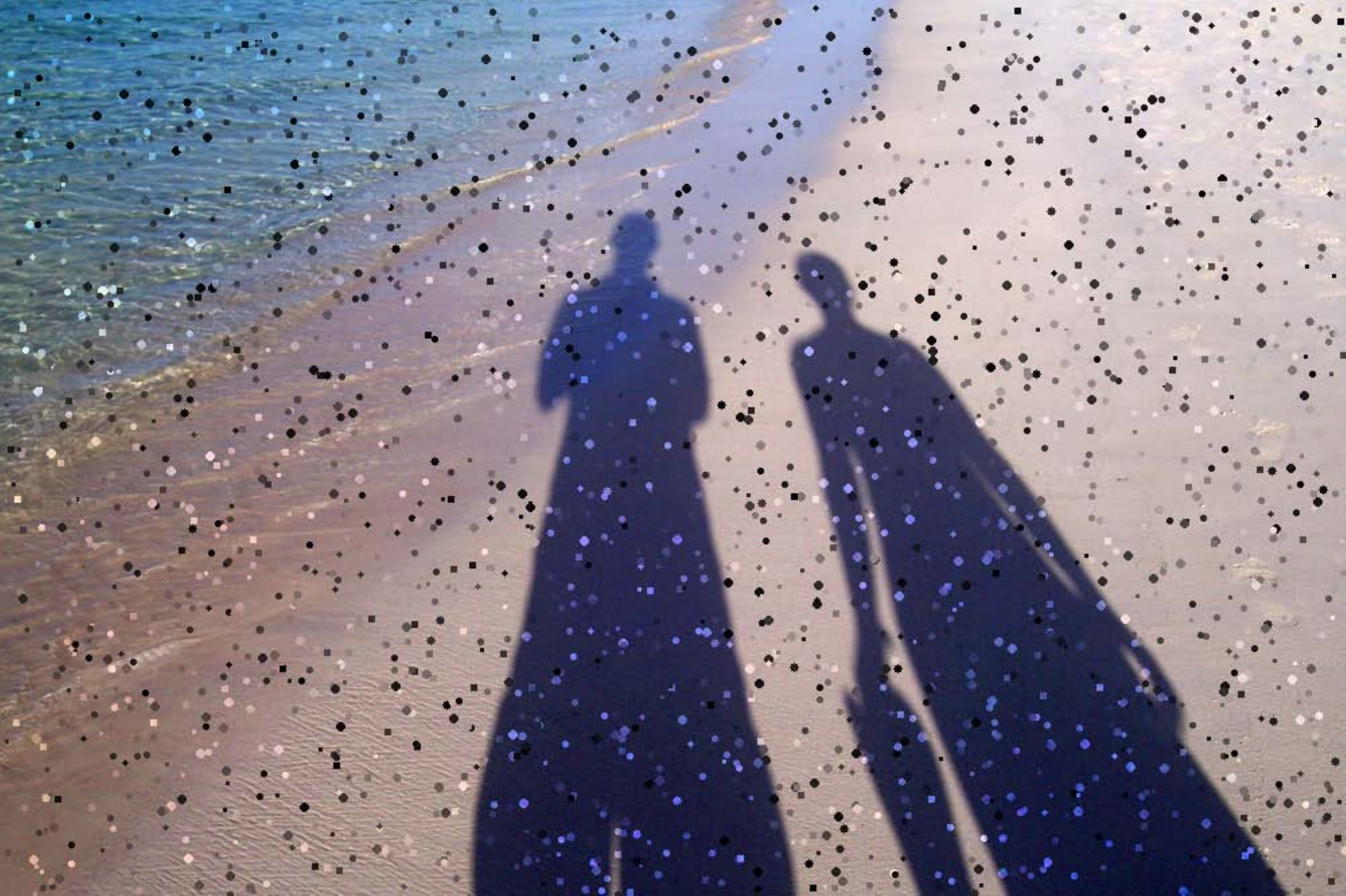}&\includegraphics[height=1.2in]{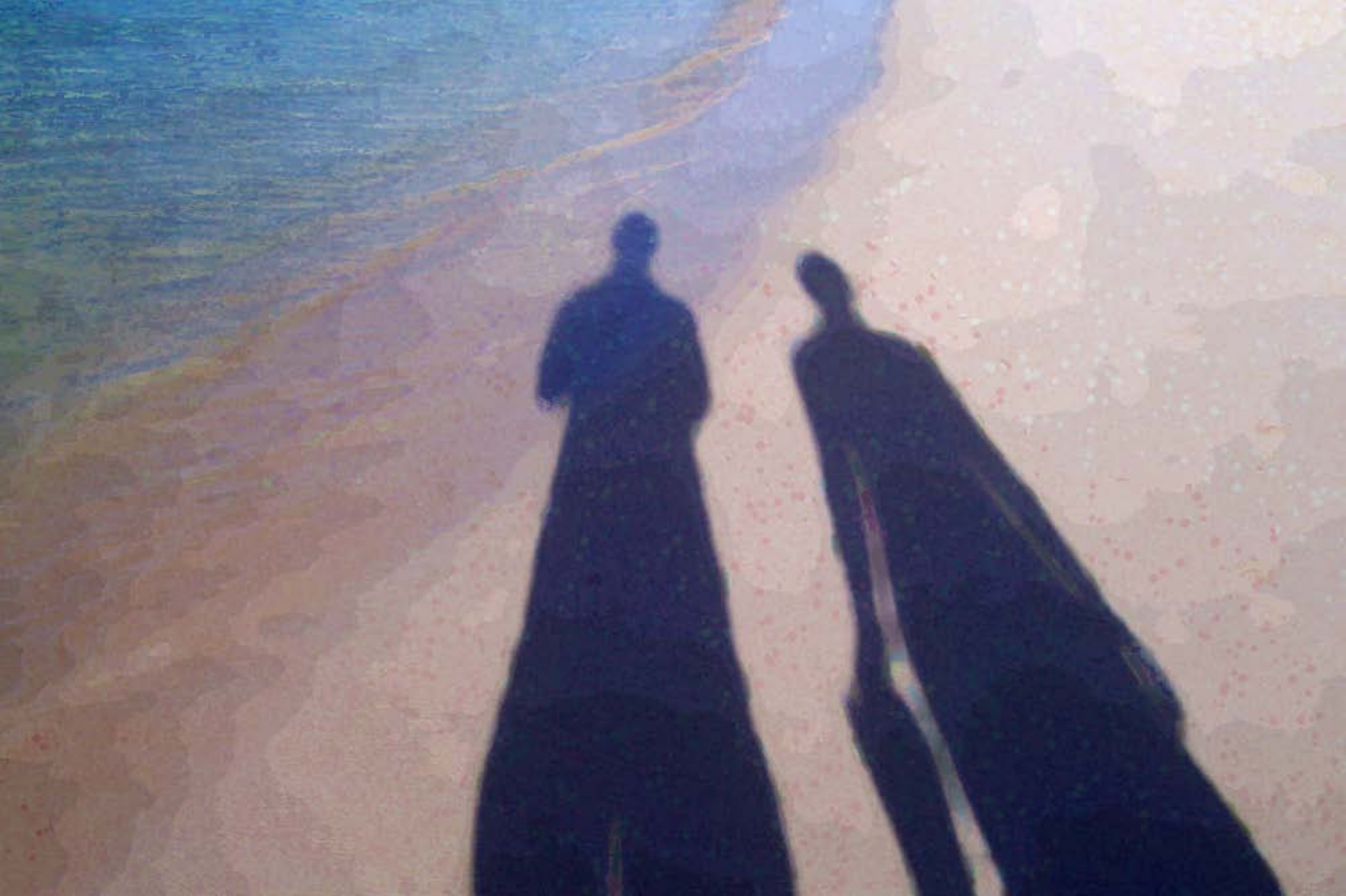}&\includegraphics[height=1.2in]{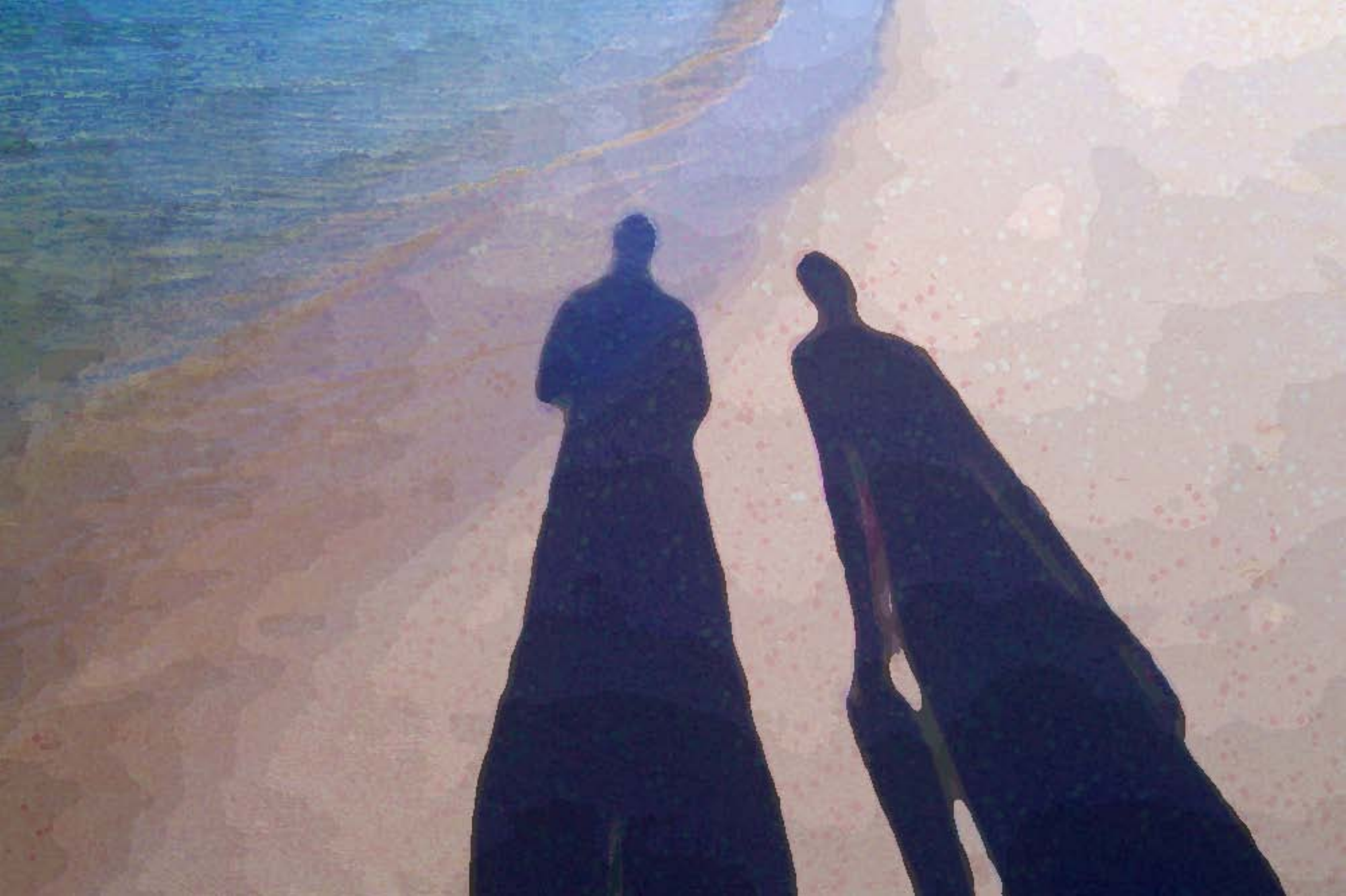}\\
\hspace{10mm}&\includegraphics[height=1.2in]{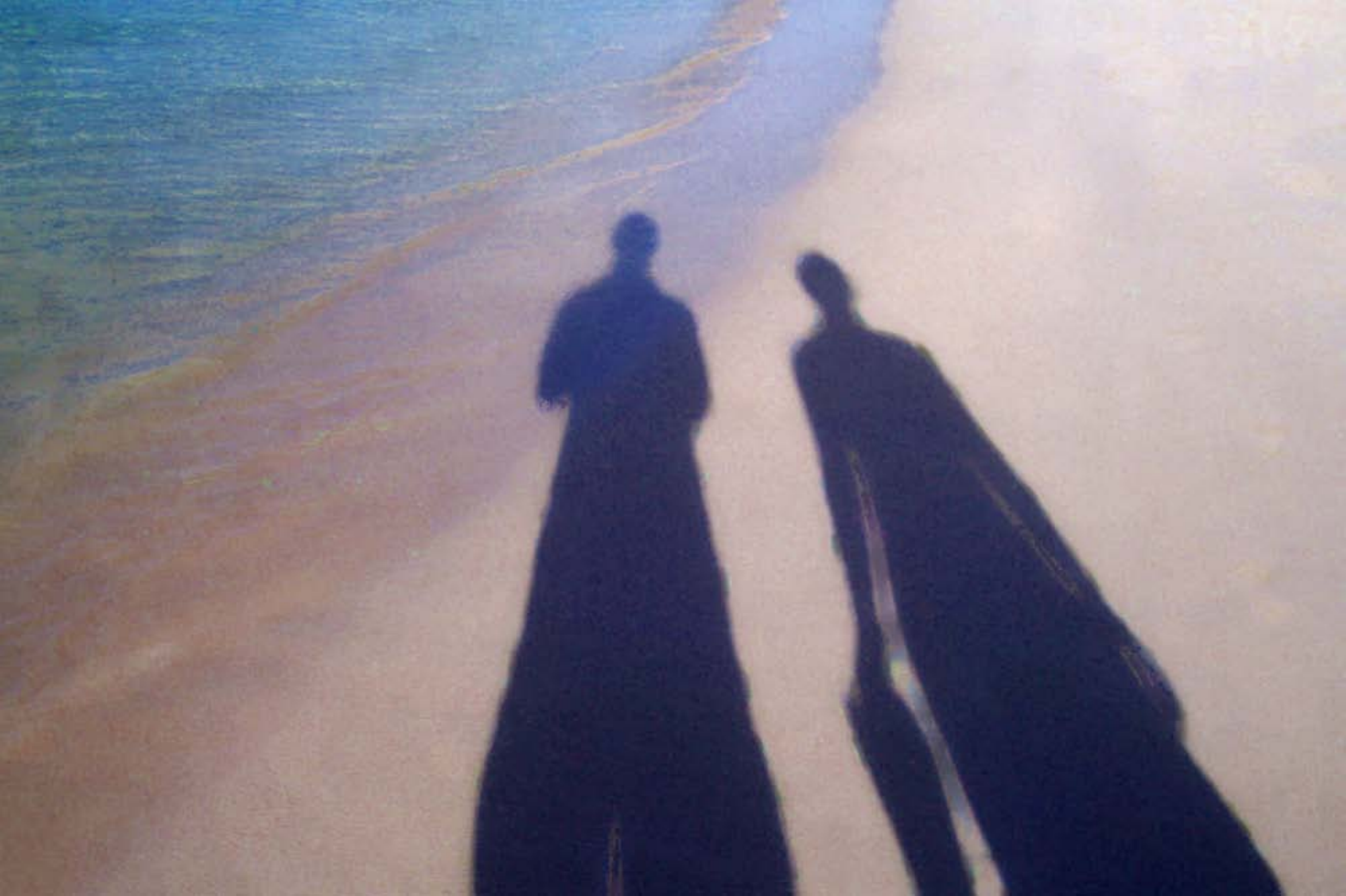}&\includegraphics[height=1.2in]{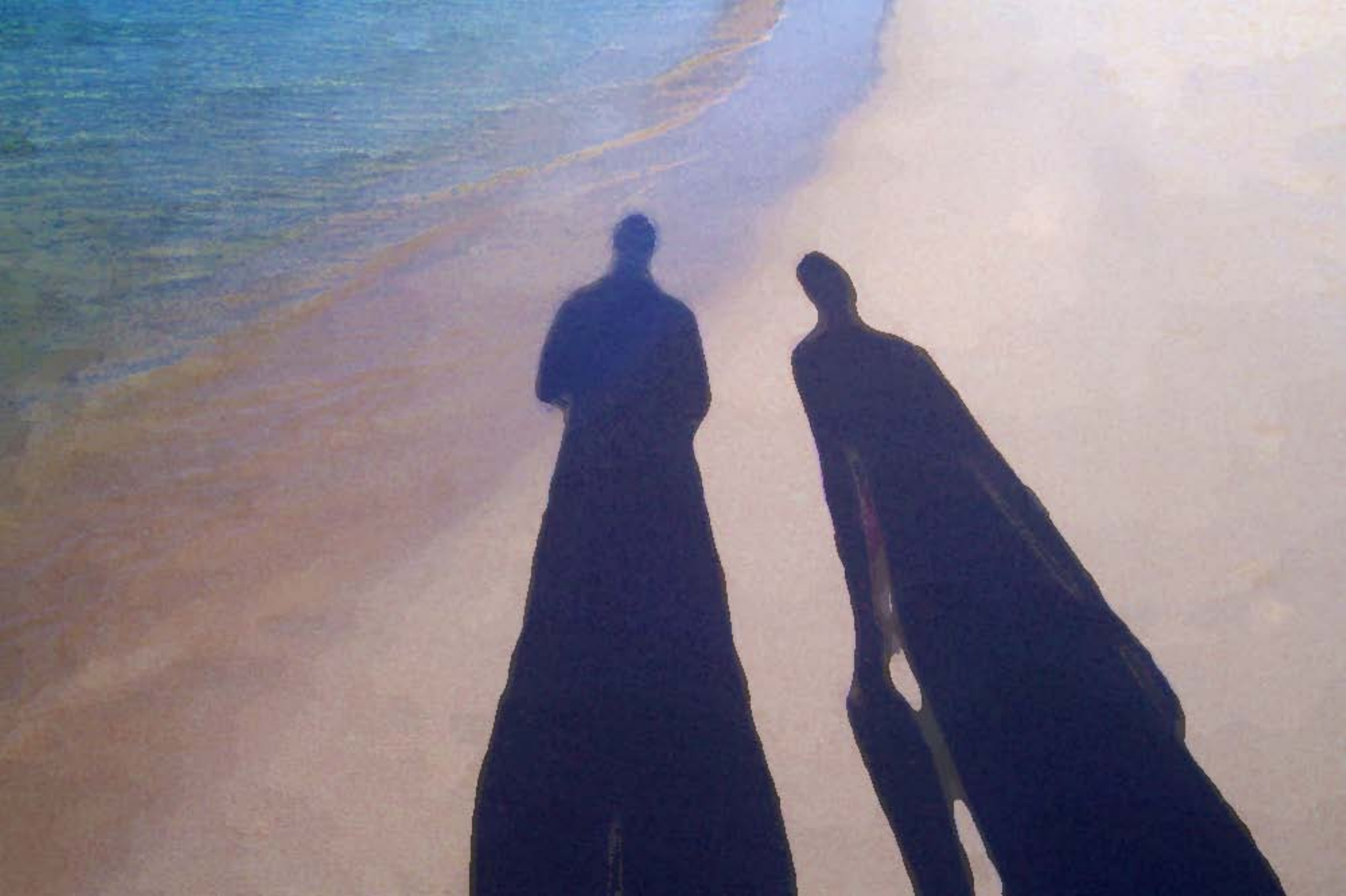}
\end{tabular}
\caption{Left: original image. Middle: $p=1.0$, $r=10$. Right: $p=0.1$, $r =10$. First row: approximate solution of (\ref{eq6}) using brute force search. Second row: followed by a guided diffusion.
}\label{fig6}
\end{figure*}

We take an example image from~\cite{Solomon:2010:TOG}. We avoid the outliers by first using brute force search to approximate the global solution of (\ref{eq6}) at a few discrete values~\cite{Ahuja:2009:CVPR}. This intermediate result has a quantized look. (Fig~\ref{fig6}, row 1 columns 2\&3) We calculate the diffusity at this approximate solution / use this as the guidance image and use the one pass weighted average filtering (\ref{eq7}) to output a smoothed image. (Fig \ref{fig6}, row 2)

\subsection{HDR Compression}
\label{sec4-3}

\begin{figure*}
\centering
\subfigure[Original]
{\includegraphics[height=1.2in]{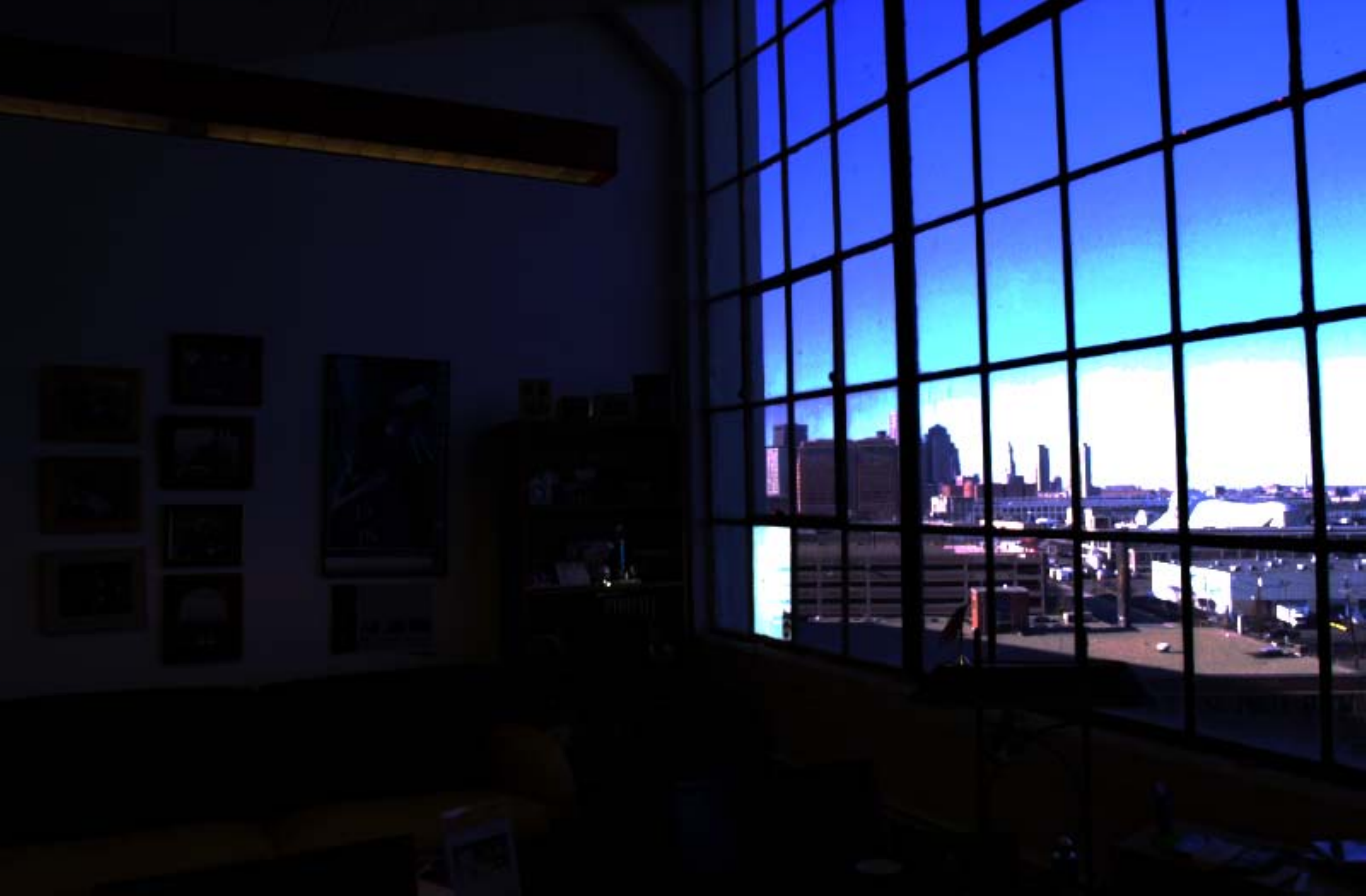}}
\subfigure[Farbman08]{\includegraphics[height=1.2in]{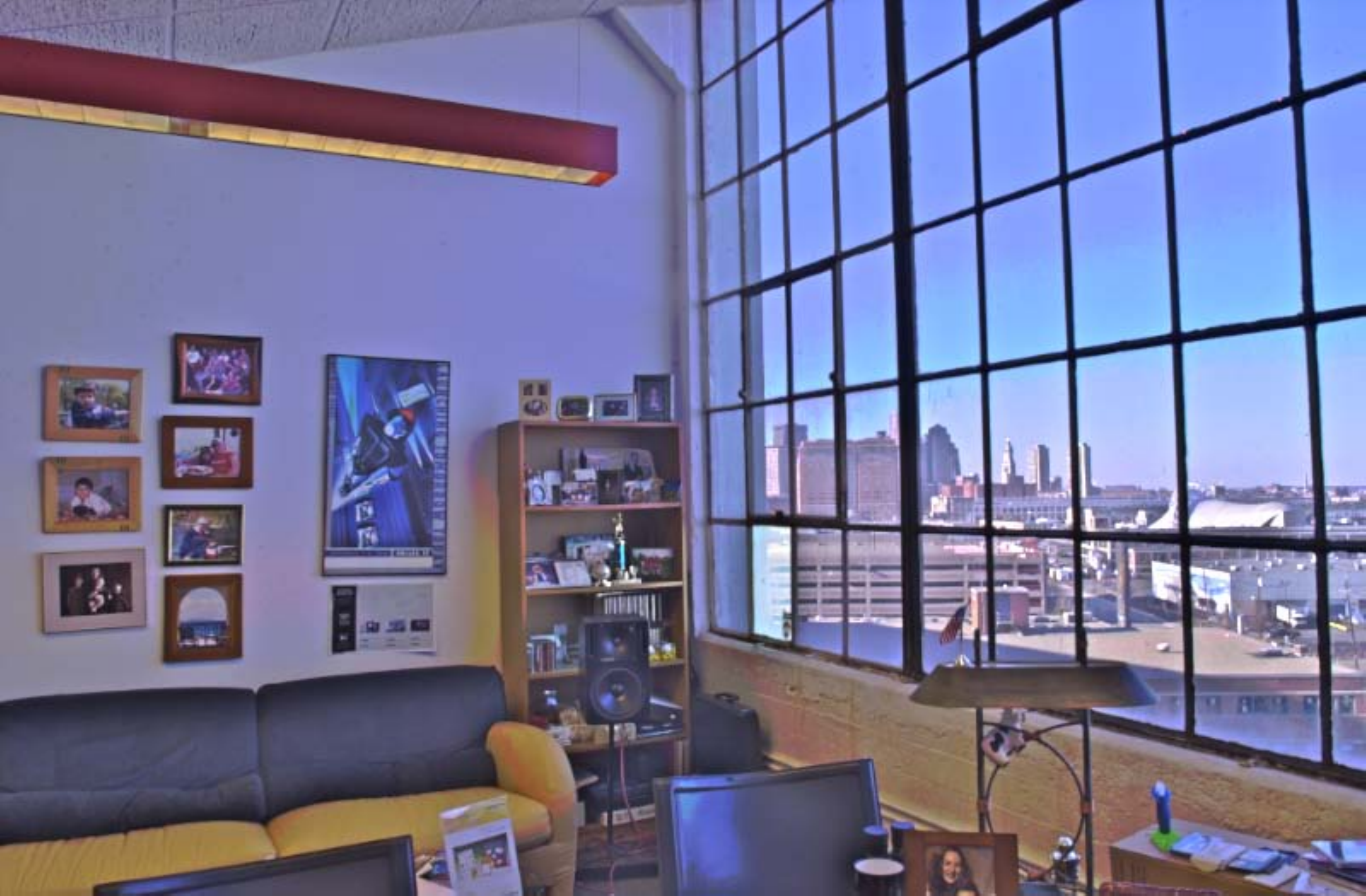}}
\subfigure[Durand02]{\includegraphics[height=1.2in]{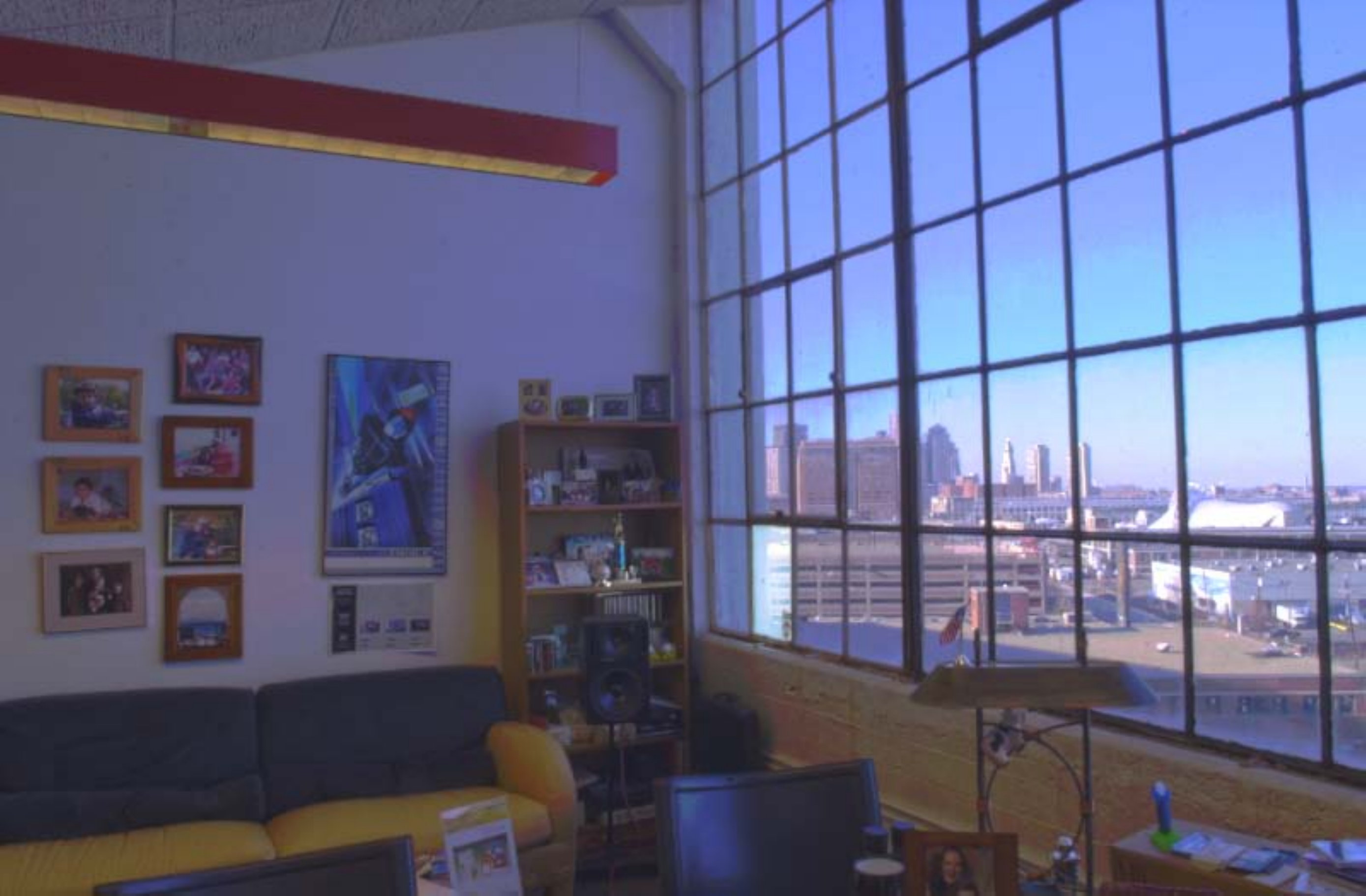}}
\subfigure[Drago03]{\includegraphics[height=1.2in]{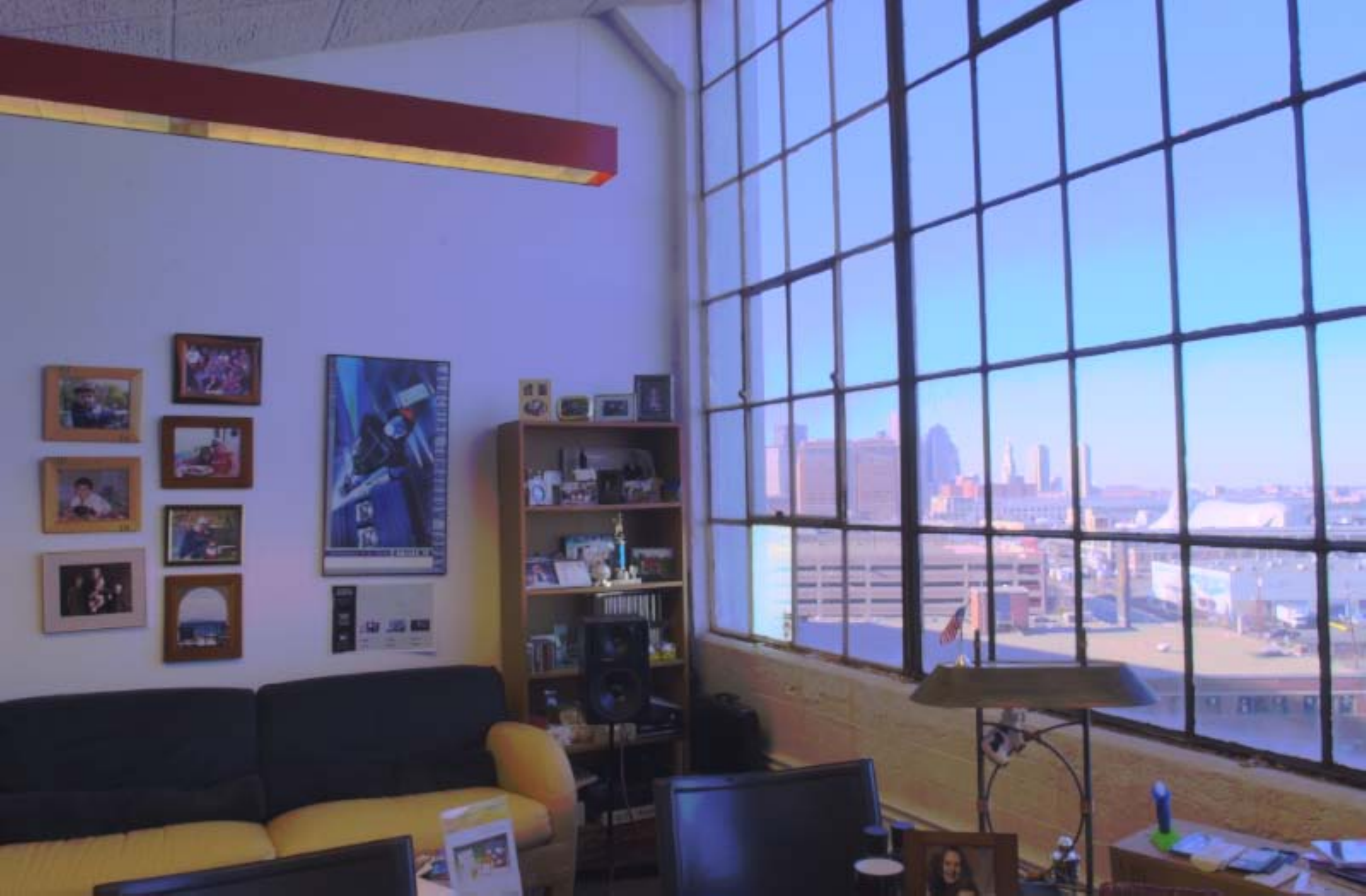}}
\subfigure[Fattal02]{\includegraphics[height=1.2in]{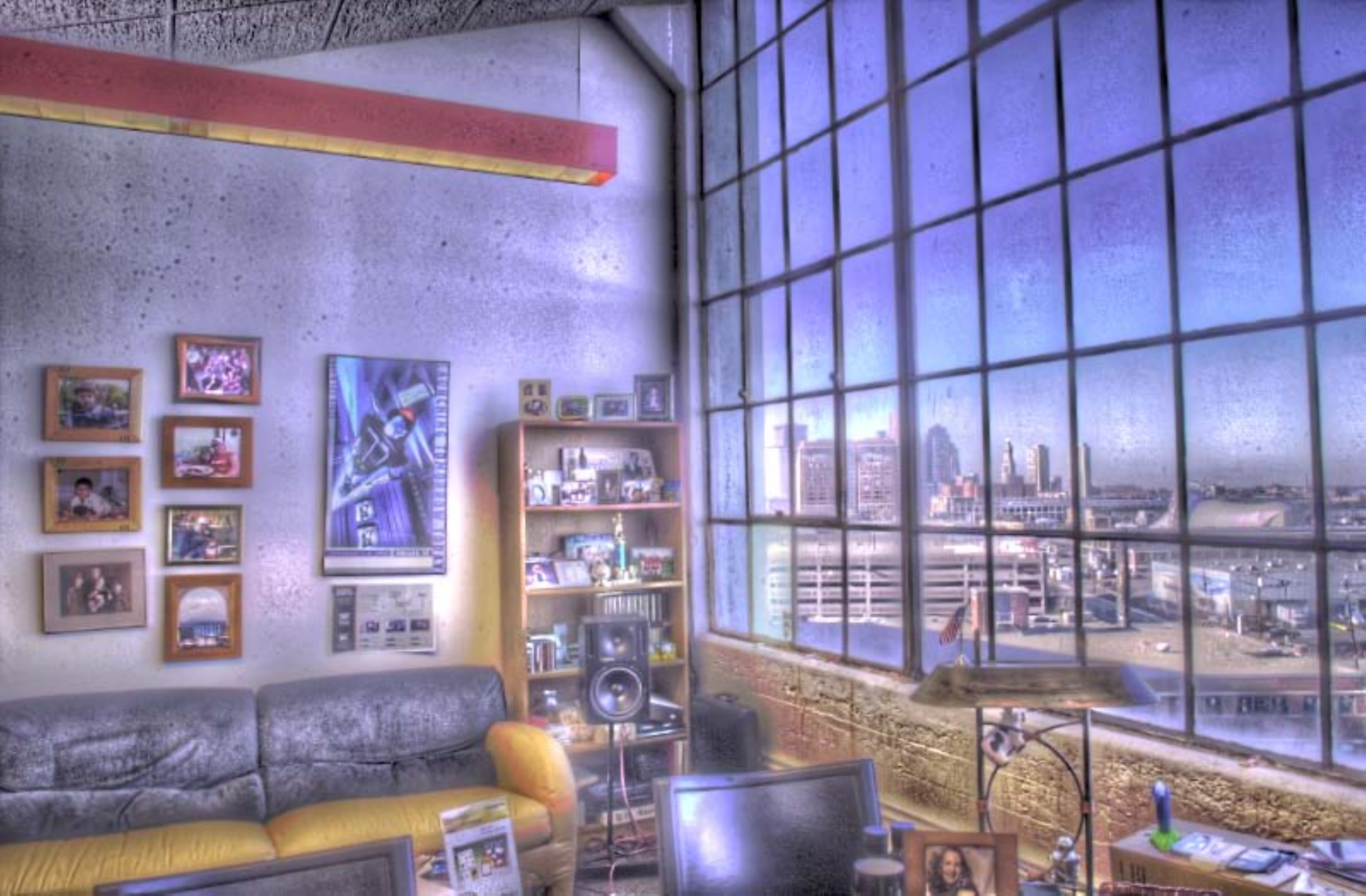}}
\subfigure[Pattanaik00]{\includegraphics[height=1.2in]{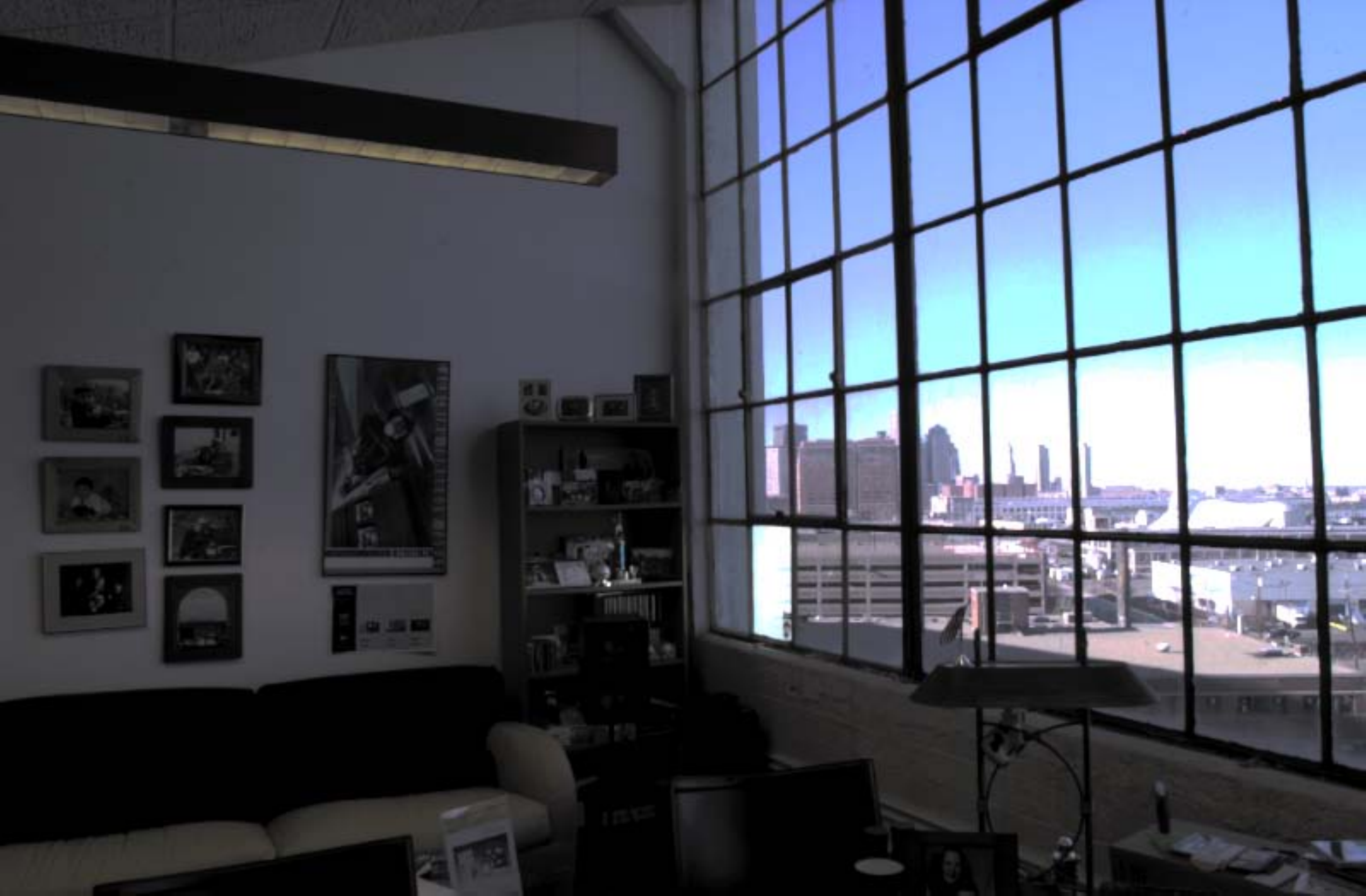}}
\subfigure[Mantiuk06]{\includegraphics[height=1.2in]{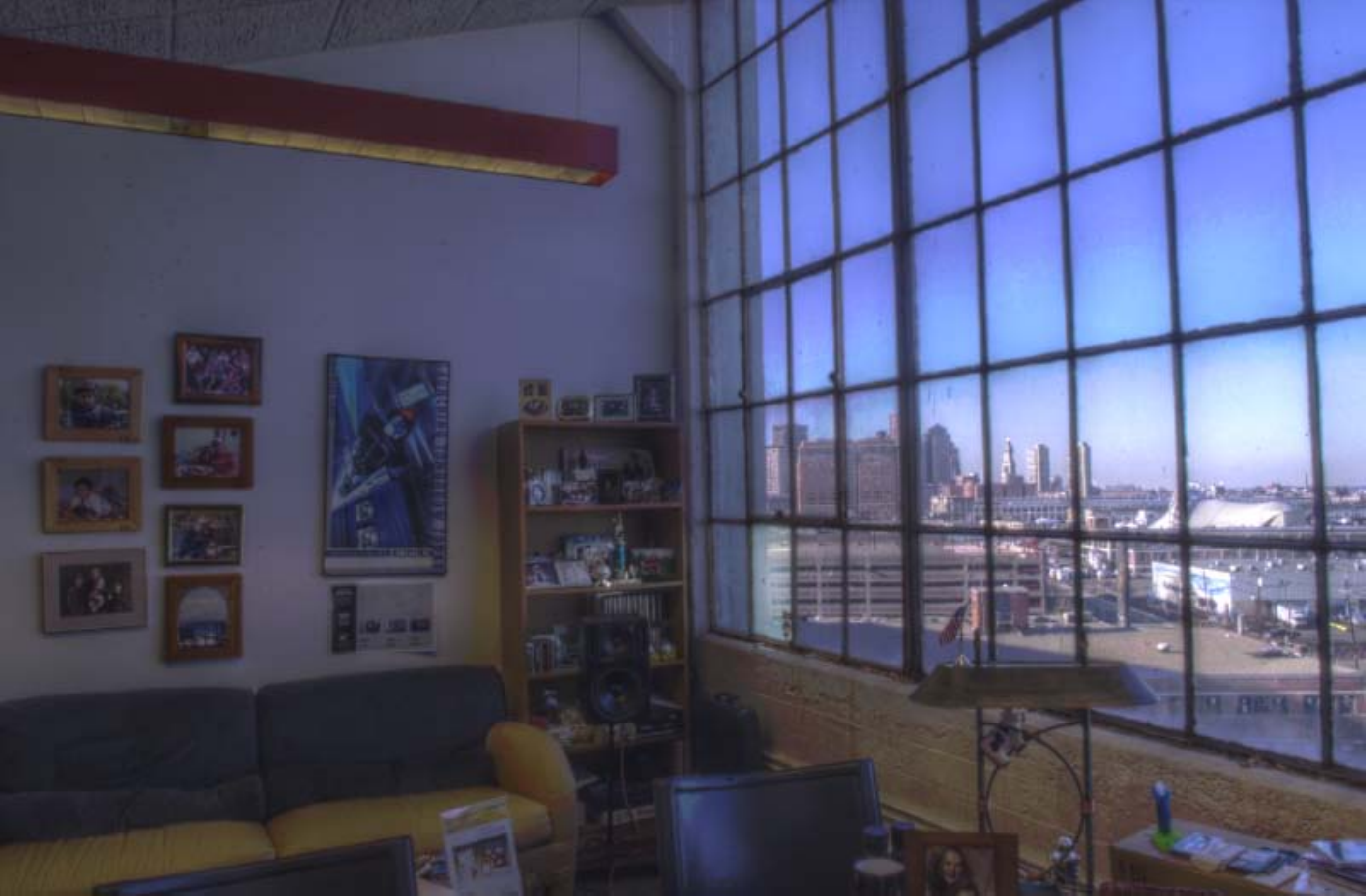}}
\subfigure[Reinhard05]{\includegraphics[height=1.2in]{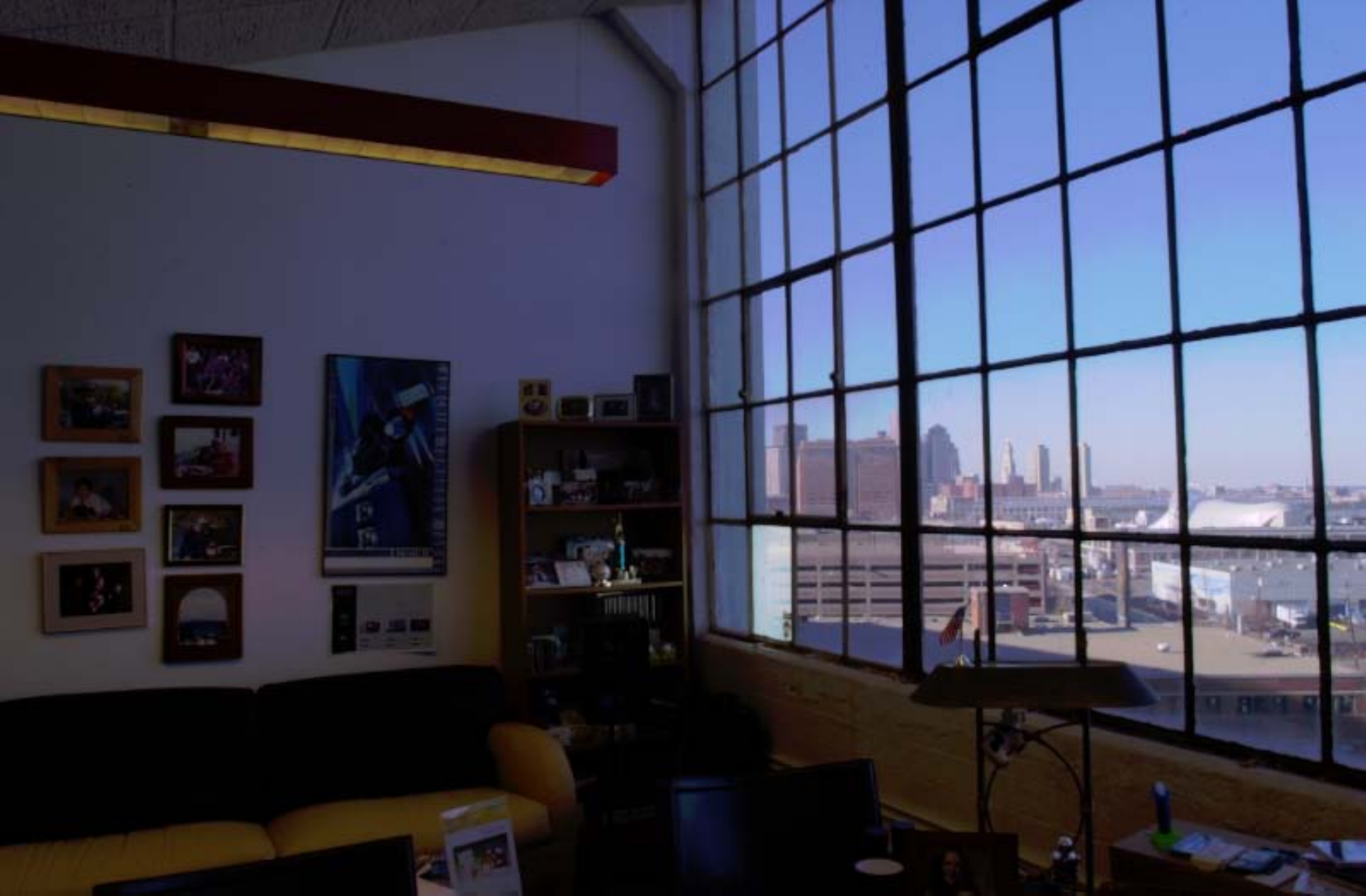}}
\subfigure[SNF]{\includegraphics[height=1.2in]{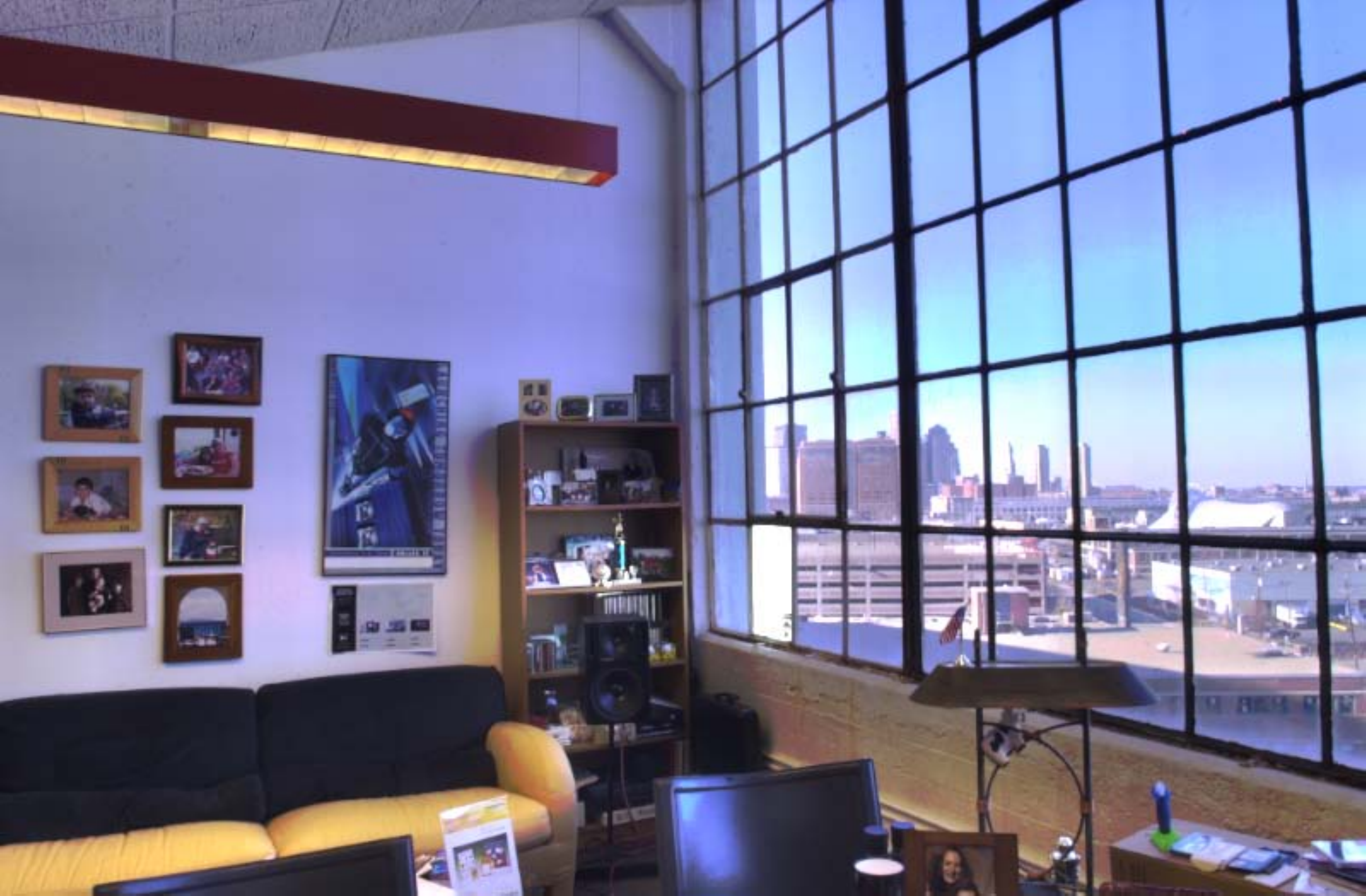}}
\caption{HDR tone compression comparison. }\label{fig7}
\end{figure*}

HDR tone mapping is a popular application which can be achieved by compressing the base layer $B$ while keeping the detail layer $D$. In the following comparison (Fig \ref{fig7}) we can see that the weighted least square (WLS) \cite{Szeliski:2008:TOG}, Fattal02 \cite{Werman:2002:TOG}, Durand02 \cite{Dorsey:2002:TOG} have visible halos near the strong edges. For the sparse norm filter, we set $p=0.2$  and radius to be 1/6 of the image height to conduct one pass of non-local diffusion to extract the base layer. We observe under the same $p=0.2$ the WLS method seems to be trapped in a local minimum because the cost function is non-convex. Although Drago03 \cite{Chiba:2003:CGF}, Pattanaik00 \cite{Greenberg:2000:TOG}, Mantiuk06 \cite{Seidel:2006:TAP}, Reinhard05 \cite{Devlin:2005:TVCG} try to reduce the halo, they fail to make some details visible in the results.

\subsection{Non-blind Deconvolution}
\label{sec4-4}

\begin{figure*}
\centering
\subfigure[]
{\includegraphics[height=1in]{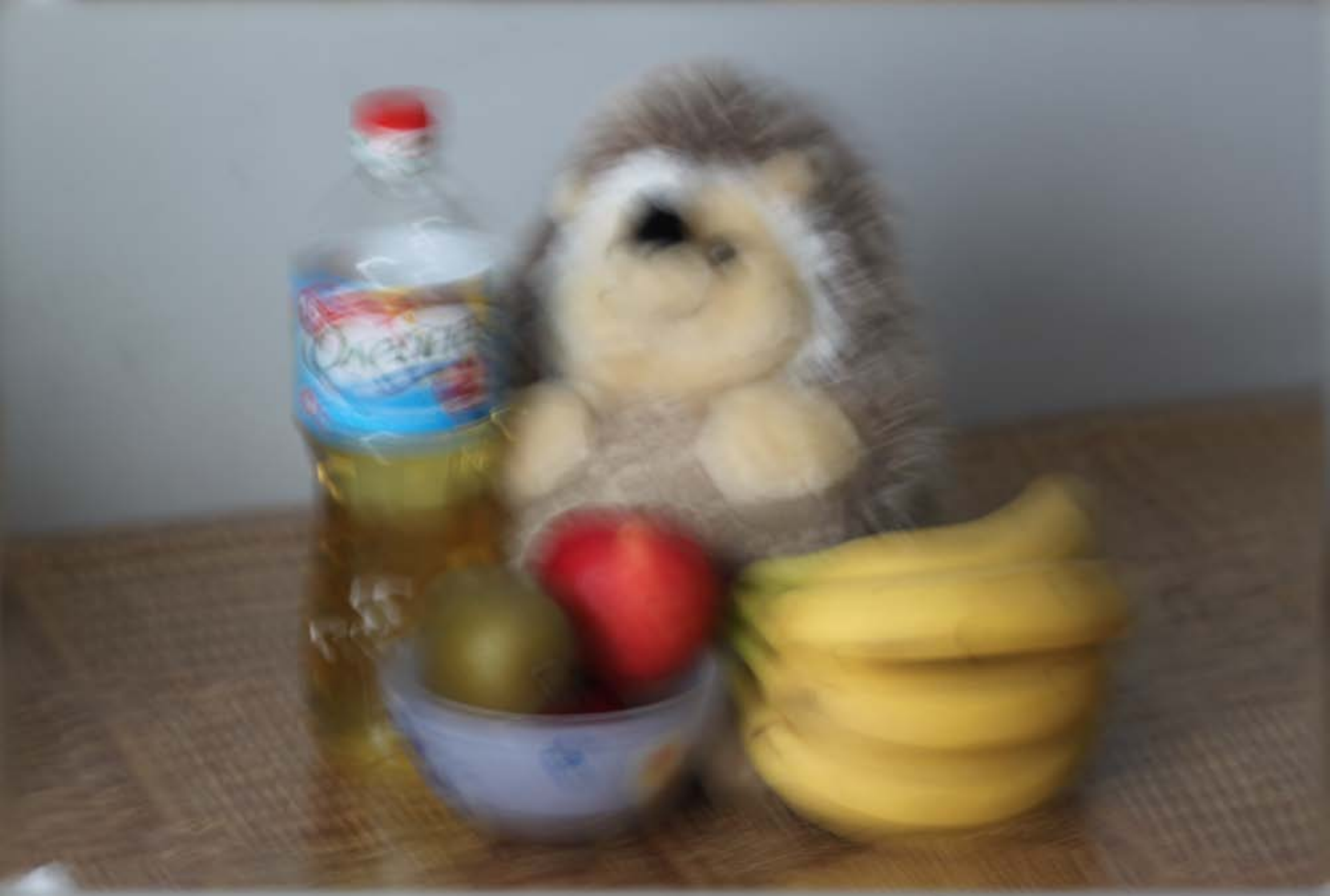}}
\subfigure[]{\includegraphics[height=1in]{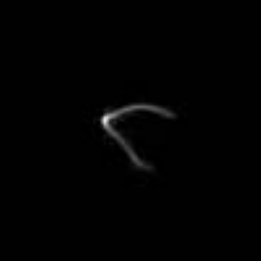}}
\subfigure[]{\includegraphics[height=1in]{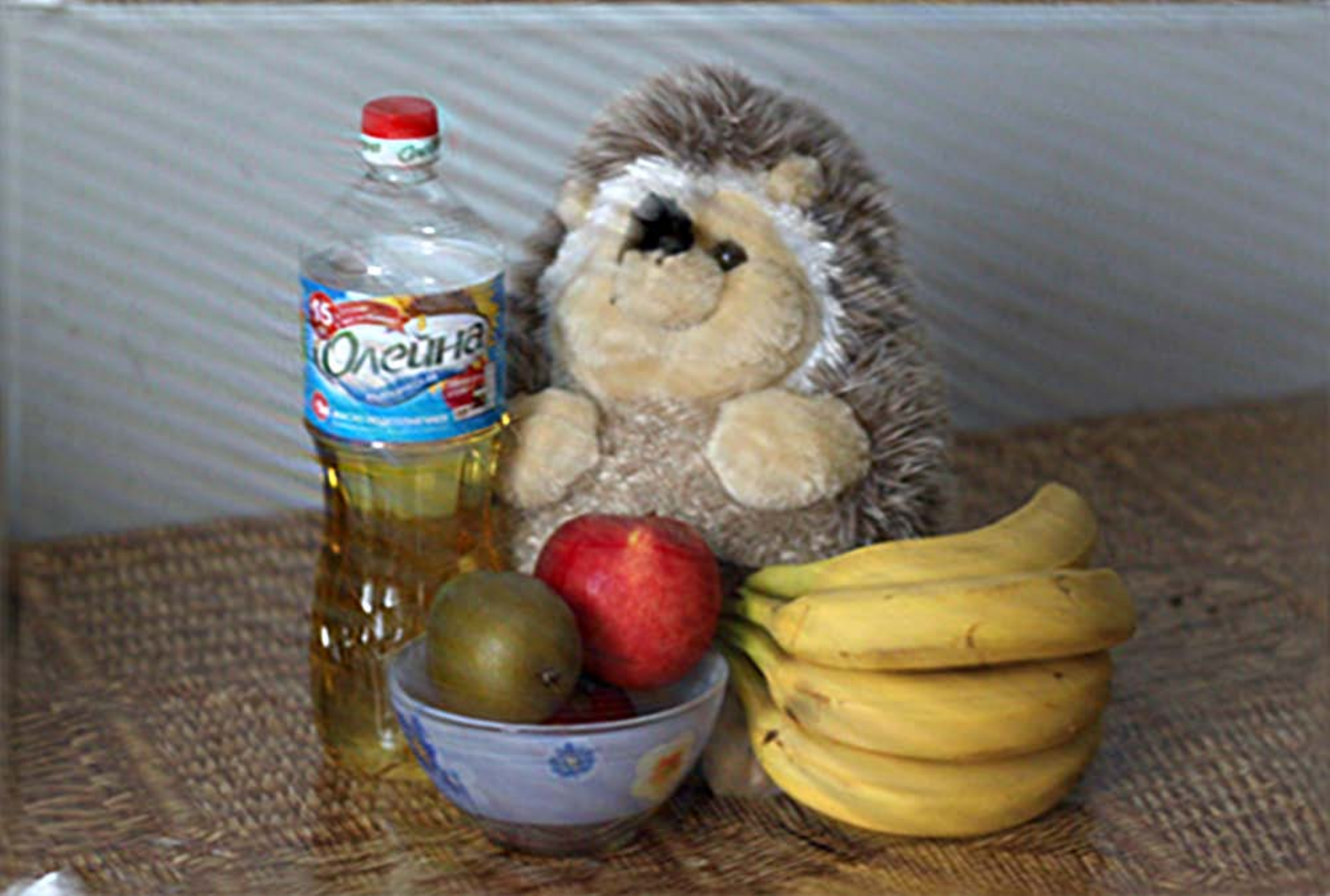}}
\subfigure[]{\includegraphics[height=1in]{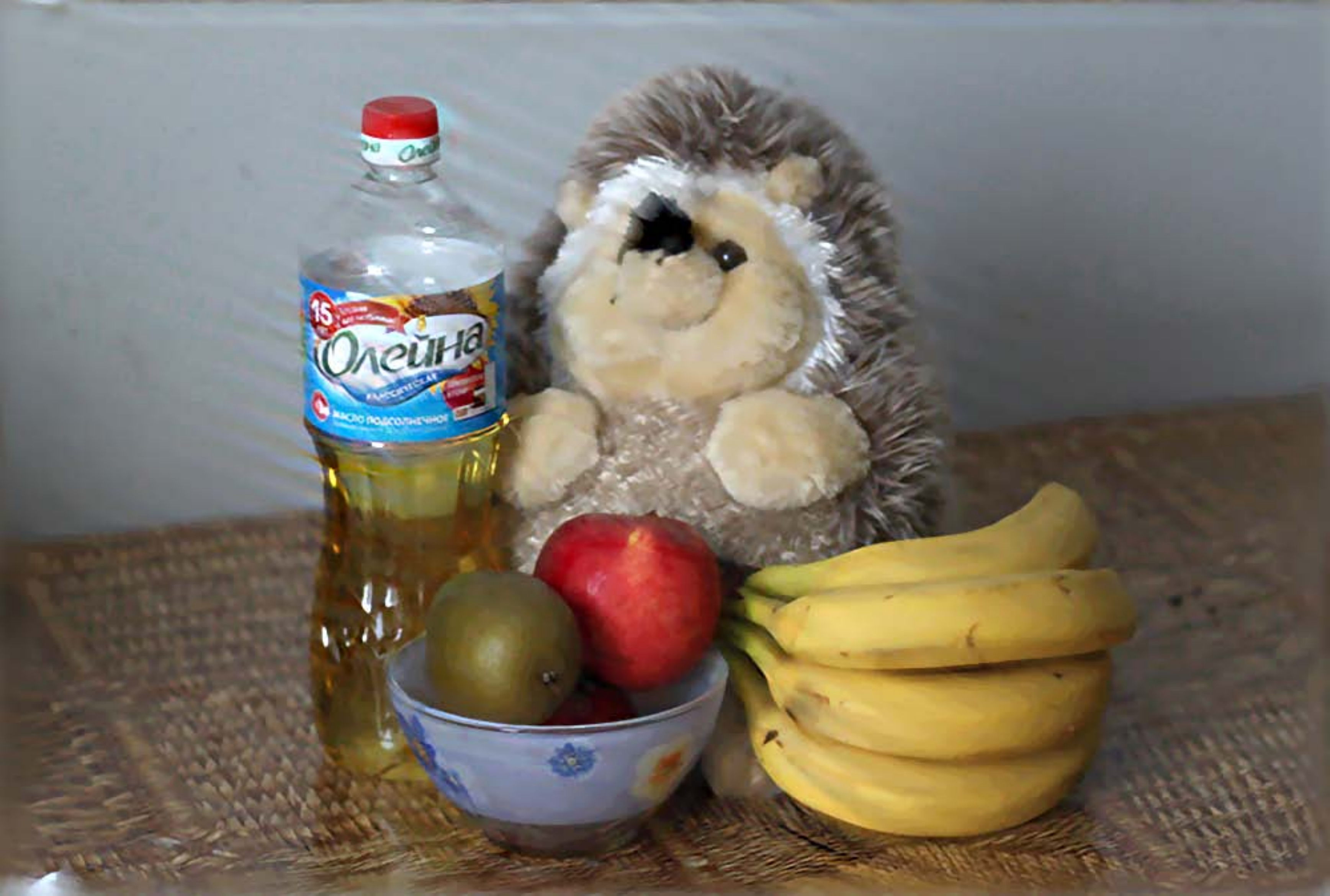}}
\caption{Deconvolution comparison. (a) Input. (b) Estimated kernel. (c) Tikhonov regularization result. (d) Sparse norm deconvolution using $p=0.5$, $r=5$.}\label{fig8}
\end{figure*}

Ringing artifacts are common in deconvolution when the kernel estimation is not accurate or when frequency nulls occur. The ringing artifacts can be significantly reduced by putting a sparse norm prior on the gradient term \cite{Freeman:2007:TOG,Fergus:2009:NIPS}. Similarly we put a non-local sparse norm on the gradients
\begin{equation}
\min_{I_{i}^{new}}\|I_{i}^{new}-k\bigotimes I(i)\|^{2}+\frac{\lambda}{N_{i}}\sum_{j\in N_{i},j\neq i}\|I_{i}^{new}-I_{j}\|^{p}, 0<p<2
\label{eq9}
\end{equation}
and use an alternative minimization~\cite{Zhang:2008:SIAM} technique to deconvolve the blurry image (Fig \ref{fig8}). In the following comparison we compare our result with the standard Tikhonov regularization which uses an  $l^{2}$ penalty on the gradient term. We notice SNF can produce crisper results with fewer ringing effects.

\subsection{Joint Filtering}
\label{sec4-5}

\begin{figure*}
\centering
\subfigure[]
{\includegraphics[height=1.5in]{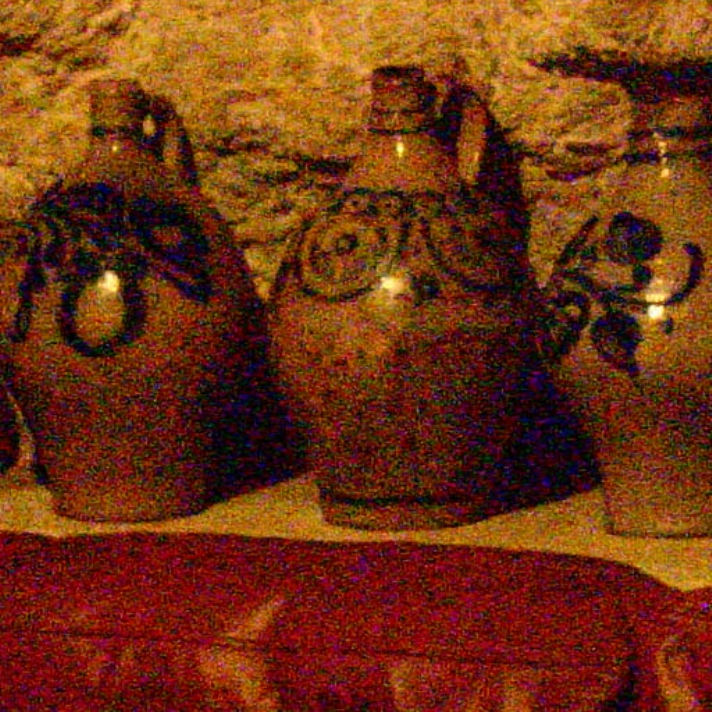}}
\subfigure[]{\includegraphics[height=1.5in]{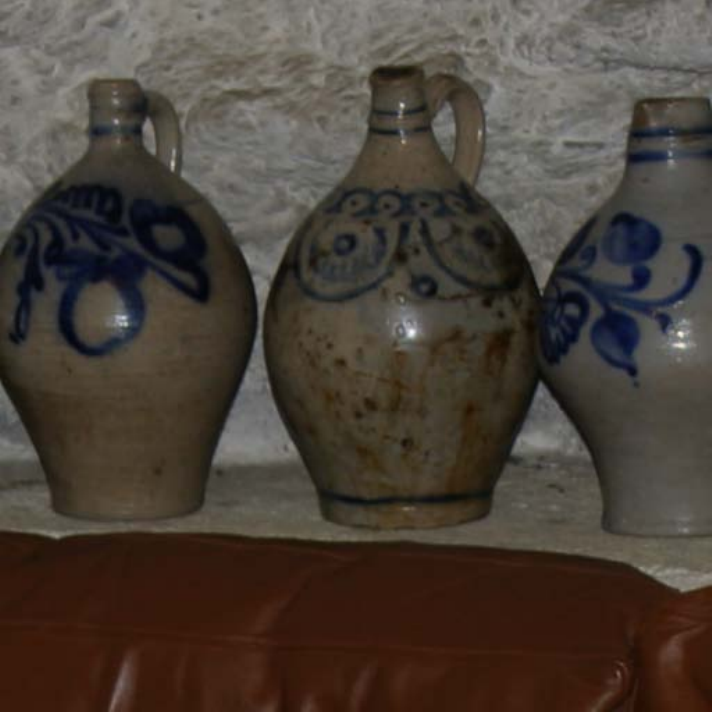}}
\subfigure[]{\includegraphics[height=1.5in]{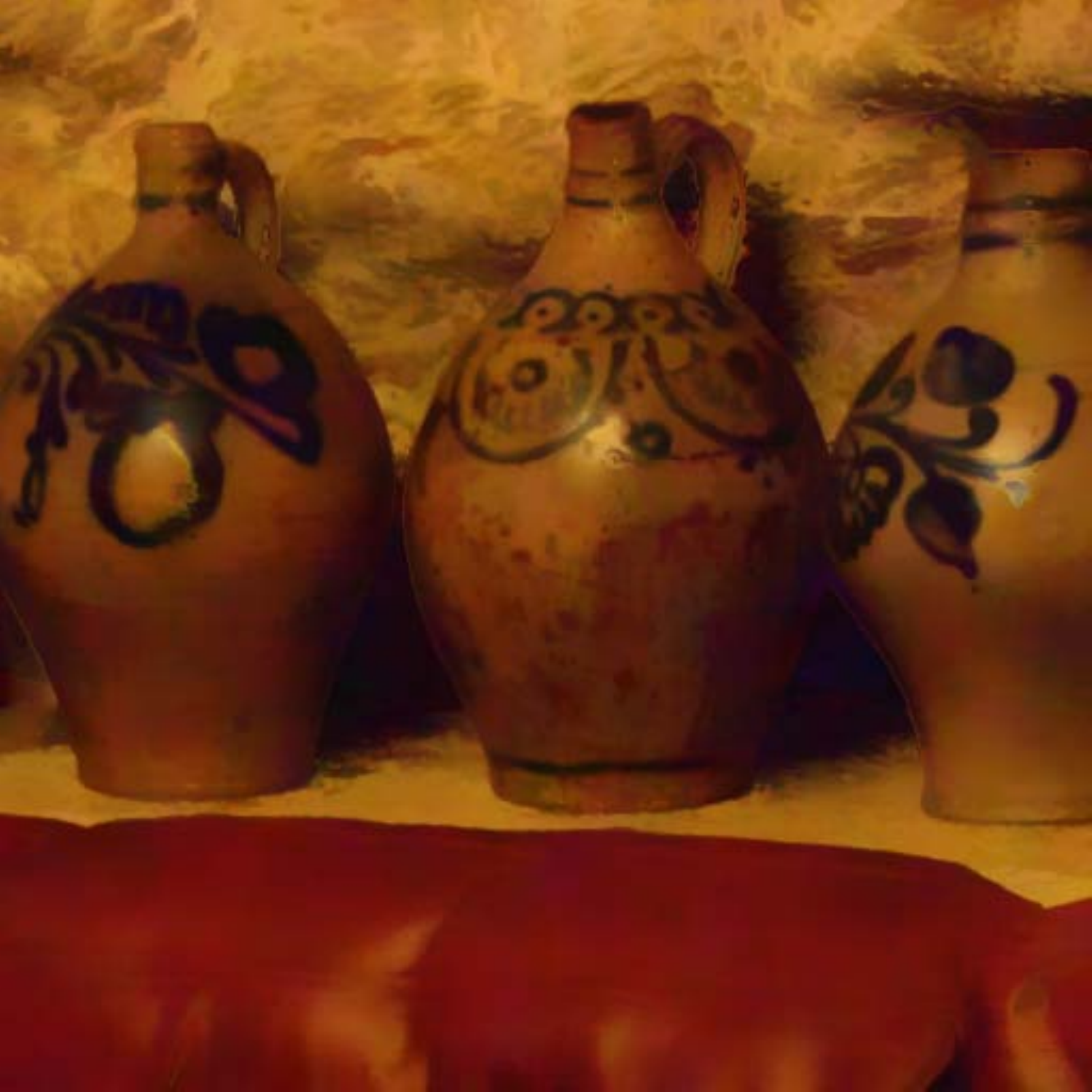}}
\subfigure[]{\includegraphics[height=1.5in]{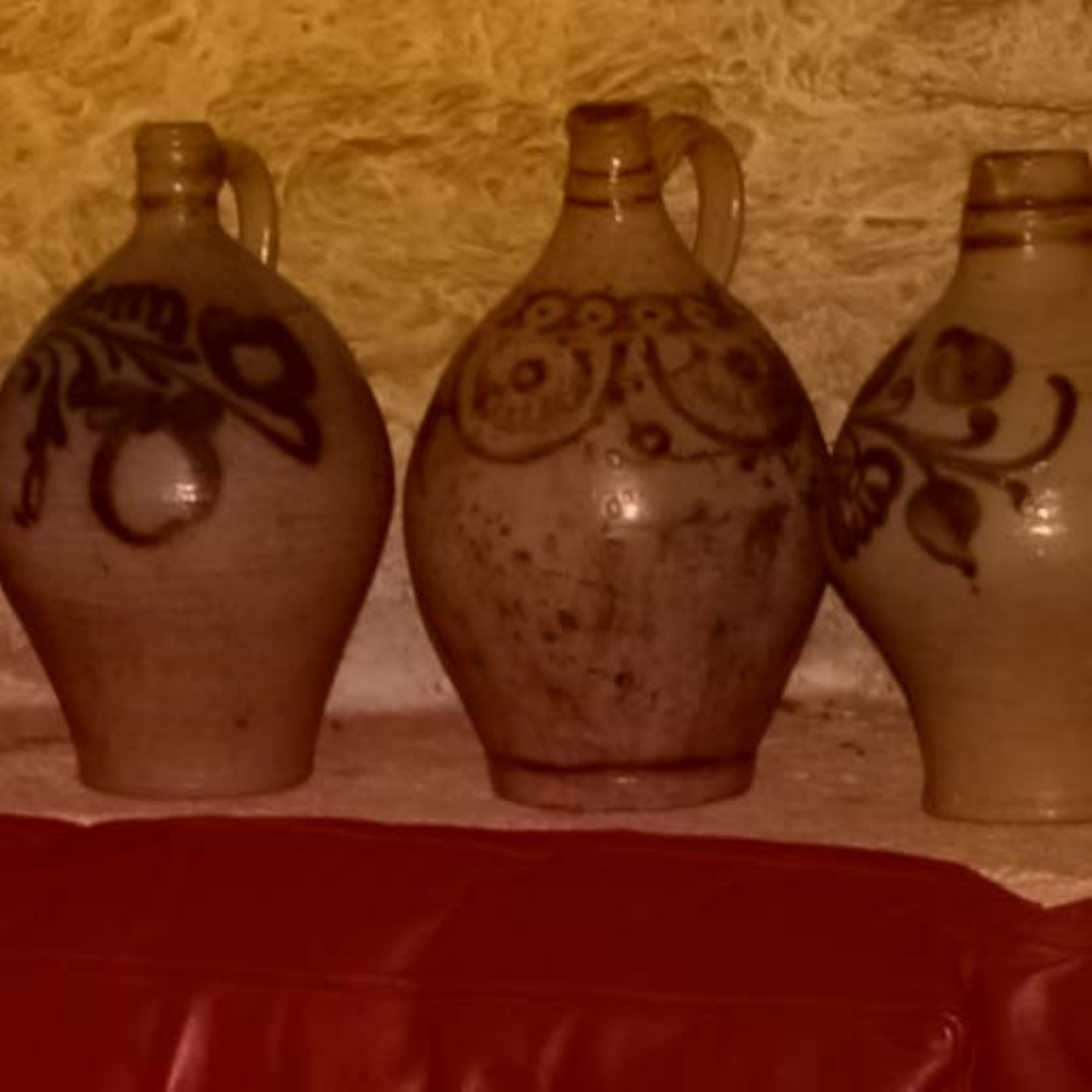}}
\caption{ (a) Noisy image. (b) Flash image. (c) Joint filtered image. (d) Recolored image.}\label{fig9}
\end{figure*}

The sparse norm filter can naturally incorporates a guidance or joint image \cite{Toyama:2004:TOG} to provide the filtering weight or diffusity. Below we show the result of flash/No-flash denoising taking the image using flash as the joint image to remove the noise in the non-flash image, using $p=0.2$, $r=11$ (Fig \ref{fig9}).

\subsection{Image Segmentation}
\label{sec4-6}

\begin{figure*}
\centering
{\includegraphics[height=1.1in]{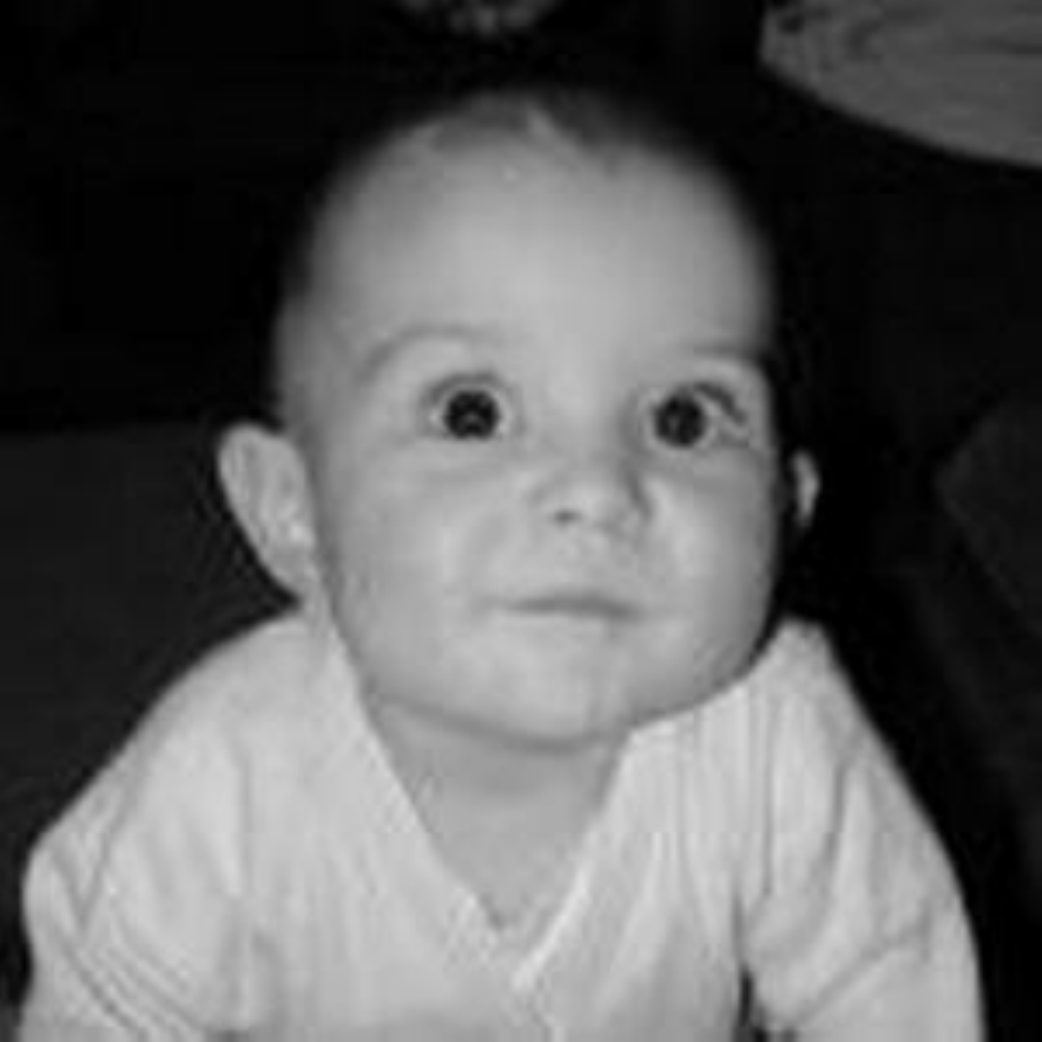}}
{\includegraphics[height=1.1in]{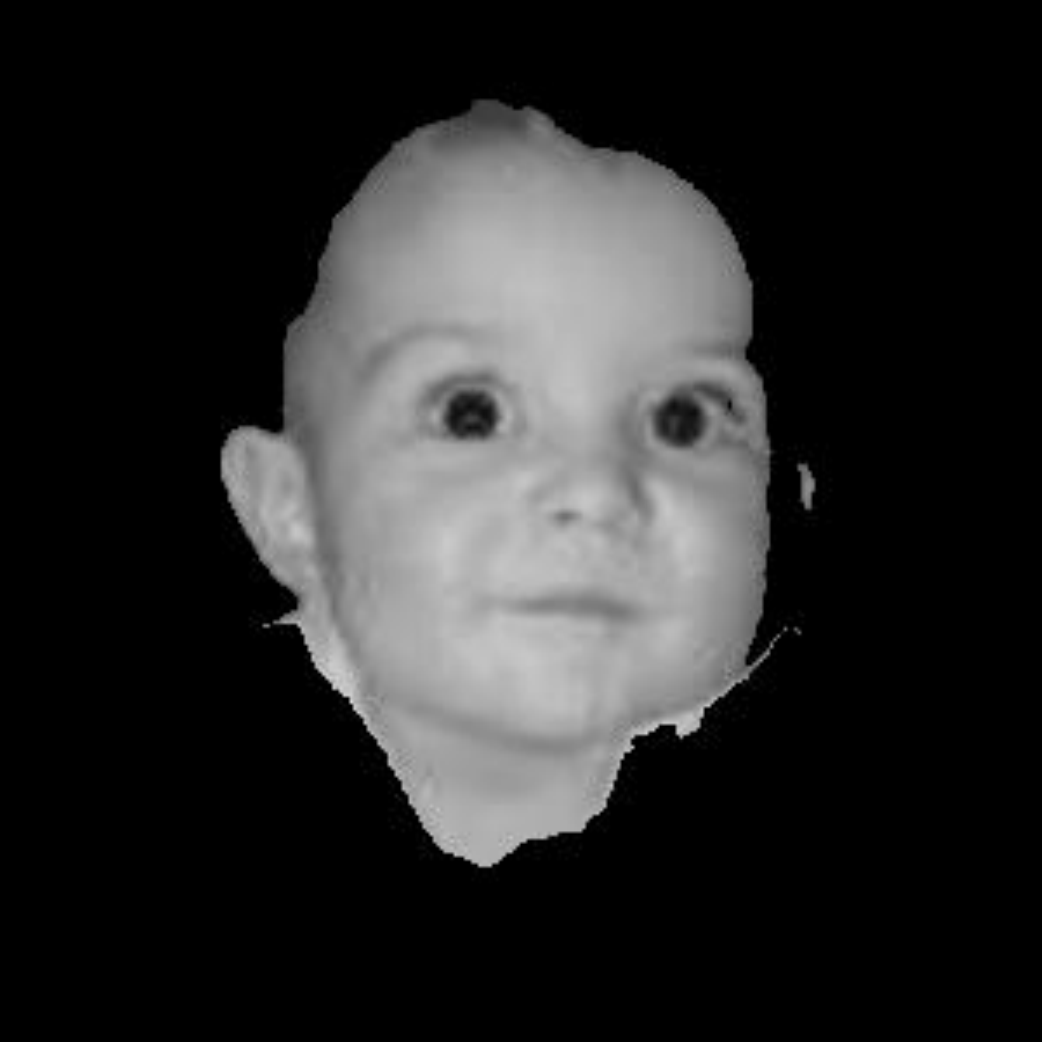}}
{\includegraphics[height=1.1in]{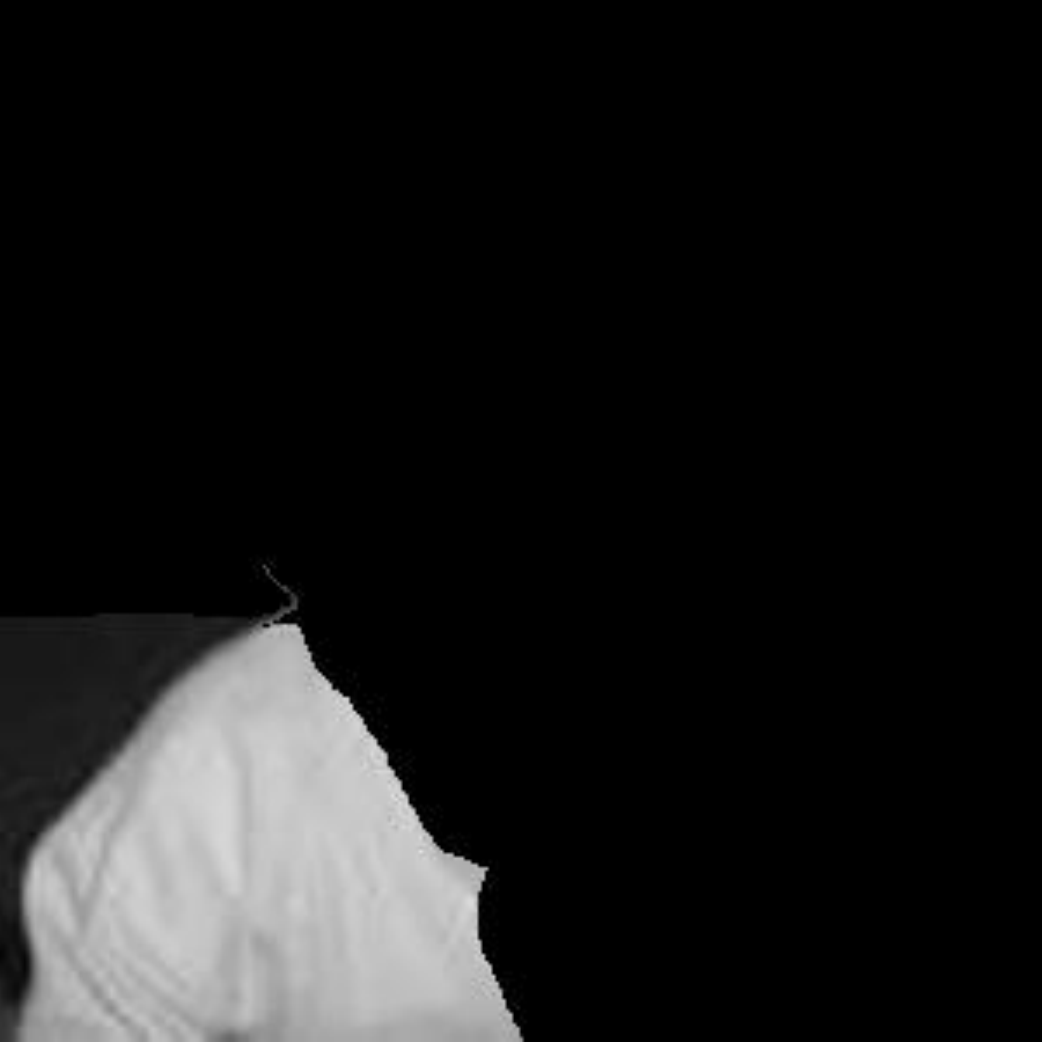}}
{\includegraphics[height=1.1in]{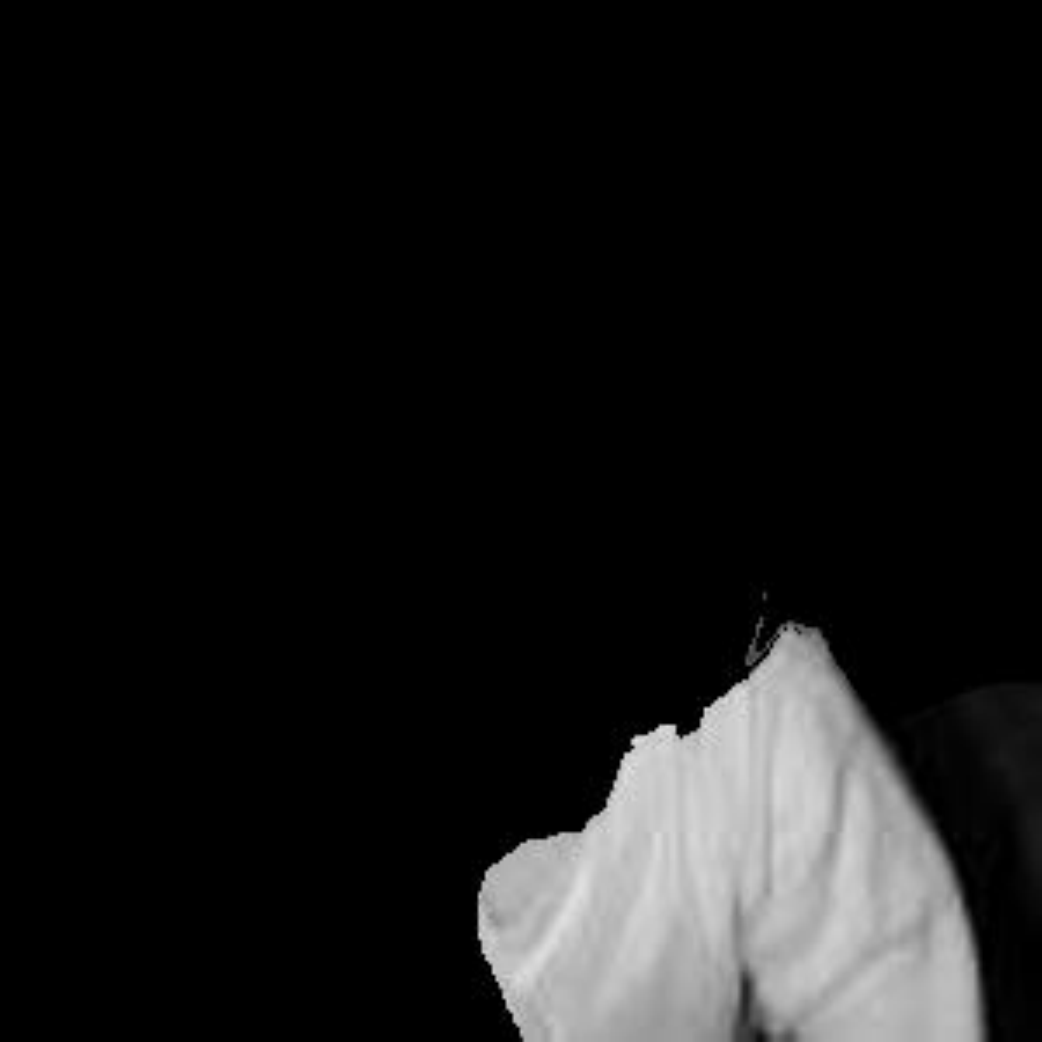}}
{\includegraphics[height=1.1in]{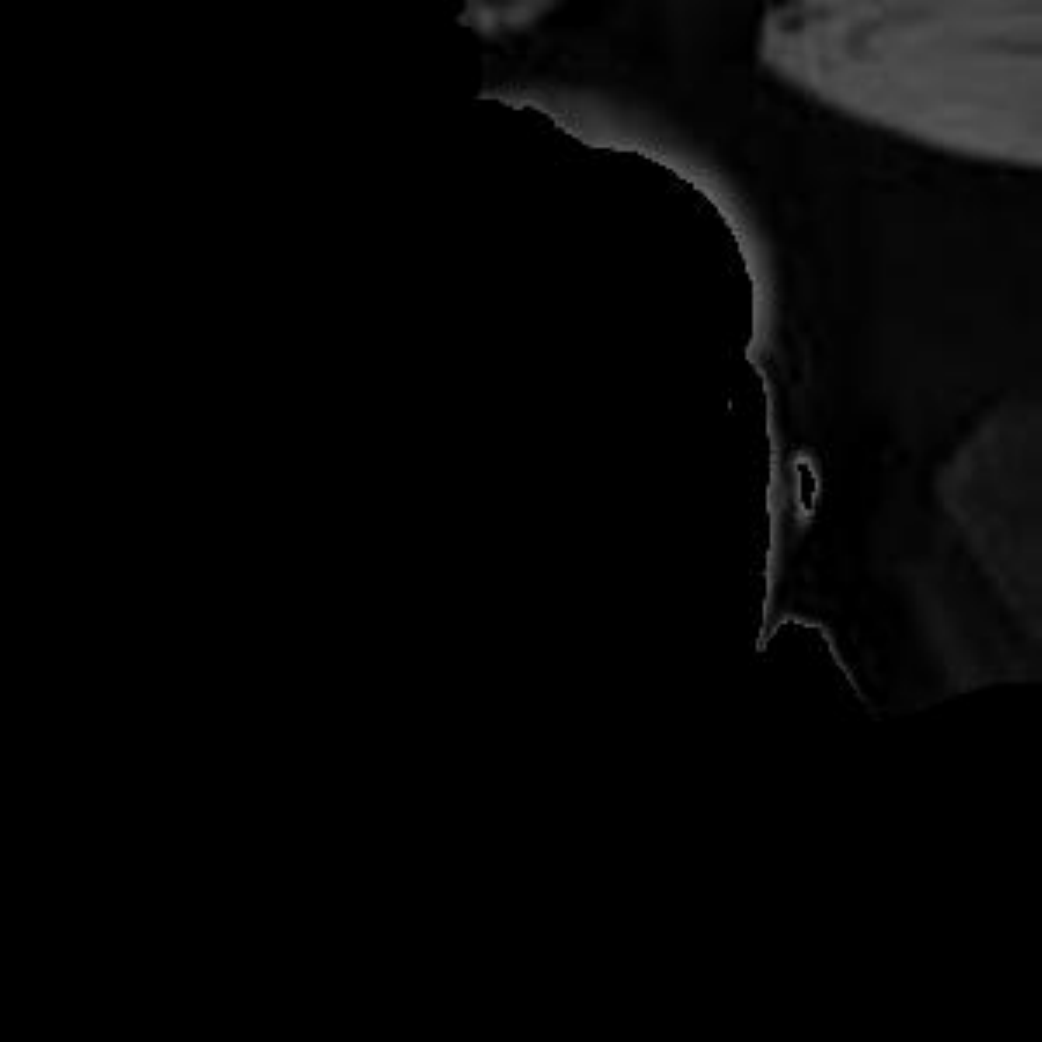}}
{\includegraphics[height=1.1in]{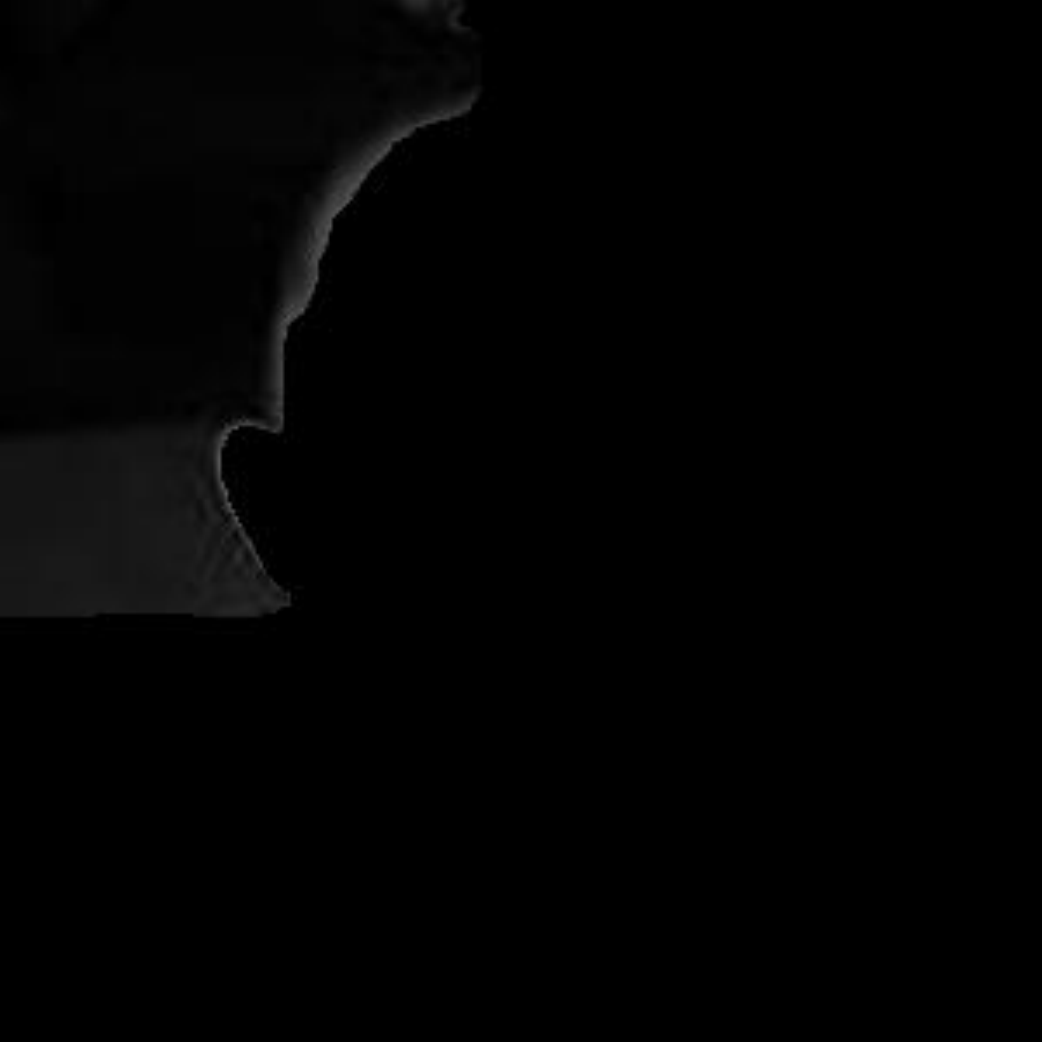}}
\caption{Column1: input image from [Shi and Malik 2009]. Columns 2-6: segments using the $p=0.3$, $r=1/16$ of the image.}\label{fig10}
\end{figure*}

The sparse norm is usually used to model the gradient profile of natural images in various computer vision models. In the following experiment (Fig \ref{fig10}) we show it can be used to accelerate the normalized cut \cite{Malik:2000:TPAMI} using the joint filtering techniques. Since normalized cut finds the eigenvectors of a diffusion/affinity matrix, we replace the slow matrix multiplication (which is quadratic to the neighborhood radius) in the eigensolver with our joint SNF which takes constant time with any neighborhood size \cite{Ye:2012:arXiv}. We use the original image as the guidance image and easily provides 10x-100x acceleration depending on the filtering radius. Moreover, we can extend the technique to explain and accelerate other normalized cut related algorithms \cite{Tang:2010:ECCV,Weiss:2004:TOG,Lischinski:2008:TPAMI,Levin:2008:TPAMI,He:2011:CVPR,He:2010:CVPR}.

\subsection{Colorization}
\label{sec4-7}

\begin{figure*}
\centering
\subfigure[]
{\includegraphics[height=1.3in]{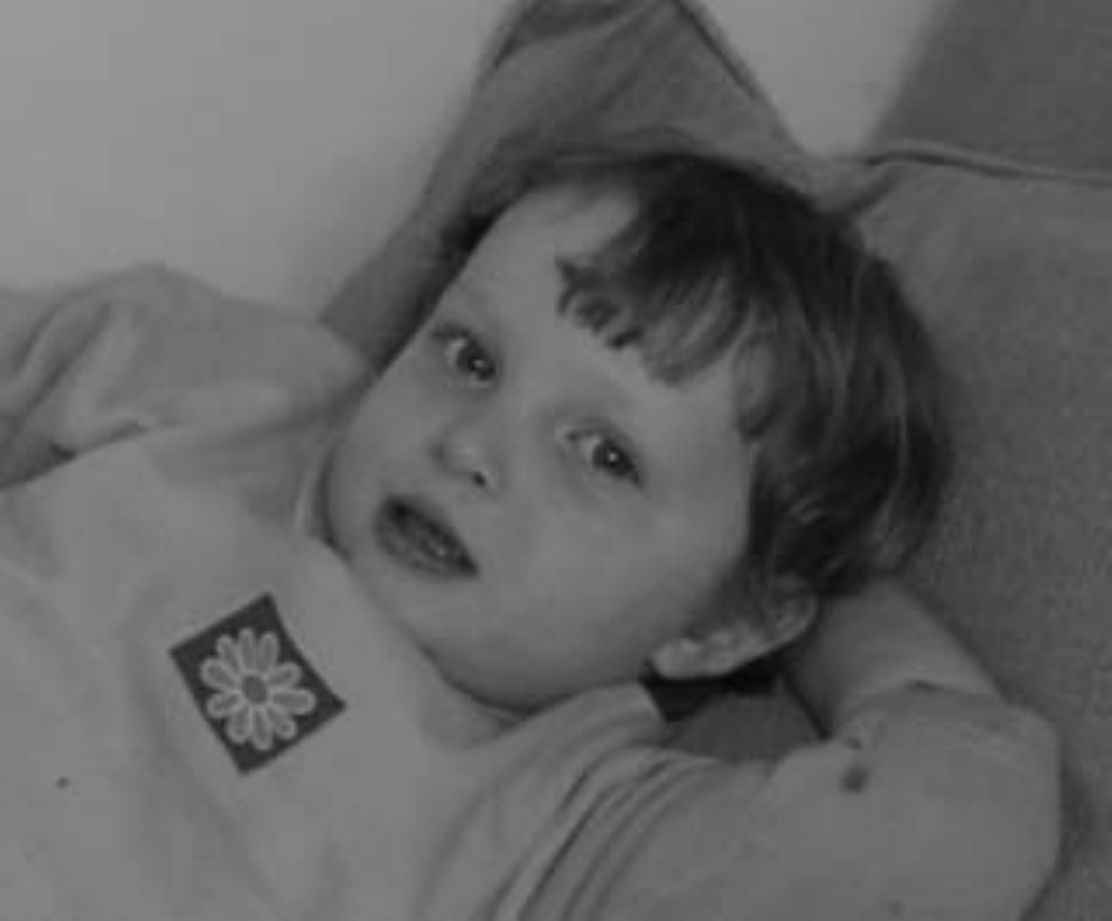}}
\subfigure[]{\includegraphics[height=1.32in]{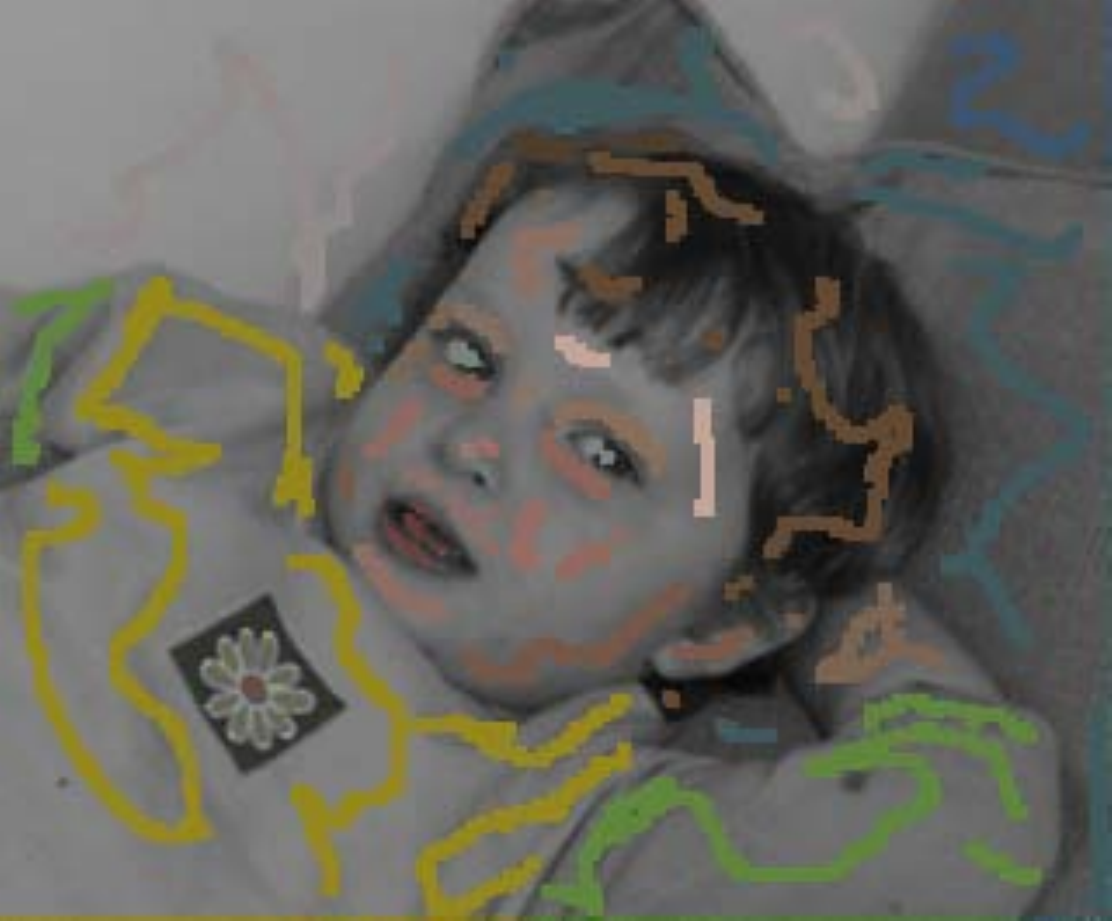}}
\subfigure[]{\includegraphics[height=1.32in]{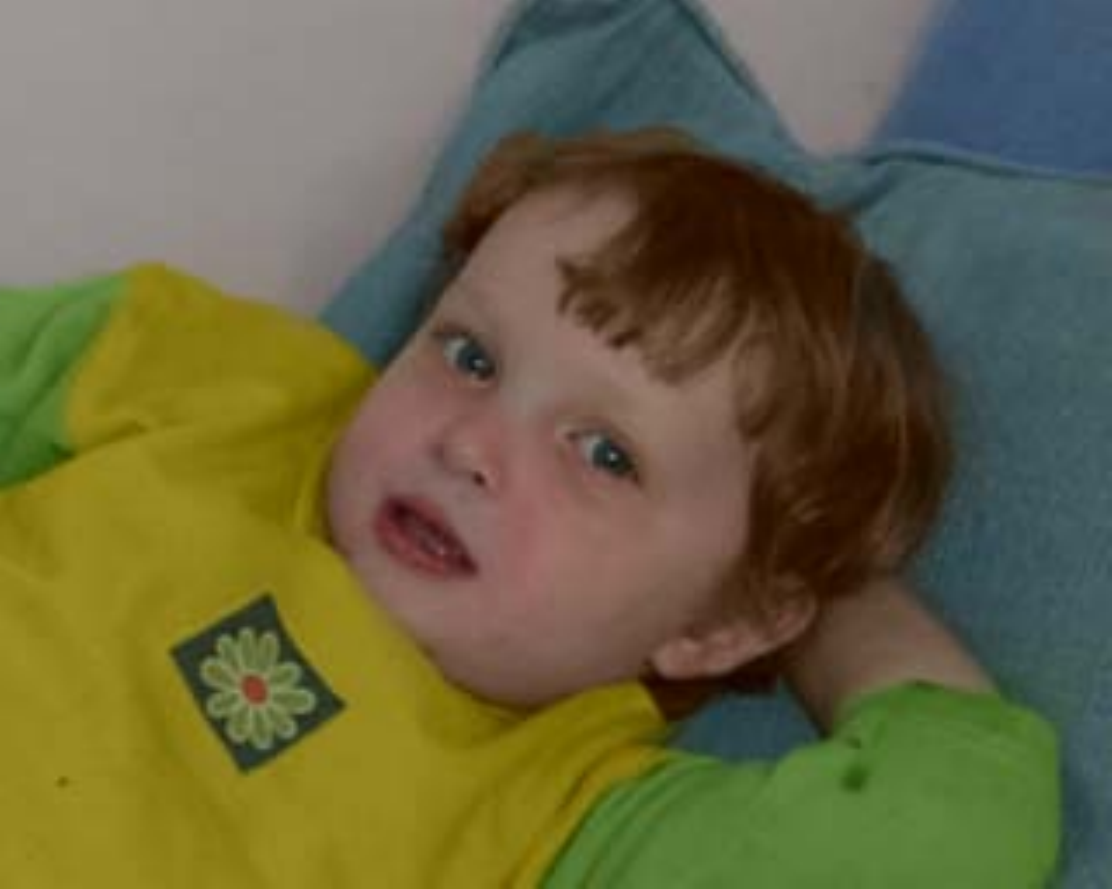}}
\subfigure[]{\includegraphics[height=1.32in]{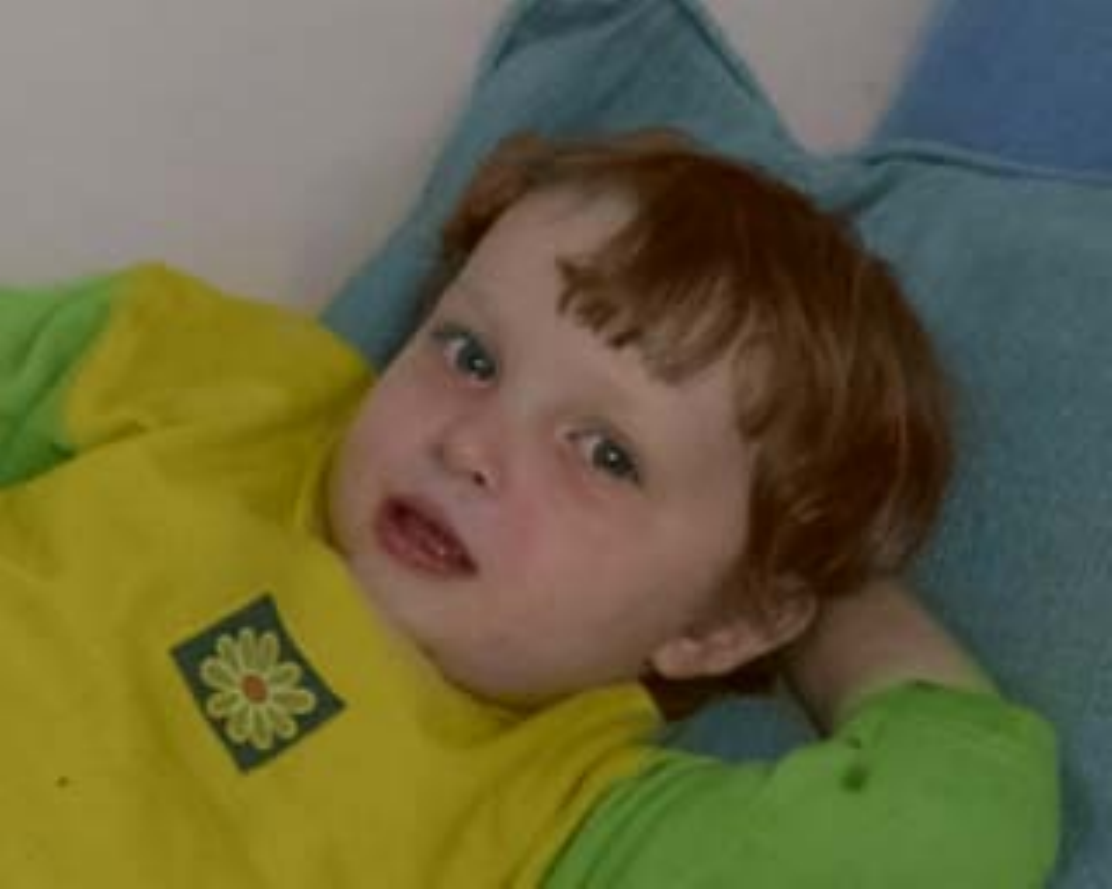}}
\caption{Colorization. (a) Input gray scale image. (b) Input color strokes. (c) Result by [Levin et al. 2004]. (d) Our result using $p=0.1$, $r=1/4$ of the image height.}\label{fig11}
\end{figure*}

We demonstrate an application in colorization \cite{Weiss:2004:TOG} as an example using joint filtering. Colorization can also be achieved by finding the stable distribution of an edge-preserving filter \cite{Ye:2012:arXiv}. The guiding weight of this filter is calculated from the gray scale image, and similar nearby pixels are assigned higher weights, which is naturally enforced using the sparse norm. To promise pixels similar in gray scale intensities are assigned similar colors, we use this guiding weight/diffusity to spread the color cues obtained from the input color strokes. We use a straightforward gradient descent algorithm to update the diffusion system. With less than 10 iterations, we can obtain high quality results (Fig \ref{fig11}(d)). This algorithm can also be used to re-color the flash image, shown Fig \ref{fig9}(d).

\subsection{Seamless Photo Editing}
\label{sec4-8}

This acceleration technique enabled by the sparse norm filter can be extended to the non-sparse norm. Seamless editing is a popular feature in image processing. Due to inconsistent color between the source and target, simple drag-and-drop editing is known to create artificial boundaries. The Poisson equation is widely used to seamlessly fill in a target region using a source region. In this framework, guided interpolation is conducted via solving $\Delta I=div (\nabla J)$  in the fill-in area  $\Omega$, subject to the Dirichlet boundary condition \cite{Blake:2003:TOG}. If we extend this equation by taking non-local gradients, the high dimensional Poisson equation, hints at a new system.
\begin{equation}
|N_{i}|I_{i}^{new}-\sum_{j\in N_i} I_{j}=|N_j|J_i -\sum_{j\in N_{i}}J_{j}
\label{eq10}
\end{equation}

\begin{figure*}
\centering
\subfigure[]
{\includegraphics[height=1.1in]{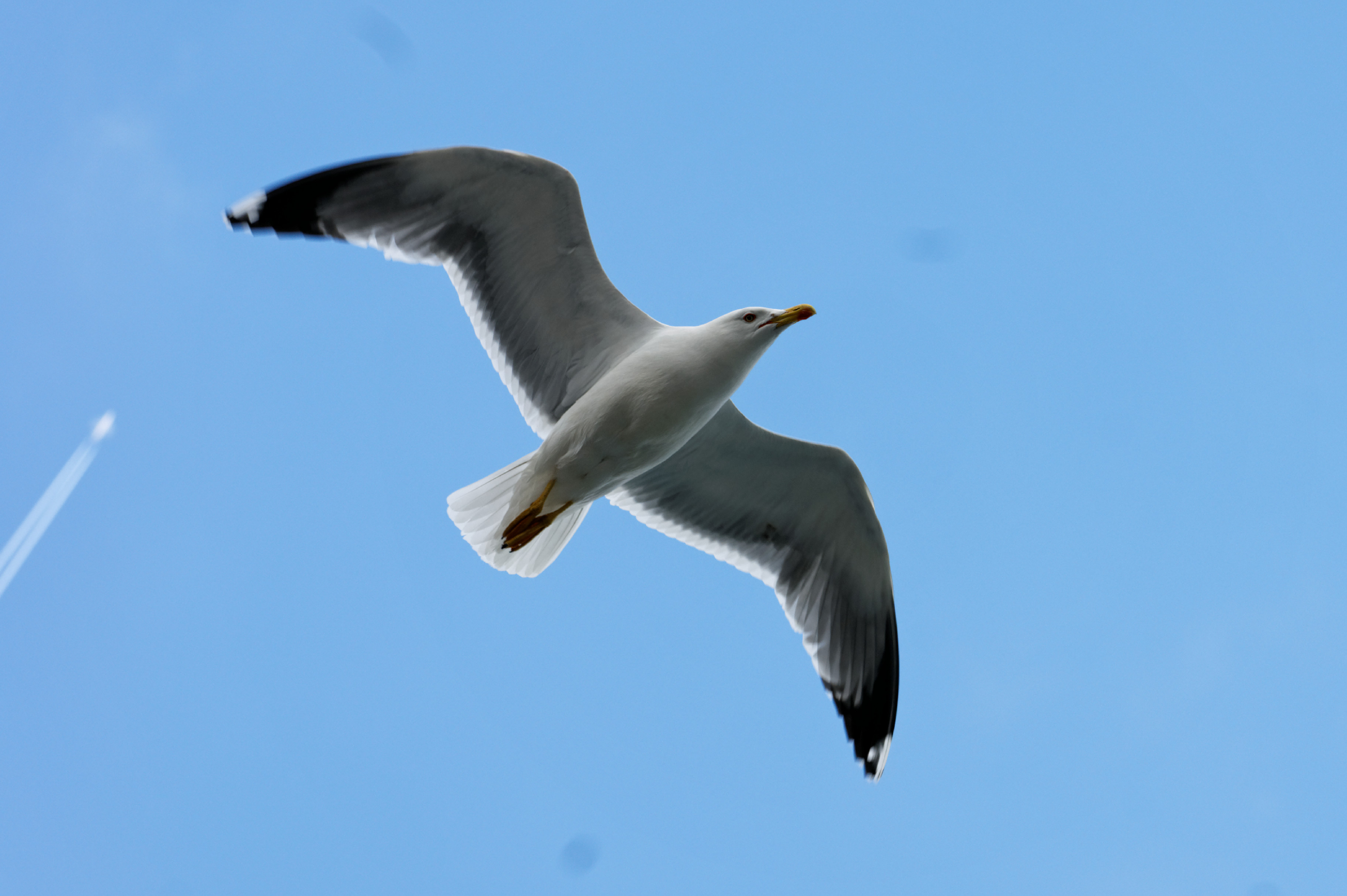}}
\subfigure[]{\includegraphics[height=1.1in]{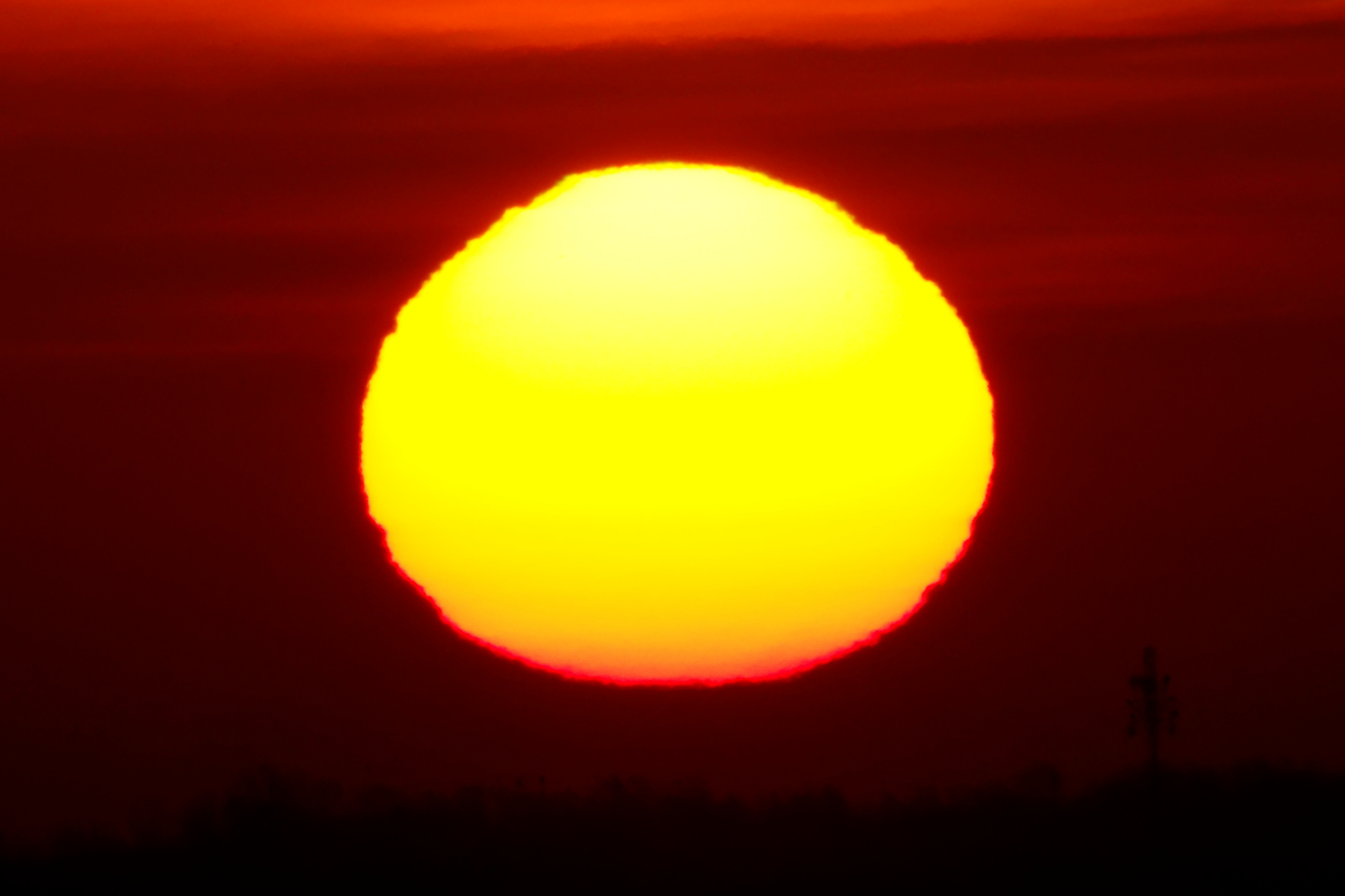}}
\subfigure[]{\includegraphics[height=1.1in]{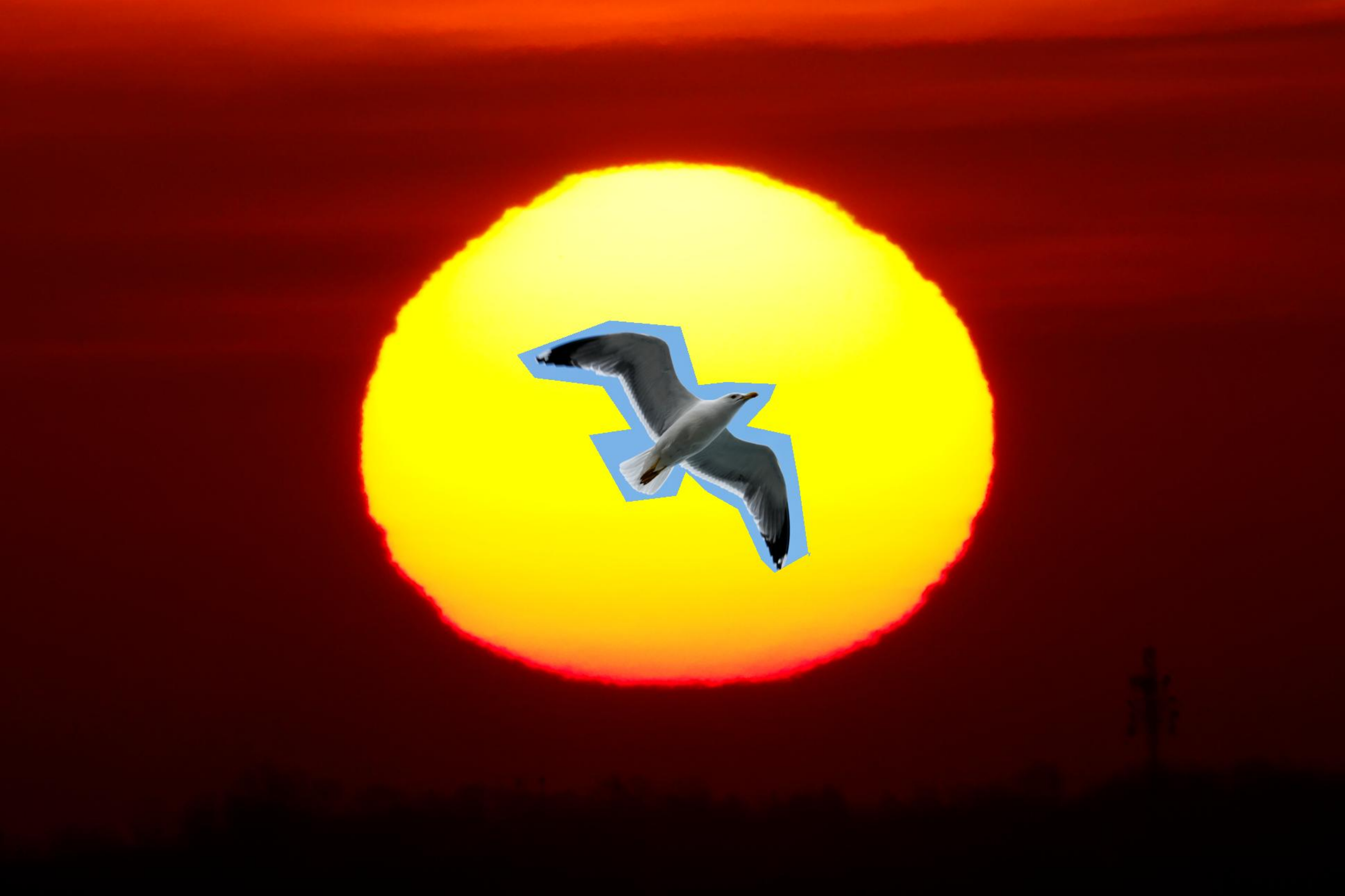}}
\subfigure[]{\includegraphics[height=1.1in]{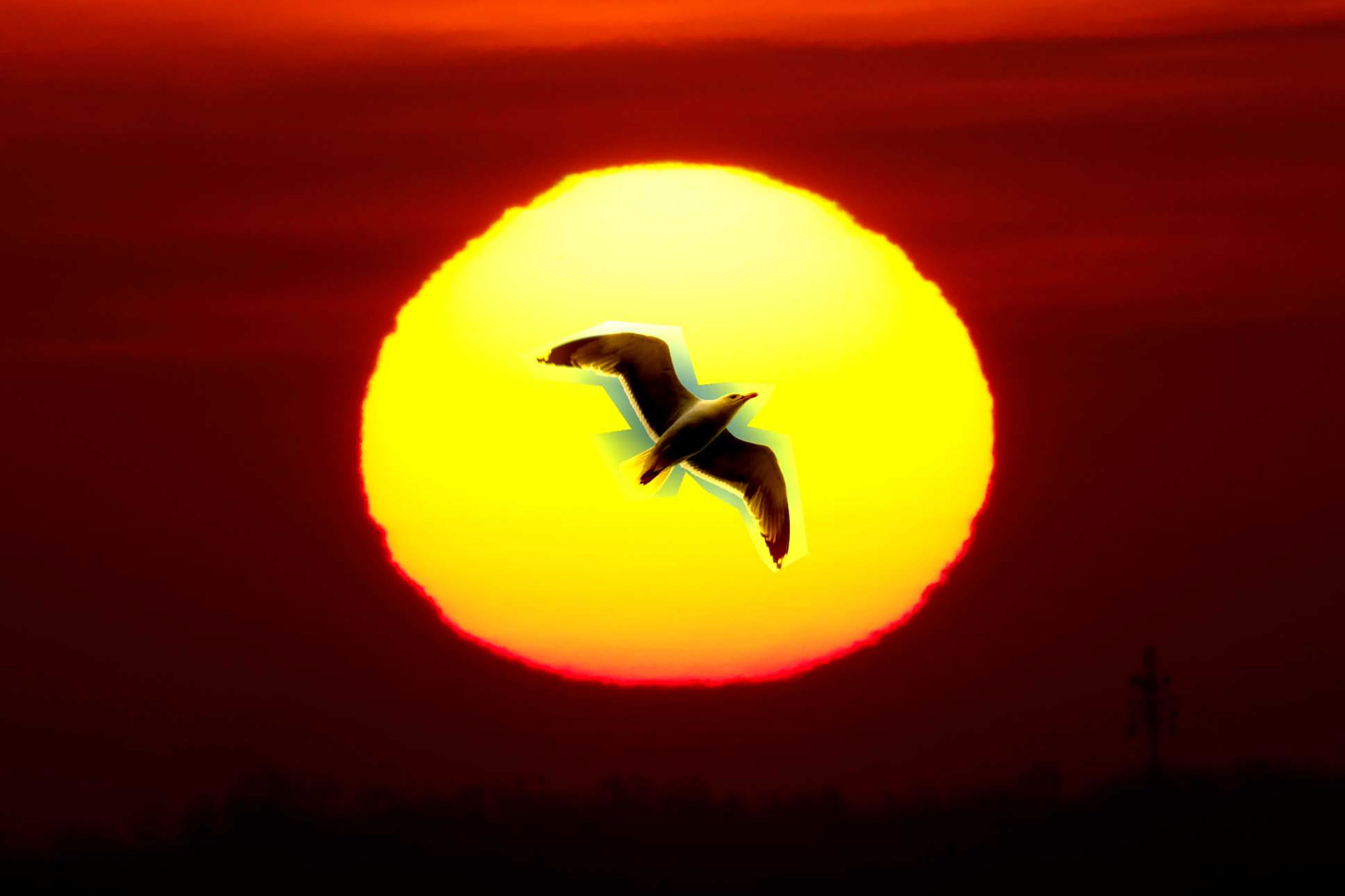}}
\subfigure[]{\includegraphics[height=1.1in]{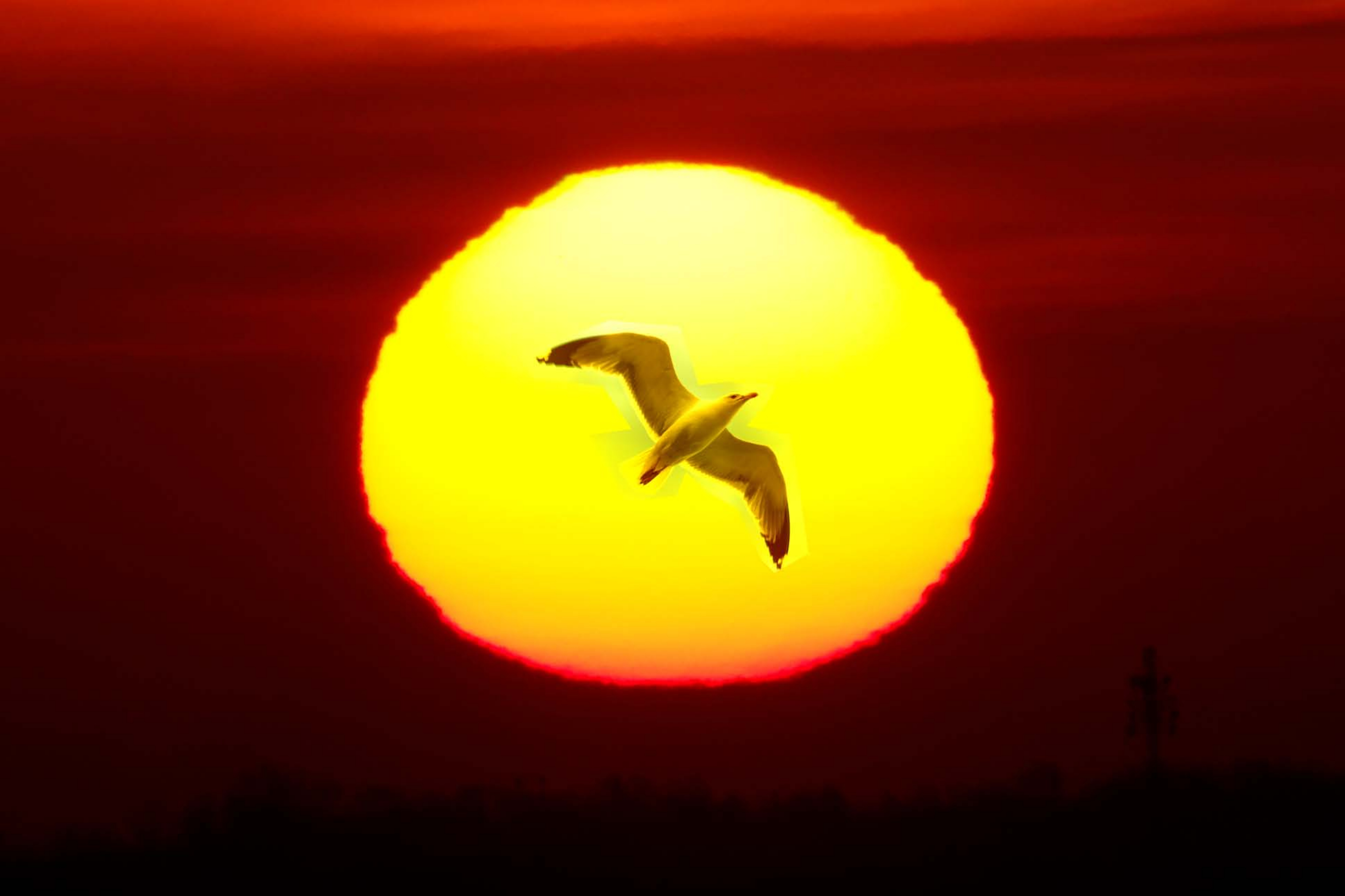}}
\subfigure[]{\includegraphics[height=1.1in]{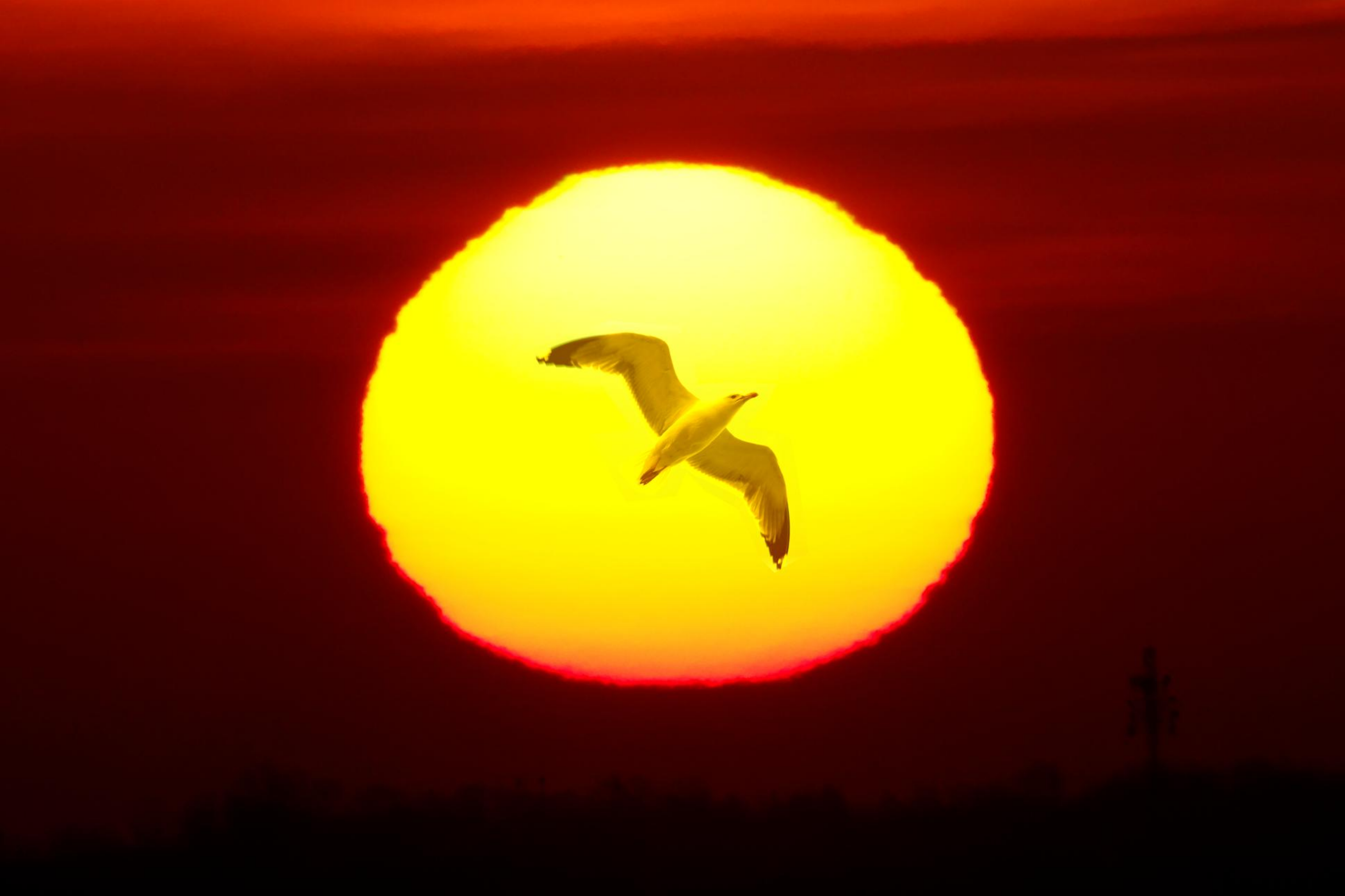}}
\subfigure[]{\includegraphics[height=1.1in]{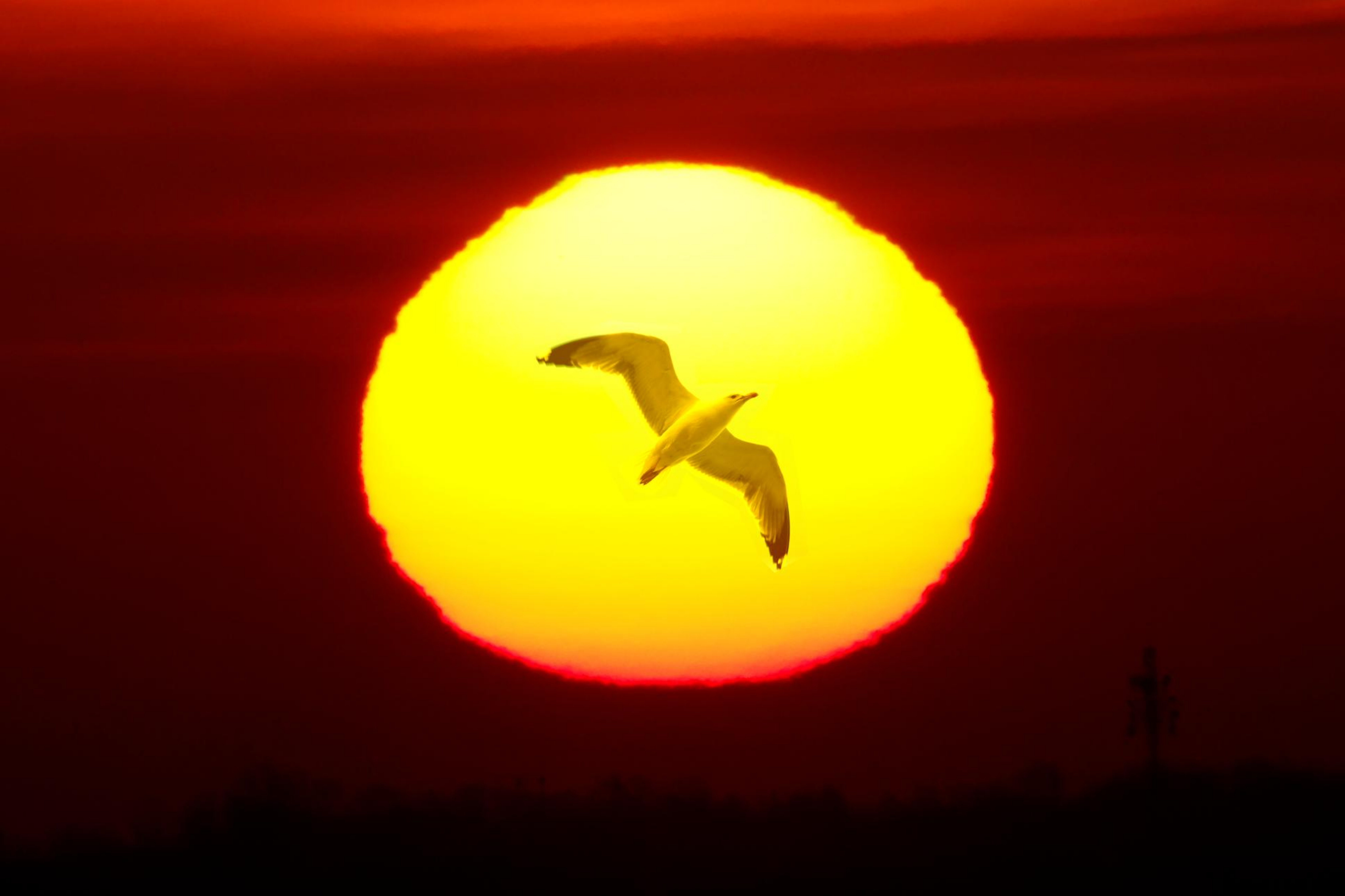}}
\caption{Seamless editing. (a) Source image. (b) Target image. (c) Drag-and-drop result. (d) 1st iteration of our algorithm. (e) 3rd iteration of our algorithm. (f) Our algorithm output. (g) [P\'{e}rez et al. 2003] output.}\label{fig12}
\end{figure*}

We also solve the above system also with the gradient descent algorithm. Since the diffusion is non-local, the algorithm converges within 10 iterations. Only $\sum_{j\in N_{i}}I_{j}$ needs to be updated in each iteration, which can be calculated using the box filter. We compare our algorithm with the original Poisson solver. In our environment, we use the backslash operation in Matlab to solve the Poisson equation, which takes 3 seconds per mega-pixel, excluding the time required to construct the sparse linear system. As reported above, the box filter takes only 0.04 seconds per mega-pixel in Matlab, or 0.01 seconds in C++. The results are comparable in quality (Fig \ref{fig12}).

\section{Conclusion}

In this work we present a simple but fundamental filter that builds connections with various classic smoothing techniques. The sparse norm filter can be regarded as a non-local extension of the optimization-based smoothing methods, which allows one-pass approximate solution via filtering. Through a variety of applications in image processing and computer vision, we demonstrate that the sparse norm filter gives new insights into popular applications and provides high quality accelerations.

\bibliographystyle{acmsiggraph}
\bibliography{template}
\end{document}